# Landscaping Systematic Mapping Studies in Software Engineering: A Tertiary Study

Muhammad Uzair khan, Salman Sherin, Muhammad Zohaib Iqbal, Rubab Zahid
Software Quality Engineering and Testing (QUEST) Laboratory,
National University of Computer and Emerging Sciences, Islamabad, Pakistan
{uzair.khan, salman.sherin, zohaib.iqbal, rubab.zahid}@questlab.pk

## Abstract

Context: A number of Systematic Mapping Studies (SMSs) that cover Software Engineering (SE) are reported in literature. Tertiary studies synthesize the secondary studies to provide a holistic view of an area.

Objectives: We synthesize SMSs in SE to provide insights into existing SE areas and to investigate the trends and quality of SMSs.

Methodology: We use Systematic Literature Review protocol to analyze and map the SMSs in SE, till August 2017, to SE Body of Knowledge (SWEBOK).

Results: We analyze 210 SMSs and results show that: (1) Software design and construction are most active areas in SE; (2) Some areas lack SMSs, including mathematical foundations, software configuration management, and SE tools; (3) The quality of SMSs is improving with time; (4) SMSs in journals have higher quality than SMSs in conferences and are cited more often; (5) Low quality in SMSs can be attributed to a lack of quality assessment in SMSs and not reporting information about the primary studies.

Conclusion: There is a potential for more SMSs in some SE areas. A number of SMSs do not provide the required information for an SMS, which leads to a low quality score.

## 1. Introduction

Software engineering is an active area of research with a large number of studies published every year. These studies include research papers presenting primary works, secondary studies that summarize and classify published research papers [5, 6], and tertiary studies that provide an assessment of secondary studies [8, 9]. Secondary research studies are well-established in software engineering and are useful in providing an overview and a mapping of the published papers. Tertiary studies provide an assessment of secondary studies and provide the researchers with an insight into the research landscape of a particular research area. Such studies help in the assessment and efficient interpretation of available knowledge. These studies also help in directing the focus of primary research by indexing the existing works and identifying potential gaps for further research. With the increase in the number of secondary studies in software engineering, there is a need of more tertiary research studies that aggregate and summarize the existing secondary studies. This aggregation of information is possible by conducting studies that focus on indexing, classifying, mapping and assessing the secondary studies.

Secondary studies in SE can be Systematic Literature Reviews (SLRs), systematic mapping studies (SMSs), and regular surveys. Regular surveys often suffer from selection bias and the results are often not repeatable as these are not conducted in a systematic way with an explicitly defined protocol [4]. An SLR aims to provide an in-depth review of the primary studies and to aggregate and describe the research methodologies and results [1, 2]. While SLRs bring several benefits including a well-defined systematic research protocol, these are often time-consuming and require a significant effort. Another type of systematic study – Systematic Mapping Studies have become quite popular among researchers. These studies, similar to SLRs, follow a systematic process and a well-defined research protocol to avoid biases, however unlike SLRs, the aim of SMSs is to answer broader sets of questions of a given area. There has been a rapid increase in the number of SMSs published in various SE disciplines in recent years. For example, over 90 SMSs have been published between 2015 and 2017, with a total of 210 published SMSs till August 2017. For a new researcher or practitioner, it is laborious to go through all published secondary studies in order to gain a high-level view of the research landscape in software engineering. Such a high-level view is well captured in tertiary studies.

Tertiary studies are a form of meta-studies and cover secondary studies. Such meta-level studies are necessary to investigate, classify, analyze and interpret existing secondary studies. These studies aid new researchers and practitioners in identifying the papers that have been published in a given area, the topics that have been addressed and the gaps and potential future research topics of the area. Tertiary studies help researchers and practitioners by reducing individual effort of gathering and summarizing relevant literature [2]. In our opinion, tertiary studies play an important role in supporting future research and providing evidence regarding the impact of secondary studies as well as the growing areas in SE. Such studies support new researchers by clearly identifying the research gaps and highlighting interesting future research directions. These studies also provide evidence to reinforce or reject widely held beliefs about the research domain under study.

In this paper, we present a tertiary study that covers the SMSs published in SE up to August 2017. The paper integrates and classifies all published SMSs and will serve as a high-level catalogue of research in SE for both researchers and practitioners. The paper identifies the current trends that are observed in the published SMSs, reports on the quality and demographics of the SMSs, identifies the top researchers in SE, analyze the trends relating to the primary studies, maps the existing literature to identify gaps and evaluates the impact of the papers (measured in number of citations and normalized citations). To assess the quality of the SMSs we have used Database of Abstracts of Reviews of Effects (DARE) criteria [15] by Centre for Reviews and Dissemination, University of York. We build a systematic map of the published SMSs by using the thematic analysis approach [16]. To provide insights on how existing literature maps to the standard SE curriculum, we map the identified SMSs on the knowledge areas defined by the Software Engineering Body of Knowledge (SWEBOK) [17]. An important aspect of an SMS is the quality assessment of existing studies. For this purpose, a number of quality assessment guidelines have been recommended in literature for assessing quality of an SMS. The most widely used and comprehensive [8, 18, 19] quality assessment guidelines in the area of SE is by Center for Reviews and Dissemination [15] . In our study, we also conduct a quality assessment of published SMSs according to the guidelines given in [15]. There are a number of guidelines available for conducting an SMS. We also identify the most-often cited guidelines for conducting an SMS to guide the researchers who are new to secondary studies. To date, this is the only study that aims to map all existing SMSs across the entire SE spectrum.

The rest of the paper is structured as follows: Section 2 presents the related work; Section 3 provides the details of the research method followed in this study. The Results of the study are presented in Section 4. Section 5 discusses the implication of our tertiary study for EBSE researchers and on software engineering education as well as practitioners. Section 6 provides a comparison of our study with the results of previous tertiary studies. The validity to threats are addressed in Section 7 and Section 8 provides limitations of the study. Finally, Section 9 discusses future directions and concludes the paper.

## 2. Related Work

Our paper is a tertiary study that follows the methodology of an SLR. In this section, we review the other tertiary studies published in SE and position our work.

Tertiary studies are conducted in the research areas when there are a large number of existing secondary studies that answer broader research questions [2]. The tertiary studies are designed to answer more coarse-grained questions and serve as indexes of published secondary studies [2]. A number of tertiary studies that are published in SE are presented in Table 1. We identified 17 tertiary studies in SE published between 2009 to August 2017. Out of these 17 studies, seven are broader studies that focus on SE in general, whereas ten tertiary studies cover specific sub-areas of SE.

The tertiary study by Kitchenham *et al.* [6] in 2009 is largely credited as the first tertiary study in SE. The study provides an overview of SLRs published in the domain of SE. The study looked at 19 published SLRs and provided guidelines for performing an SLR in SE.

Silva *et al.* performed a critical appraisal of SLRs in SE based on the research questions posed by studies. The study analyzed 53 SLRs and found that most of the research questions posed in the papers are exploratory and only a few studies contained causality questions [20]. The authors suggested a need for the reliable use of terminology to categorize the secondary studies and that the studies should follow well-defined reporting guidelines for assessment and comparison. Marques *et al.* [21] presented a tertiary study of SLRs in the sub-area of distributed software development. The study included 14 SLRs and identify the potential topics for SLRs to be conducted as well as the limitations of existing SLRs. The authors presented a classification of areas related to Distributed Computing in addition to the classification proposed by SWEBOK.

Hanssen *et al.* [22] reported results of a tertiary study in the area of Global Software Development (GSD). Their study includes 12 SLRs. The study emphasized the need of developing a common agenda for research in GSD community. Verner *et al.* [9] performed another tertiary study in the area of GSD that included 37 SLRs. The work identified project execution, environmental organization, and project planning and control as major topics covered by SLR in GSD. Moreover, the study mapped the geographic location of more prominent researchers in GSD.

Zhang *et al.* [23] performed a tertiary study covering 38 secondary studies in order to provide a systematic search strategy for identification of relevant literature in software engineering. The strategy helps to automatically search for relevant papers in different domain-specific conferences and journals. The strategy is evaluated on two case studies and the results show that the approach is effective to improve the process of searching for related papers while conducting an SLR. Cruzes and Dyba *et al.* [24] performed a tertiary study on data synthesis methods used in SLR. The paper reported that more than half of the SLRs did not contain data synthesis. Santos *et al.* [25] performed a tertiary study of 20 SLRs in the area of distributed software development (DSD) and identified the factors that impact the effectiveness of communication in DSD. Imtiaz *et al.* [26] reported experiences of performing SLR by covering 116 SLRs and identified the search process and data extraction and planning as the most challenging tasks while performing an SLR. Zhou et al. [27] conducted a tertiary study on the quality assessment conducted in SLRs. The work discusses the features of the quality assessment carried out in 127 SLRs and classifies the questions related to quality assessment. The study found that most of the SLRs do not mention the purpose of quality assessment and do not include it as a part of an SLR.

Goula *et al.* [18] conducted a tertiary study in the area of Model Driven Engineering (MDE). The study discussed the quality attributes used in MDE based on its review of 22 SLRs. The paper identified maintainability as the most widely studied attribute in MDE. Nurdiani *et al.* [28] carried out a tertiary study that covered 13 SLRs to analyze the impact of applying agile and lean development practices. The study found Test Driven Development as the most followed practice among the 14 practices found in literature. Hoda *et al.* [29] performed a tertiary study that included 28 SLRs published on ten different research topics in Agile Software Development (ASD). The study found that research and industry relevance in ASD still remains a challenge and more models are required to combine both the communities. Garousi *et al.* [19] conducted a tertiary study of 101 different SLRs in software testing and concluded that most of the works focus on model-based approaches.

Khan *et al.* [30] presented guidelines for improving the reporting quality of empirical studies in Model Based Testing (MBT). The authors analyzed the reporting quality of 87 empirical studies and found that most of the empirical studies fail to follow the existing reporting guidelines. Furthermore, the study proposes MBT specific reporting guidelines to support reproducibility of the results for empirical studies in MBT.

Peterson *et al.* [10] provides an updated version of the guidelines for conducting an SMS and mapped 52 SMSs up to the year 2012. The study identifies significant improvements in the process of systematic mappings, the topics covered by SMSs and explored other demographic information. The paper answers questions related to the search protocol and reporting structure of SMSs with the aim of providing guidelines. The paper assesses the search strategies, the activities performed, criteria used for quality assessment and diagrams used for visualization of information in different SMSs. In our study, we aim to answer a different set of research questions that focus more on current trends and research gaps in SE as well as demographic information, rather than focusing on developing guidelines for mapping studies.

Table 1. An overview of existing tertiary studies.

| S.no | Title | Included secondary studies | Year | Ref |
|---|---|---|---|---|
| 1 | Systematic literature reviews in software engineering | 20 | 2009 | [6] |
| 2 | Critical appraisal of SLRs in SE from the perspective of the research questions | 53 | 2010 | [20] |
| 3 | Signs of Agile Trends in Global Software Engineering Research | 12 | 2011 | [22] |
| 4 | Identifying relevant studies in SE | 38 | 2011 | [23] |
| 5 | Research synthesis in software engineering | 49 | 2011 | [24] |
| 6 | Systematic Literature Reviews in Distributed Software Development | 14 | 2012 | [21] |
| 7 | Systematic Literature Reviews in Global Software Development | 37 | 2012 | [9] |
| 8 | Communication in Distributed Software Development Projects | 20 | 2012 | [25] |
| 9 | Experiences of Conducting Systematic Literature Reviews in Software Engineering | 116 | 2013 | [26] |
| 10 | Risks and risk mitigation in global software development | 37 | 2013 | [31] |
| 11 | Quality Assessment of Systematic Reviews in Software Engineering | 127 | 2015 | [27] |
| 12 | Guidelines for conducting systematic mapping studies in software engineering: An update | 52 | 2015 | [10] |
| 13 | Quality in model-driven engineering | 22 | 2016 | [18] |
| 14 | The Impacts of Agile and Lean Practices on Project Constraints | 13 | 2016 | [28] |
| 15 | A systematic literature review of literature reviews in software testing | 101 | 2016 | [19] |
| 16 | Systematic literature reviews in agile software development: a tertiary study | 28 | 2017 | [29] |
| 17 | Empirical studies omit reporting necessary details: A systematic literature review of reporting quality in model based testing | 87 | 2017 | [30] |

Our tertiary study is significantly different from the above mentioned tertiary studies in many ways. Mainly, these tertiary studies are conducted in specific areas (e.g. Agile Software Engineering and MDE) and their scope is limited to specific sub-areas in SE. Unlike others, the scope of our study is much broader and includes all the SMSs in different areas of SE. Consequently, we pose a different set of questions to build an overall map of the domain and to identify key trends by including the papers up to August 2017. Our study provides results and insights on a higher level of abstraction than the already published tertiary studies due to being broader in nature. Moreover, we assess the quality of each included SMS and provide in-depth analysis and insights on the quality of SMSs published in journals and in conferences. We build a systematic map of SMSs in SE using both a thematic analysis and by mapping the identified results on SWEBOK knowledge areas. We provide insights on trends observed in the published SMSs to guide new researchers in Evidence based Software Engineering (EBSE) domain. We also discuss implications of identified trends on SE education and for SE practitioners. Finally, we compare our findings with the findings of previous studies and discuss where our results confirm previously published results and where they differ.

## 3. Research Method

Our paper is a tertiary study that follows the same methodology as an SLR, following the guidelines proposed by Kitchenham *et al.* [2] and Peterson *et al.* [1]. These guidelines are widely accepted in the *evidence-based software engineering* (EBSE) community to conduct SMSs and SLRs in SE [21] [19] [31].

Figure 1(a) shows the review protocol followed in this study. We define six phases for completing this study. In phase 1, we formulate the research questions according to the stated objectives of the study. In phase 2, we define our search strategy by identifying different keywords and the relevant online databases. In phase 3, we apply the inclusion and exclusion criteria for article selection. In phase 4, quality assessment of selected SMSs is performed using the widely used DARE guidelines [15]. In phase 5, we identified the attributes and extracted the relevant information to build our systematic classification map. Finally, in phase 6, we synthesize the data and report our results. Figure 1(b) shows our complete search process. Following, we discuss the various phases of the study.

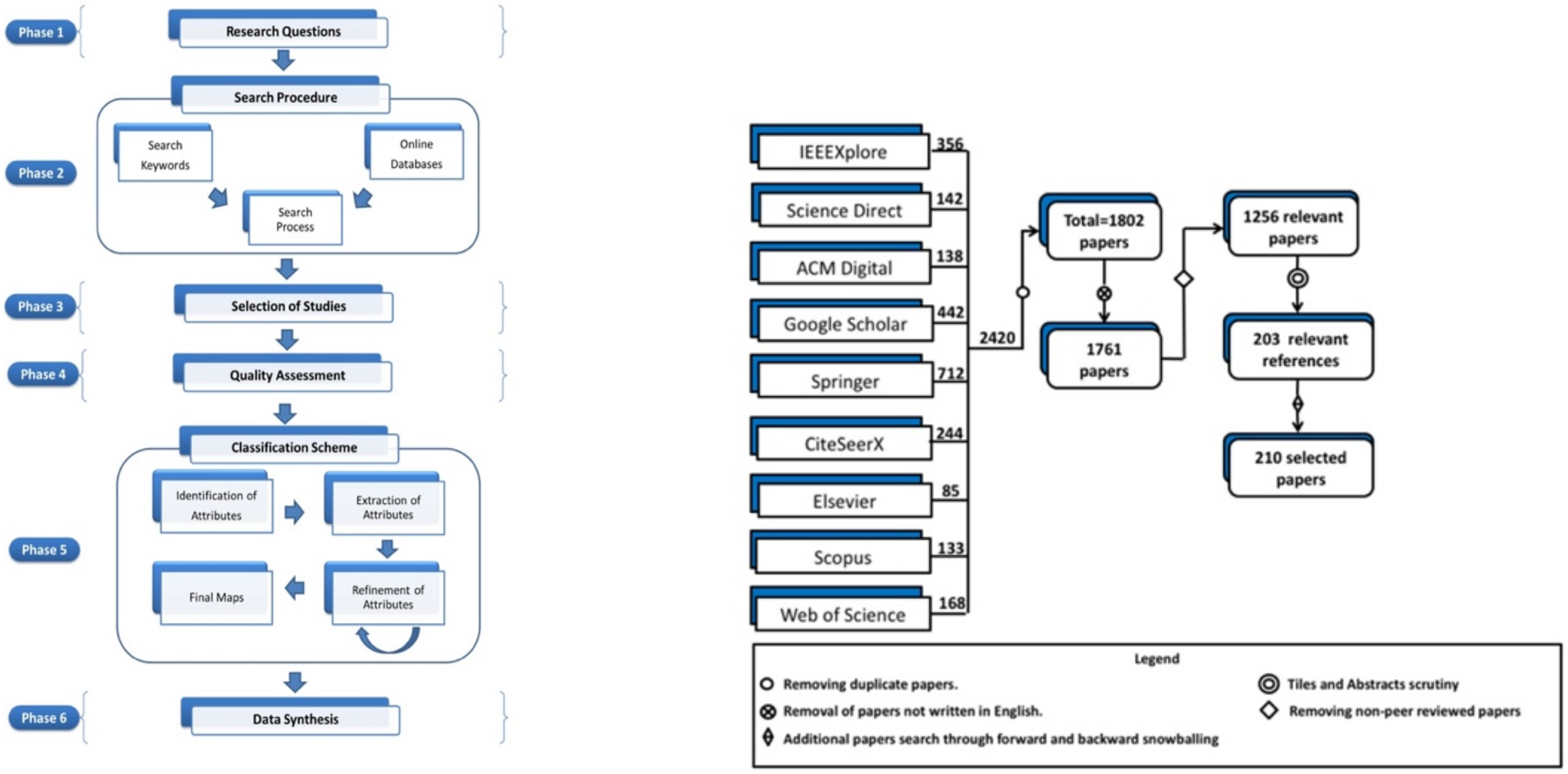

(a) (b)

*Figure 1. (a) Review Protocol, (b) Search Process*

### 3.1. Research questions

We formulate our research questions in light of the objectives of our study and after comprehensive review of research questions in other tertiary studies, such as [8, 29, 31]. Table 2 presents our research questions.

Table 2. Research questions

| SNo. | Research questions |
|---|---|
| RQ1 | **Which areas in SE are addressed by SMSs?**<br>**Motivation:** This research question aims to classify the existing SMSs to identify the targeted areas. The findings of the question will help to identify gaps for researchers to conduct systematic mapping studies in the areas that are overlooked by existing studies. It will help practitioners to identify relevant SMSs conducted in an area. To answer this question, we break it down into two sub questions, in the first sub-question, we identify different themes and categories covered by the included SMSs by reviewing their title, abstract and keywords, following the widely accepted guidelines [1, 3, 7]. In the second sub-question, we use the standard SWEBOK classification and map the studies according to the knowledge areas described in it. |
| RQ1.1 | Which areas in SE are addressed by SMSs based on a thematic analysis? |
| RQ1.2 | Which areas in SE are addressed by SMSs based on the SWEBOK classification of knowledge areas in SE? |
| RQ2 | **What are the trends relating to quality of published SMSs?**<br>**Motivation:** We evaluate the quality of SMSs as per the quality assessment guidelines by DARE criteria [15, 27]. The quality of SMSs is determined by the quality of all the activities followed in the protocol, i.e., search for relevant papers, selection of studies, synthesis of data, quality assessment of included studies, and review of the included studies. A large number of SMSs are published in the literature on different topics of SE since 2007. However, one of the main concerns in systematic mappings is the level of certainty to be placed in the recommendations and conclusions drawn from an SMS [27]. One of the approach to increase the level of certainty of the results of an SMS is to meticulously measure its quality. Therefore, this research question aims to rigorously assess the quality of systematic mapping studies in SE [27]. To analyze quality and answer this question, we break the main research question down into further sub questions. |
| RQ2.1 | Is the quality of published SMSs improving over time? |
| RQ2.2 | Do the SMSs published in Journals have better quality than SMSs published in conference papers? |
| RQ2.3 | What are the strengths and weaknesses of published SMSs in terms of quality? |
| RQ2.4 | How does the quality of SMSs vary with publication venues? |
| RQ2.5 | What is the distribution of quality w.r.t different SE areas? |
| RQ3 | **What are the current trends in SMSs relating to guidelines, data sources, types of questions and number of included studies?**<br>**Motivation:** This research question aims to identify most followed guidelines for conducting systematic mapping studies, the online databases that are used for searching primary studies, the types of questions posed in SMSs, and the number of included primary studies in each SMS. To answer this question, we break it down into more refined questions. |
| RQ3.1 | Which guidelines are most frequently cited for conducting SMSs? |
| RQ3.2 | Which online databases are most frequently included as sources? |
| RQ3.3 | What types of research questions are most frequently addressed by SMSs? |
| RQ3.4 | How many primary studies are included on average by published SMSs? |
| RQ3.5 | Is there an increase in the number of included studies over the years? |
| RQ4 | **What are the demographics of published SMSs in SE?**<br>**Motivation:** This research question will help to identify the key researchers, the top venues and the highly cited papers in SE. To answer this question, we break it down into the following sub questions. |
| RQ4.1 | How many SMSs are published annually? What is the publication trend? |
| RQ4.2 | Which venues publish SMSs most frequently? |
| RQ4.3 | Which are the most cited papers in the area? |
| RQ4.4 | Are SMSs published in journals cited more often than SMSs published in conferences? |
| RQ4.5 | Who are the active researchers with most published SMSs? |

## 3.2. Search procedure

The search procedure used in this research is as follows:

### 3.2.1. Search strings

We use the guidelines provided by Kitchenham *et al.* [2] to finalize search keywords and the search string as follows:

- Derivation of keywords from research questions, relevant papers and books
- Identification of synonyms and alternate spellings
- Using Boolean operator, OR, for synonyms and alternative spellings
- Using Boolean operator AND to connect search keywords

We slightly altered the search strings according to the format required by each online database. Table 3 provides the search strings for each of the ten included databases.

Table 3. Search strings for each online database

| Search engine | Query string |
|---|---|
| **Google scholar** | "Systematic mapping" OR "systematic map" OR "mapping study" OR "systematic mapping study" OR "literature review" OR "scoping study" OR "systematic mapping review" OR "systematic review" AND "software" |
| **ACM portal** | ("Systematic mapping" OR "systematic map" OR "mapping study" OR "systematic mapping study" OR "literature review" OR "scoping study" OR "systematic mapping review" OR "systematic review") OR AND ("software") |
| **IEEExplore** | ((((((((("Abstract":"systematic mapping") OR "Abstract":"systematic map") OR "Abstract":"mapping study") OR "Abstract":"systematic mapping study") OR "Abstract":"scoping study") OR "Abstract":"systematic mapping review") OR "Abstract":"systematic review") AND "software") NOT "Publication Title":"systematic literature review") |
| **CiteSeerX Library** | (abstract: "Systematic mapping" OR abstract: "systematic map" OR abstract:" mapping study" OR abstract:" systematic mapping study" OR abstract: "literature review" OR abstract: "scoping study" OR abstract:" systematic mapping review" OR abstract: "systematic review") AND (text: "software") |
| **Elsevier** | (("Systematic mapping" OR "systematic map" OR "mapping study" OR "systematic mapping study" OR "literature review" OR "scoping study" OR "systematic mapping review" OR "systematic review") AND "software") |
| **Science Direct** | ((Systematic mapping OR systematic map OR mapping study OR systematic mapping study OR literature review OR scoping study OR systematic mapping review OR systematic review)) and (software). |
| **Wiley** | ("Systematic mapping" OR "systematic map" OR "mapping study" OR "systematic mapping study" OR "literature review" OR "scoping study" OR "systematic mapping review" OR "systematic review") AND ("software") |
| **Scopus** | ("Systematic mapping" OR "systematic map" OR "mapping study" OR "systematic mapping study" OR "literature review" OR "scoping study" OR "systematic mapping review" OR "systematic review") AND ("software") |
| **Web of Science** | ("Systematic mapping" OR "systematic map" OR "mapping study" OR "systematic mapping study" OR "literature review" OR "scoping study" OR "systematic mapping review" OR "systematic review") AND ("software") |
| **Springer Link** | ("Systematic mapping" OR "systematic map" OR "mapping study" OR "systematic mapping study" OR "literature review" OR "scoping study" OR "systematic mapping review" OR "systematic review") AND ("software") |

### 3.2.2. Online Databases

We searched in a total of ten electronic databases to extract relevant SMSs. These include: Google scholar, ACM Digital Library, IEEEXplore, CiteseerX, Elsevier, Science Direct, Wiley, Scopus, Web of Sciences, and Springer Link. The search includes title, abstract, and keywords of published papers. These databases form a comprehensive set as they are likely to index all the published SMSs and are identified from published secondary and tertiary studies.

### 3.2.3. Search process

Systematic search of databases for identifying relevant studies is a key activity in a systematic review. First we searched major online databases to retrieve an initial set of relevant studies and then we applied our inclusion and exclusion criteria for further selection of relevant studies. Additionally to reduce the risk of missing relevant studies, we then applied the snowballing technique to our selected set of studies to identify additional relevant studies. Snowballing is the process of identifying relevant papers from the references and citations of the existing study pool. The newly identified studies were then filtered based on our inclusion and exclusion criteria. The same process was then iteratively applied on newly identified papers until all the relevant papers were identified. We followed the guidelines presented by Wohlin *et al.* to conduct the process of snowballing [32]. In backward snowballing, we examined the references of the selected studies. For example, S15 has a total of 84 references in which there was one mapping study, which was not included in our selected pool of SMSs. Therefore, we added the new study to our selected pool for the next iteration. In forward snowballing, we used Google Scholar to analyze the citations of the papers and search for more relevant studies. For example, S14 has 42 citations (by August 2017) and we found one relevant study which was already present in our selected set of SMSs. In both, forward and backward, snowballing we looked at the title, venue, and abstract of the papers cited or citing the paper. We also applied snowballing to the tertiary studies discussed in related work section. This process resulted in a total of seven new studies which raised the number of included studies to 210. The same technique was applied to the newly added studies (07) and no new study was discovered. Hence, we were left with the final set of 210 SMSs for further analysis.

## 3.3. Study selection

As shown in Figure 1(b), initially, we retrieved 2420 papers from online databases. We applied the inclusion and exclusion criteria for selecting the final set of studies. First, we removed duplicate papers (EC3) by using the reference management software, Mendeley [33]. After removing duplicate papers, there were 1802 papers for further scrutiny. We discarded the papers which were non-peer reviewed and for which the full text was not available in English (EC1, EC2, IC1 and IC2 are applied respectively). This step reduced the number of papers to 1256. To remove irrelevant papers, two of the authors read the title and abstract of each paper and selected 203 papers that were considered relevant to our study by at least one author (EC5, EC6, IC4 and IC5). After completing the formal search process, we applied forward and backward snowballing in the bibliography of the selected 203 studies which led to identification of seven additional relevant studies, raising the final number of studies to 210. Our protocol required that the papers for which both researchers reach consensus should be included in the study directly. The two authors managed to reach consensus on most of papers (187 SMSs, 89%) to include without requiring arbitration. The papers where the two researchers disagree, required the third author to act as an arbitrator. Out of 30 such studies, 23 were included after the arbitration process and seven were rejected.

The final selection for this study includes 210 SMSs.

**Inclusion criteria:**

IC1: The secondary study is reported in English.
IC2: The secondary study is peer-reviewed.
IC3: Full text is available.
IC4: The secondary study reported a systematic mapping in SE.
IC5: The systematic mapping study included a systematic review process and primary studies.

**Exclusion criteria:**

EC1: The secondary study is not reported in English.
EC2: The secondary study is not peer-reviewed.
EC3: The secondary study is duplicate of an already included study.
EC4: The full text of secondary study is not available.
EC5: The secondary study does not report a systematic mapping in SE.
EC6: The systematic mapping study does not include a systematic review process and primary studies.

### 3.4. Quality assessment

Quality assessment of included studies is an essential part of secondary and tertiary studies [2, 3, 7]. The reliability of results and conclusions drawn in an SMS is associated to its quality [27]. Hence, we use the well-accepted DARE criteria [15, 18, 29] to assess the quality of SMSs. The DARE criteria are presented by Center for Reviews and Dissemination, University of York, for assessment of systematic reviews to be included in their database, called Database of Abstracts of Reviews of Effects (DARE) (http://www.crd.york.ac.uk/CRDWeb/). We use the following five questions as a criterion to evaluate the quality of included SMSs, we refer to them as Quality Criteria (QC).

QC1. Is the inclusion and exclusion criteria defined and appropriate?
QC2. Is the search adequate, i.e., does it cover all the relevant papers?
QC3. Are the included papers synthesized?
QC4. Does the paper evaluate the quality of included primary studies?
QC5. Did the study provide data about the primary studies?

For quality assessment, we extracted the attributes given in Table 4 for each included SMS. Each attribute has three possible values: 1, 0.5 or 0. The total quality score of each SMS is the sum of the values of all five attributes for that study and is ranged from *low* (0.5 ≤ quality score ≤ 2) to medium (2.5 ≤ quality score ≤ 3) and to *high* (3.5 ≤ quality score ≤ 5).

Table 4. Parameters extracted for assessing quality of each SMS

| Attribute | Score |
|---|---|
| **Inclusion and exclusion criteria (QC1)** | • If the inclusion and exclusion criteria are explicitly defined, the score is 1<br>• If the inclusion and exclusion criteria are partially defined, the score is 0.5<br>• If the inclusion and exclusion criteria are not defined, the score is 0 |
| **Search adequacy (QC2)** | • If four or more reputed digital libraries are searched, the score is 1<br>• If 3 or 4 digital libraries are searched with no appropriate search strategy, the score is 0.5<br>• If two or less than two online databases are searched, the score is 0 |
| **Synthesis method (QC3)** | • If the paper presents explicit synthesis method and a reference to the method is given, the score is 1<br>• If the paper presents synthesis method but no reference is given, the score is 0.5<br>• If the paper presents no synthesis method, the score is 0 |
| **Quality assessment of primary studies (QC4)** | • If the quality assessment criterion is explicitly defined and assessed in the paper, the score is 1<br>• If quality assessment is conducted but not reported, the score is 0.5<br>• If no effort is made to assess the quality of included papers, the score is 0 |
| **Information about included studies (QC5)** | • If data is provided about each primary study, the score is 1<br>• If only summary information of primary studies is given, the score is 0.5<br>• If the information about the primary studies is not given than the score is 0 |

Performing mathematical operations on quality scores is a widespread practice followed in recent SMSs published in software engineering [8, 18, 28, 29, 34]. However, using mathematical operations on DARE criteria values for quality assessment of a paper poses a threat to validity. This results in treating ordinal scale values as ratio scale values. In order to mitigate this risk, where possible, we rely more on presence or absence of answers to the DARE criteria questions to assess the quality of a paper. We make limited use of "total quality score" to broadly categorize the quality of the included studies. We do not use quality scores as a criterion to exclude papers.

In decision making process, Multiple Criteria Decision Aid (MCDA) instruments are widely used to facilitate decision makers, as it is based on unstructured and potentially conflicting goals and objectives [35,36]. MCDA approaches allow decision making keeping in view the preferences and judgements of the decision makers who may have different preferences for similar decision criteria. There are MCDA methods that allow evaluators to model their preferences, here we use the well-established DARE criteria in SE and the prevailing scoring scheme because it is simple, efficient and a common practice in similar studies [8,18, 28, 29, 34].

The quality scores of all 210 SMSs are given in Table A *3* in the Appendix. Two authors conducted the quality assessment independently and noted any disagreements in spread sheets. The disagreements were resolved through meetings where a third author acted as an arbitrator. For example, S5 did not report explicit criteria for the selection of studies, but implicitly explained the selection of studies and used voting process among the authors for increasing confidence in the selection of studies. So one of the authors of our work assigned a quality score of 0.5 and another author assigned a score 1 for inclusion and exclusion criteria. In such cases, the third author was used as an arbitrator to listen to the arguments of both the authors and decide which score to assign to each quality criterion. Such conflicts among the authors occurred several times in around 40 SMSs which were solved through several review meetings.

### 3.5. Data Extraction and Synthesis

The information about the identified attributes is extracted from each of the SMS and is stored in a spreadsheet for analysis. Table 5 lists all the 23 attributes extracted for each SMS. The non-subjective attributes, such as publication year, list of databases used, and number of authors were extracted by single author of this study. For the extraction of subjective attributes, such as, review topic, knowledge area and quality assessment, two of the authors were involved. A third author acted as an arbitrator when there were disagreements.

1. First, the two authors independently extracted the subjective attributes for each SMS and provided reasoning for the value of each attribute. The reasoning was captured in the spreadsheet along with the extracted values.
2. The results were compared with each other and conflicts were noted.
3. The conflicts were resolved through meetings that involved a third author as an arbitrator.

The aim of data synthesis is to analyze and summarize the evidence from the included studies to answer the research questions of this study. We use descriptive analysis to answer the questions involving quantitative data and thematic synthesis to answer questions involving qualitative data [38].

### 3.6. Data Extraction results

The 210 SMSs from 2007 to august 2017 included in this study are shown in the *Bibliography of Included SMSs.* The Table A *1* (in appendix) shows some of the extracted attributes for all 210 papers. For each SMS, we identify the publication year, total quality score, publication type (conference/journal), number of primary studies included by the study, guidelines used for conducting the SMSs, guidelines used for quality assessment, time period covered, venue, digital databases used, number of authors, number of citations, knowledge area, type of research questions, and research areas based on keywords and thematic analysis. We only show the most significant attributes in the table instead of showing all 23 attributes due to space limitations. The studies with *high* quality score (quality score ≥ 3) are highlighted in bold. The included SMSs studies aggregate primary studies from 1966 to august 2017. Among the 210 SMSs included, 102 are journal papers, 102 are conference papers and 6 are published in workshops.

Table 5. Data extraction attributes

| No. | Data Item name | Description | Relevant RQ |
|---|---|---|---|
| **D1** | Title | The title of the study | RQ1.1, RQ1.2 |
| **D2** | Year | Publication Year | RQ4.1 |
| **D3** | Venue | Where the paper is published | RQ4.2 |
| **D4** | Publication type | Type of paper (Conference, Journal, Symposium or Workshop) | RQ2.2 |
| **D5** | Time Period | Time period covered by an SMS | RQ3.4 |
| **D6** | List of Digital databases used | Which electronic databases are used in for search? | RQ3.2 |
| **D7** | Number of included studies | How many studies were included in each SMS? | RQ3.5 |
| **D8** | Number of Authors | How many reviewers were involved in each SMS and what are their names? | RQ4.5 |
| **D9** | Keywords | The keywords listed in the SMSs | RQ1.1, RQ1.2 |
| **D10** | Review topic | Area of research | RQ1.1 |
| **D11** | Knowledge Area | Knowledge Area as per SWEBOK classification. This is done based on title, keywords, abstract and full text. | RQ1.2 |
| **D12** | Number of citations | How many times the paper is cited by other papers? | RQ4.3, RQ4.4 |
| **D13** | Guidelines used | Which existing guidelines are followed for conducting the review? | RQ3.1 |
| **D14** | Inclusion and exclusion criteria | DARE Quality Criteria 1 (QC1) | RQ2.1, RQ2.2, RQ2.3, RQ2.4, RQ2.5 |
| **D15** | Search adequacy | DARE Quality Criteria 2 (QC2) | RQ2.1, RQ2.2, RQ2.3, RQ2.4, RQ2.5 |
| **D16** | Synthesis method | DARE Quality Criteria 3 (QC3) | RQ2.1, RQ2.2, RQ2.3, RQ2.4, RQ2.5 |
| **D17** | Quality assessment of primary studies | DARE Quality Criteria 4 (QC4) | RQ2.1, RQ2.2, RQ2.3, RQ2.4, RQ2.5 |
| **D18** | Information regarding primary papers | DARE Quality Criteria 5 (QC5) | RQ2.1, RQ2.2, RQ2.3, RQ2.4, RQ2.5 |
| **D19** | Current trends | Type of research questions | RQ3.3 |
| **D20** | Demographics | Type of research questions | RQ3.3 |
| **D21** | Research gaps | Type of research questions | RQ3.3 |
| **D22** | Quality assessment | Type of research questions | RQ3.3 |
| **D23** | Research area | Type of research questions | RQ3.3 |

The quality assessment scores for all 210 included SMSs evaluated against the criteria given in Table 4 is shown in Table A 3 (in appendix). The total quality scores assigned to each SMS range from 1-2 (low), 2.5 to 3 (medium), and to 3.5-5 (high). The distribution of quality scores is shown in Figure 4, indicating how quality varies across the included SMSs. Our analysis shows that overall the published SMSs have *high* quality (mean quality score is 3.2 out of 5.0). More than half of the papers have *medium* or *high* quality scores. However, the quality of SMSs published in journals is significantly higher than the quality of SMSs published in conferences. The frequency of 1's in each quality criteria for SMSs published in journals is more than the SMSs published in conferences and can be seen from Figure 2 and Figure 3, showing the frequency of 1's in each quality criteria. For example, there are 90 out of 102 SMSs, published in journals, that have quality score of 1 in inclusion and exclusion criteria whereas 75 out of 102 SMSs in conferences have a quality score 1 in the same quality criteria. In all of our five quality criteria, the SMSs published in journals have higher frequency of 1's in each criterion than those published in conferences. Overall, 41 (33 conference and 8 journal papers) SMSs do not provide information regarding the included studies and scored zero in the corresponding criterion, information about primary studies. Majority of SMSs (60 from journals, 84 from conferences) published in journals and conferences do not measure the quality of included primary studies and scored zero in the corresponding criterion, quality assessment of included studies. Moreover, 23 SMSs (21 conference and 8 journal) do not mention any synthesis method and 87 (55 conference and 32 journal) failed to give reference of the synthesis method used for analyzing the extracted data, therefore, scored zero and 0.5 respectively. The mean quality score of SMSs in journals is 3.7 versus the mean quality score of 2.8 for conferences (the sum of quality of a paper is calculated by adding the quality scores of a given paper across all quality criteria). The mean quality score of SMSs published in workshops is 3.1, but as there are only 4 SMSs published in workshops, no conclusions can be drawn from such a small sample. While the mean quality score suffers from the drawbacks associated with taking arithmetic means of non-ratio scale values, it is a widely used approach of summarizing the quality of SMSs [6, 29] and can provide a high level view of the quality of a paper in conjunction with the frequency-based ranking against each of the quality criteria. Further discussion is provided in Section 4, when answering research question related to quality of the published SMSs.

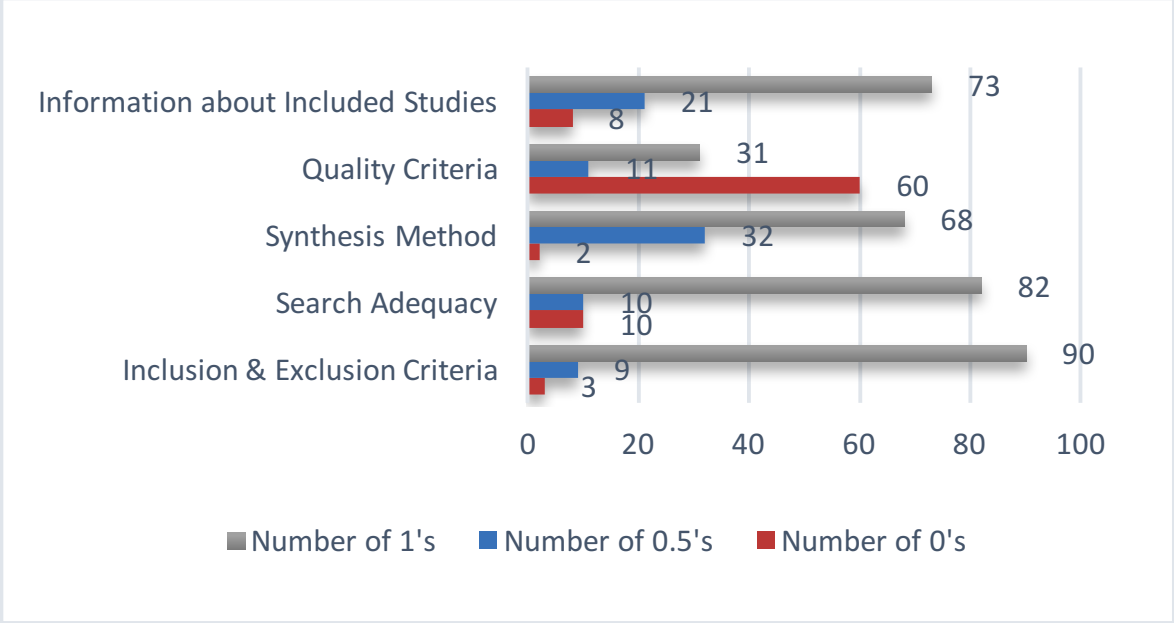

*Figure 2. Number of 1's in each quality criterion for SMSs published in journals*

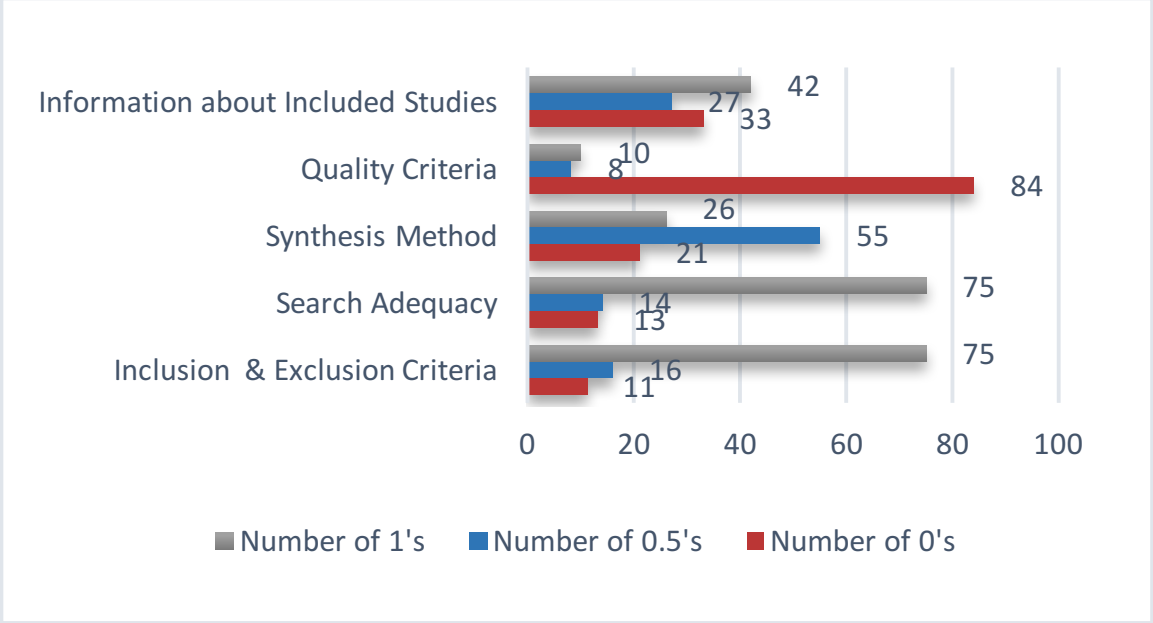

*Figure 3. Number of 1's in each quality criterion for SMSs published in conference*

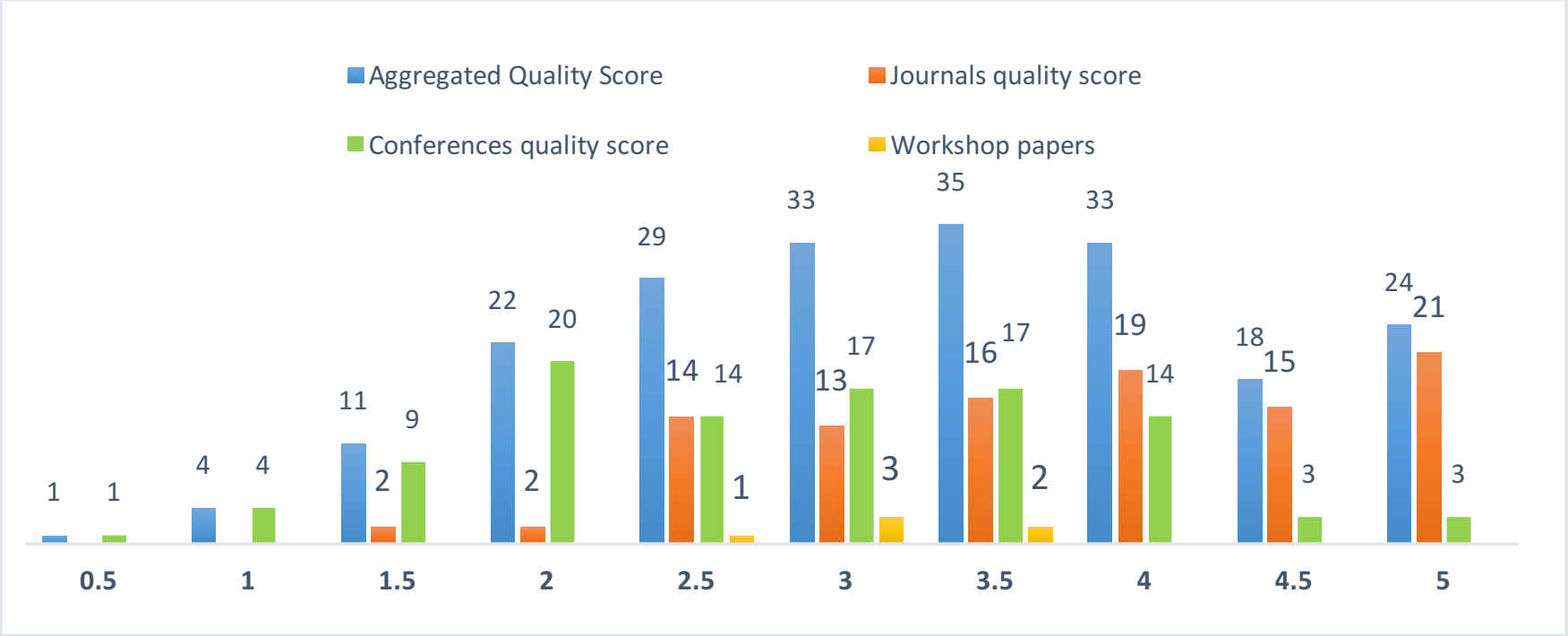

*Figure 4. Distribution of quality scores*

## 4. Results and answers to research questions

In this section, we present our findings and answer the research questions outlined in Section 2.

### RQ 1. Which areas in SE are addressed by the SMSs?

To identify the areas addressed by the included SMSs, we map the included SMSs to knowledge areas in SE using both thematic analysis and SWEBOK syllabus, as discussed in RQ1.1 and RQ1.2.

**RQ1.1 Which areas in SE are addressed by SMSs based on thematic analysis?**

We classify the selected articles by using thematic analysis approach [16]. In thematic analysis, different research areas addressed by the included studies are identified by naming and defining the key themes. Two authors independently classified the SMSs which were then discussed in the review meetings attended by all authors. The research areas are identified by looking at the title, abstract and the overall focus of the SMSs. No major conflicts were identified between the two authors and the minor disagreements were resolved in discussions during the meetings. The results of thematic analysis mapping each included SMS to a thematic area are given in Table A *1* (in the appendix). Figure 5 shows a word-cloud built from the sanitized titles of the included 210 SMSs. The titles are sanitized to remove words that are irrelevant for meta-analysis, including words such as "systematic mapping study", "literature review", and "software".

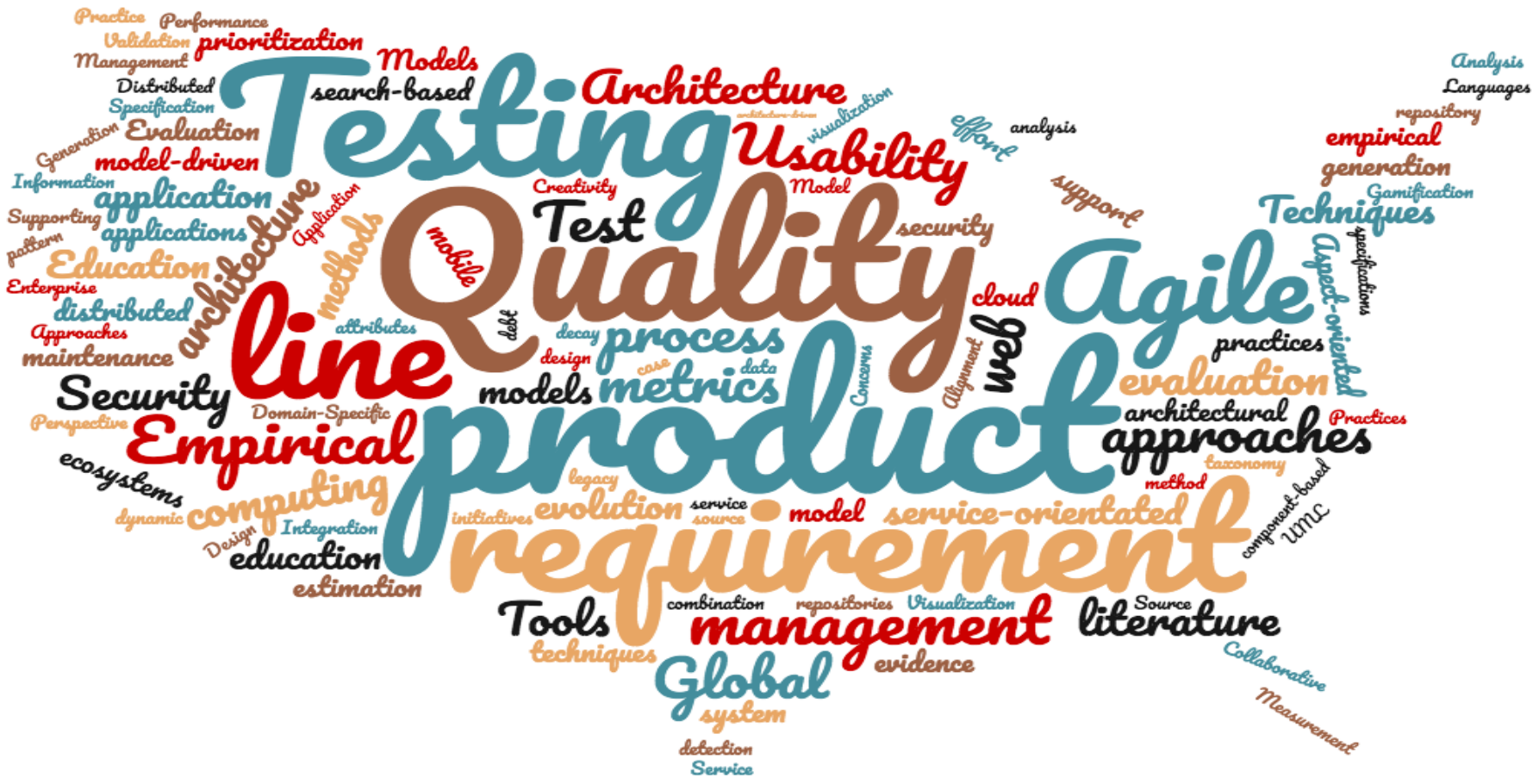

Figure 5 Word cloud by using the titles and keywords of the included papers

Some of the SE areas, such as, testing, requirements, product-lines (product line engineering), quality, architecture and empirical software engineering, have been studied more often than others. Same can be observed from Figure 6 that shows the thematic research areas that have at least five published SMSs. There are 24 SMSs that are mapped to *testing,* which is the most covered area in SE as per the thematic analysis. There are 18 SMSs that are mapped to requirements engineering while 18 SMSs are mapped to *product line engineering*. The categories are not mutually exclusive and one SMS can be placed in multiple areas. For example, S167 discusses testing in software product-line engineering and is classified under the category of both the product-line engineering and testing. *Requirement engineering* (RE) has the second highest number of SMSs as 18 out of 193 papers are related to RE. For example, S1 is classified under RE because it focuses on the improvements in the area. Similarly, 13 SMSs that focus on the models used in different areas of software engineering are placed in *model driven engineering (MDE)* category. There are ten SMSs that cover *Global Software Development (GSD)* and ten SMSs that cover *Agile Software Development (ASD)*. We identified ten SMSs which performed meta-analysis (e.g. S8) and are classified in *meta-studies* class.

Thematic analysis is a well-accepted way of building systematic maps of a research domain [16]. However, the results of thematic analysis are subjective because these are not mapped to any established standard taxonomy. Additionally, though the thematic analysis reveals the areas addressed by the included SMSs, there is no information on which areas have not been covered by these studies.

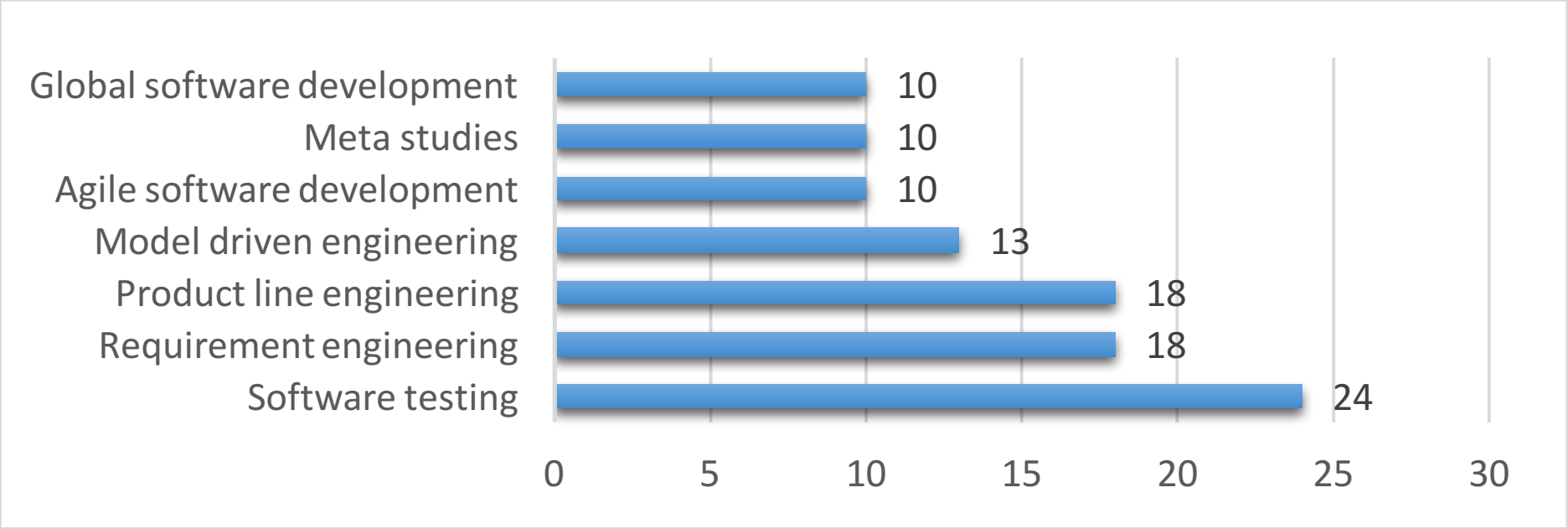

Figure 6. Thematic areas with at least 5 SMSs

**RQ1.2 Which areas in SE are addressed by SMSs based on SWEBOK classification of knowledge areas in SE?**

The classification of SE areas provided by SWEBOK is widely acknowledged and has been used in a number of secondary and tertiary studies in SE [6][S8]. The classification is formalized for academia and practitioners to present a consistent view of each knowledge area. SWEBOK provides a broad classification with some overlap between the knowledge areas, therefore a single study may potentially be mapped to multiple knowledge areas. In the second step of our classification process, we map the 210 SMSs to the SWEBOK knowledge areas. Same as the first step, mapping is done for each of included SMSs independently by two authors using the results of thematic analysis. The mappings were presented in a meeting of all authors and any disagreements in mapping between the two authors were resolved through discussions. The output of the mapping process is a systematic map of SE using SWEBOK knowledge areas, developed through consensus of all authors and is shown Figure 7.

Table 6 provides the classification of the SMSs based on SWEBOK. According to our findings, *Software design* has the highest count of 49 SMSs, whereas, *Software Engineering models and methods* is the second most frequent area of study with 31 SMSs. Other prominent areas include *Software Construction with 28 SMSs*, *Software quality* with 26 SMSs and *Software Testing* with 24 SMSs. Furthermore, *Software Engineering Management*, *Software Engineering Process* and *Computing Foundations* has 21 SMSs each. *Software Engineering Economics* has 20 SMSs, while *Software Requirements* and *Engineering foundations* has 18 SMSs each. *Software engineering professional* practices and *Software Maintenance* have 15 and 13 SMSs respectively. On the other hand, areas such as *Software configuration management and Mathematical Foundations* are relatively neglected with six and zero SMSs respectively.

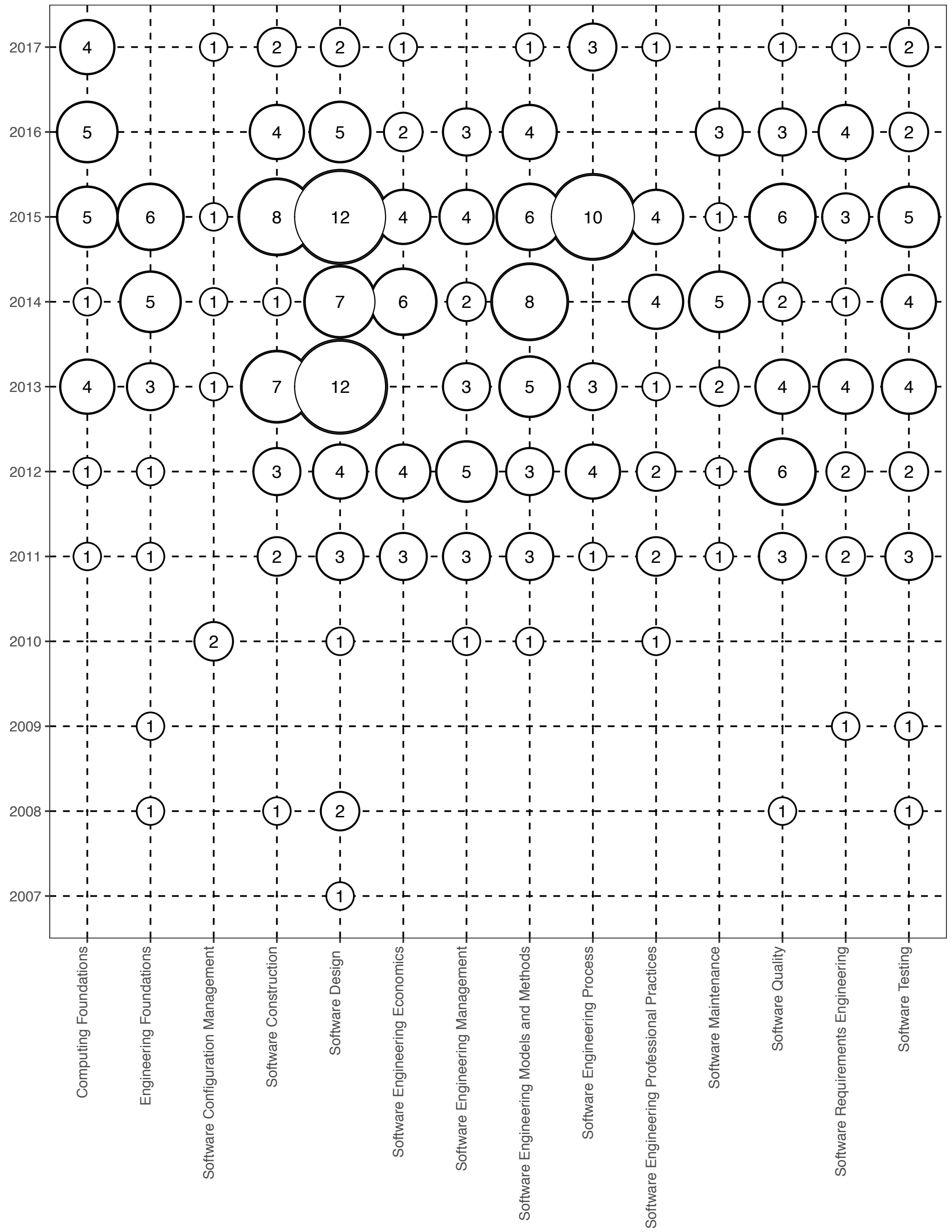

Figure 7 Systematic map of SE using SWEBOK knowledge areas

Table 6. Detailed distribution of SMSs within different areas of SE as per SWEBOK

| Chapter No | Chapter Name | Sub Areas | SMSs Count | SMSs References | Total |
|---|---|---|---|---|---|
| **1** | **Software Requirements** | Software Requirement Fundamentals | 3 | S1, S157, S164 | 18 |
| | | Requirement Process | 3 | S82, S147, S186 | |
| | | Requirement Elicitation | 6 | S33, S38, S41, S63, S121, S141 | |
| | | Requirement Analysis | 1 | S133 | |
| | | Requirement Specification | 4 | S58, S172, S182, S203 | |
| | | Requirement Validation | 0 | | |
| | | Practical Consideration | 1 | S36 | |
| | | Software Requirement Tools | 0 | | |
| 2 | **Software Design** | Software Design Fundamentals | 2 | S58, S60 | 48 |
| | | Key Issues in Software Design | 3 | S78, S125, S132 | |
| | | Software Structure and Architecture | 18 | S2, S21, S39, S43, S68, S70, S79, S96, S98, S111, S118, S122, S130, S152, S170, S180, S185, S210 | |
| | | User Interface Design | 3 | S80, S136, S165 | |
| | | Software Design Quality and Evaluation | 6 | S9, S42, S73, S131, S144, S150 | |
| | | Software Design Notations | 3 | S117, S143, S184 | |
| | | Component Based Design | 1 | S196 | |
| | | Software Design Strategies and Methods | 11 | S48, S61, S67, S76, S77, S113, S131, S140, S150, S158, S173 | |
| | | Software Design Tools | 1 | S100 | |
| 3 | **Software Construction** | Software Construction Fundamentals | 11 | S2, S3, S28, S34, S74, S103, S135, S139, S173, S183, S201 | 28 |
| | | Managing Construction | 1 | S187 | |
| | | Practical Consideration | 1 | S134 | |
| | | Construction Technologies | 15 | S17, S32, S50, S62, S77, S78, S81, S84, S125, S127, S151, S159, S169, S199, S200 | |
| | | Software Construction Tools | 0 | | |
| 4 | **Software Testing** | Software Testing Fundamentals | 5 | S19, S42, S136, S153, S172 | 24 |
| | | Test Levels | 0 | | |
| | | Test Technique | 17 | S48, S54, S59, S75, S90, S98, S99, S108, S116, S124, S129, S142, S145, S167, S168, S181, S194 | |
| | | Test Measures | 0 | | |
| | | Test Process | 2 | S88, S71 | |
| | | Software Testing Tools | 0 | | |
| 5 | **Software Maintenance** | Software Maintenance Fundamentals | 4 | S18, S34, S36, S176 | 13 |
| | | Key Issues in Software Maintenance | 4 | S57, S94, S103, S143 | |
| | | Maintenance Process | 0 | | |
| | | Evolution of Software | 1 | S197 | |
| | | Techniques for Maintenance | 3 | S69, S113, S150 | |
| | | Software Maintenance Tools | 1 | S138 | |
| 6 | **Software Configuration Management** | Management of SCM Process | 2 | S3, S178 | 6 |
| | | Software Configuration Identification | 0 | | |
| | | Software Configuration Control | 1 | S94 | |
| | | Software Configuration Status Accounting | 0 | | |
| | | Software Configuration Auditing | 0 | | |
| | | Software Release Management and Delivery | 2 | S82, S132 | |
| | | Software Configuration Management Tools | 1 | S132 | |
| 7 | **Software Engineering Management** | Initiation and Scope Definition | 0 | | 21 |
| | | Software Project Planning | 4 | S65, S158, S169, S199 | |
| | | Software Project Enactment | 3 | S22, S94, S186 | |
| | | Review and Evaluation | 4 | S18, S76, S86, S97 | |
| | | Closure | 0 | | |
| | | Software Engineering Management | 3 | S25, S87, S187 | |
| | | Software Engineering Measurement | 5 | S148, S163, S174, S175, S177 | |
| | | Software Engineering Management Tools | 2 | S133, S159 | |
| 8 | **Software Engineering Process** | Software Process Definition | 2 | S90, S155 | 21 |
| | | Software Life Cycles | 2 | S74, S92 | |
| | | Software Process Assessment and Improvement | 10 | S3, S10, S46, S53, S87, S89, S93, S163, S174, S193 | |
| | | Software Measurement | 7 | S46, S53, S84, S126, S134, S162, S189 | |
| | | Software Engineering Process Tools | 0 | | |
| 9 | **Software Engineering Models and Methods** | Modelling | 12 | S44, S56, S63, S64, S67, S77, S101, S134, S140, S161, S169, S192 | 31 |
| | | Types of Models | 5 | S54, S113, S119, S134, S164 | |
| | | Analysis of Models | 1 | S15 | |
| | | Software Engineering Methods | 13 | S31, S63, S43, S97, S102, S112, S137, S138, S145, S179, S198, S202, S206 | |
| 10 | **Software Quality** | Software Quality Fundamentals | 16 | S11, S23, S24, S28, S53, S68, S73, S78, S81, S91, S139, S156, S172, S194, S196, S209 | 26 |
| | | Software Quality Management Process | 5 | S105, S126, S146, S162, S167 | |
| | | Practical Considerations | 4 | S141, S144, S160, S170 | |
| | | Software Quality Tools | 1 | S153 | |
| 11 | **Software Engineering Professional Practices** | Professionalism | 4 | S55, S113, S69, S166 | 15 |
| | | Group Dynamics and Psychology | 10 | S11, S47, S52, S95, S106, S127, S155, S159, S164, S178 | |
| | | Communication Skills | 1 | S95 | |
| 12 | **Software Engineering Economics** | Software Engineering Economics Fundamentals | 6 | S34, S97, S103, S188, S204, S208 | 20 |
| | | Life Cycle Economics | 3 | S11, S154, S173 | |
| | | Risk and Uncertainty | 4 | S18, S106, S158, S110 | |
| | | Economic Analysis Methods | 6 | S55, S65, S107, S148, S153, S174 | |
| | | Practical Considerations | 1 | S64 | |
| 13 | **Computing Foundations** | Problem Solving Techniques | 1 | S74 | 21 |
| | | Abstraction | 0 | | |
| | | Programming Fundamentals | 0 | | |
| | | Programming Language Basics | 0 | | |
| | | Debugging tools and Techniques | 1 | S5 | |
| | | Data Structure and Representation | 0 | | |
| | | Algorithms and Complexity | 1 | S7 | |
| | | Basic Concept of a System | 0 | | |
| | | Computer Organization | 0 | | |
| | | Compiler Basics | 0 | | |
| | | Operating Systems Basics | 0 | | |
| | | Database Basics and Data Management | 5 | S6, S29, S66, S113, S120 | |
| | | Network Communication Basics | 0 | | |
| | | Parallel and Distributed Computing | 4 | S127, S131, S135, S144 | |
| | | Basic User Human Factors | 0 | | |
| | | Basic Developer Human Factors | 0 | | |
| | | Secure Software Development Maintenance | 9 | S12, S19, S20, S33, S41, S50, S72, S92, S170 | |
| 14 | **Mathematical Foundations** | Set Relations, Functions | 0 | | 0 |
| | | Basic Logic | 0 | | |
| | | Proof Techniques | 0 | | |
| | | Basics of Computing | 0 | | |
| | | Graphs and Trees | 0 | | |
| | | Discrete Probability | 0 | | |
| | | Grammars | 0 | | |
| | | Numerical Precision, Accuracy, and Errors | 0 | | |
| | | Number Theory | 0 | | |
| | | Algebraic Structures | 0 | | |
| 15 | **Engineering Foundations** | Empirical Methods and Experimental Techniques | 16 | S51, S60, S67, S68, S76, S92, S99, S111, S112, S115, S126, S133, S143, S171, S182, S184 | 18 |
| | | Statistical Analysis | 0 | | |
| | | Measurement | 1 | S148 | |
| | | Engineering Design | 0 | | |
| | | Modelling, Simulation, and Prototyping | 1 | S97 | |
| | | Standards | 0 | | |
| | | Root Cause Analysis | 0 | | |
| | | | | | |

SWEBOK is organized in the form of chapters that provide a coarse-grained classification of knowledge areas in SE. Each chapter in turn provides a more fine-grained classification. For example, Software requirements is further refined into eight sub-areas that are *Software requirement fundamentals*, *Requirement process*, *Requirement elicitation*, *Requirement analysis*, *Requirement specification*, *requirement validation*, *practical considerations* and *software requirement tools*. For each identified knowledge area, the 'sub-area' column provides a further breakdown of SMSs mapped to that area. The sub-area shows a fine-grained distribution of SMSs as mapped to SWEBOK. It can be observed that there are a number of sub-areas that have no SMS.

Tools play an important role in SE and reports on their usage and experience are of particular importance to SE practitioners. However in general, tools in SE seem to be a neglected area and there are few SMSs aggregating the primary studies reporting tool usage experiences. Software maintenance is an important activity in SE, however the number of SMSs covering this area do not reflect the importance of the area. There is only one SMS that is conducted on role of communication skills in software engineering professional practices. Similarly, there is no SMS conducted on software engineering standards, which has implications for software engineering practitioners. Empirical methods and experiments, software structure and architecture, software construction technologies, and software testing techniques are among the most frequently addressed sub-areas.

**RQ2. What are the trends relating to quality of published SMSs?**

The result of our quality assessment (discussed in section 3.4) for all 210 studies is shown in Table A3. SMSs with *high* quality scores (greater than 3) are highlighted in bold. Overall, the quality scores reveal that almost half of the published SMSs are of high quality (quality scores >= 3.5), calculated as per the DARE criteria. The journal papers have higher frequency of one's in all the quality criteria as compared to SMSs published in conferences. Also, the mean quality score for SMSs published till august 2017 is 3.2 out of 5. Overall the journal papers have a higher mean quality score (mean score = 3.7 for 102 SMSs) as compared to the conference papers (mean quality score = 2.8 for 102 SMSs). There are only six workshop papers, which are not sufficient to draw any conclusions (mean quality score = 3.1).

**RQ2.1 Is the quality of SMSs improving over time?**

Figure 8 shows the variation of quality of SMSs in SE published from 2007 to august 2017. It can be observed from the scatter plot (Figure 8) that most of the quality scores are concentrated in the top-right quadrant. There are more SMSs with medium or high quality scores in recent years. We analyze the relationship between the variables, *time* and *quality score*, and calculate the non-parametric spearman's rank order correlation coefficient $r_s$ to determine the strength and direction of relationship [39]. Spearman's $r_s$ is interpreted as follows: correlation value of 1indicates that there is a strong positive relationship between the two variables, correlation value of -1 indicates a strong negative relationship between the two variables, while correlation value of 0 indicates that the two variables are not related. The absolute value of the correlation coefficient gives the strength of the relationship. Using the quality scores from all the included studies, we found a positive relationship between the two variables ($r_s$ = 0.59, *p-value* = 0.05). The obtained $r_s$ value can be interpreted as showing a strong positive relationship, i.e., the quality scores are improving over time.

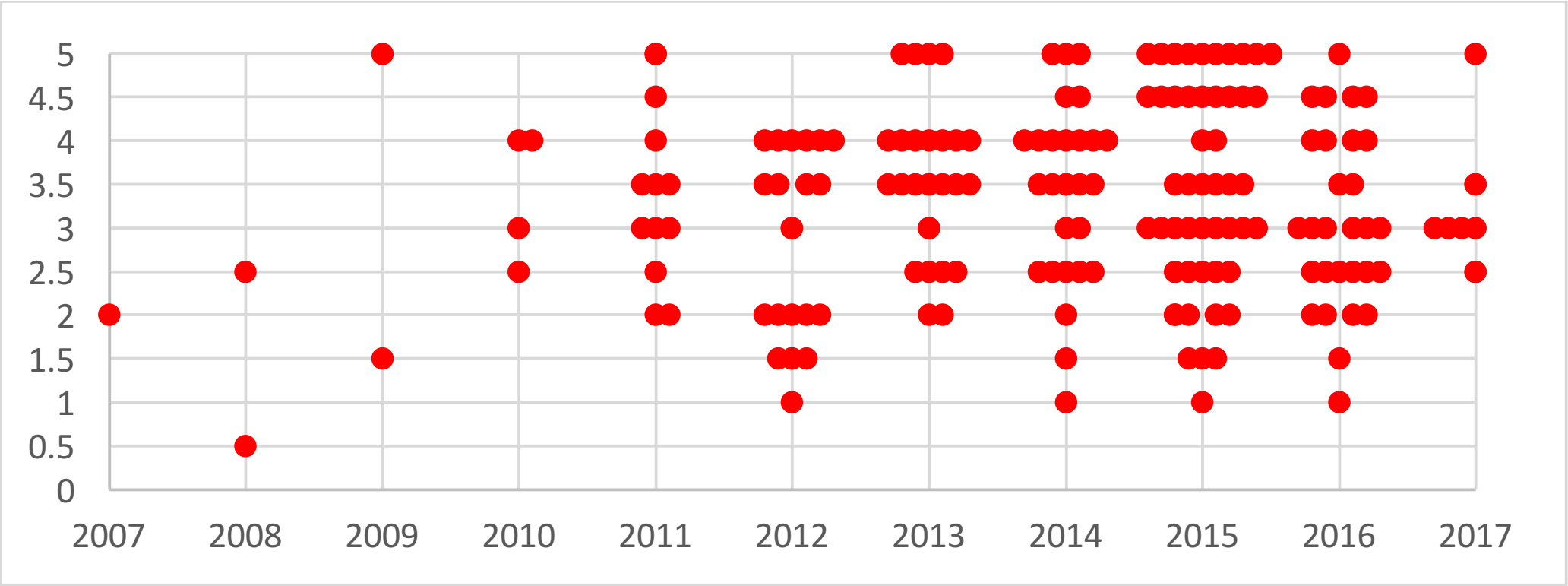

Figure 8 Quality of SMSs distributed over publication years

**RQ2.2 Do the SMSs published in Journals have better quality than SMSs published in conferences?**

Table 7 shows the number of SMSs published in journals, conferences and workshops in each of the quality ranges. It is noticeable that over 80% of the SMSs have medium or high quality scores ($2.5 \leq quality\ score \leq 5$). Figure 9-11 show the number of 1's, 0.5's and 0's, respectively, for each quality criterion of SMSs published in journals, conferences, and workshops. It can be seen from the figures that journal papers have more number of 1's in each quality criterion than conference papers. The number of one's in all the quality assessment criteria of SMSs published in journals is 344 whereas that of conference paper is 228.

Figure 9-12 provides a visualization of conference and journal papers over different quality scores. As can be observed there are fewer journal papers with *low* quality scores, i.e., there are 34 conference papers with *low* quality as compared to only four journal papers. Overall, we found that a significant number of SMSs do not include quality assessment of primary studies (QC4). We also found a significant number of SMSs published in conferences that fail to report information on included primary studies (QC5).

Table 7 Distribution of SMSs over quality range

| Quality score range | Category | No. of journal papers | No. of conference papers | No. of workshops papers | Total |
|---|---|---|---|---|---|
| **0.5 – 2** | low | 4 | 34 | 0 | 38 |
| **2.5 – 3.5** | Medium | 43 | 48 | 6 | 97 |
| **3.5 – 5** | High | 55 | 20 | 0 | 75 |

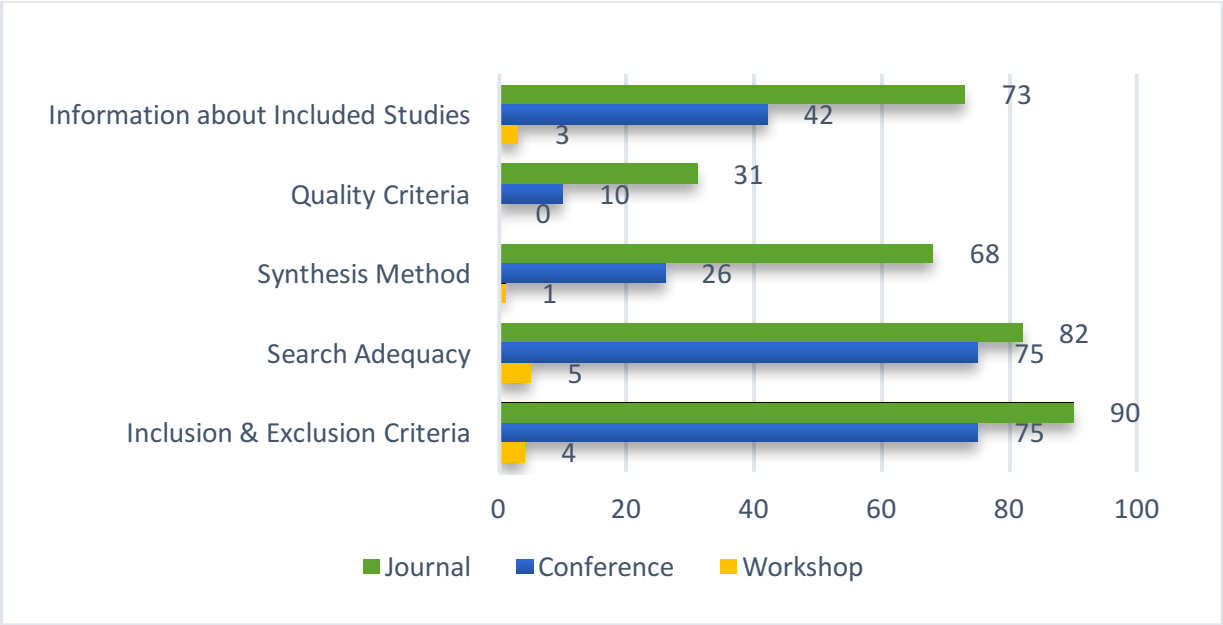

*Figure 9. Number of 1's in each quality criteria*

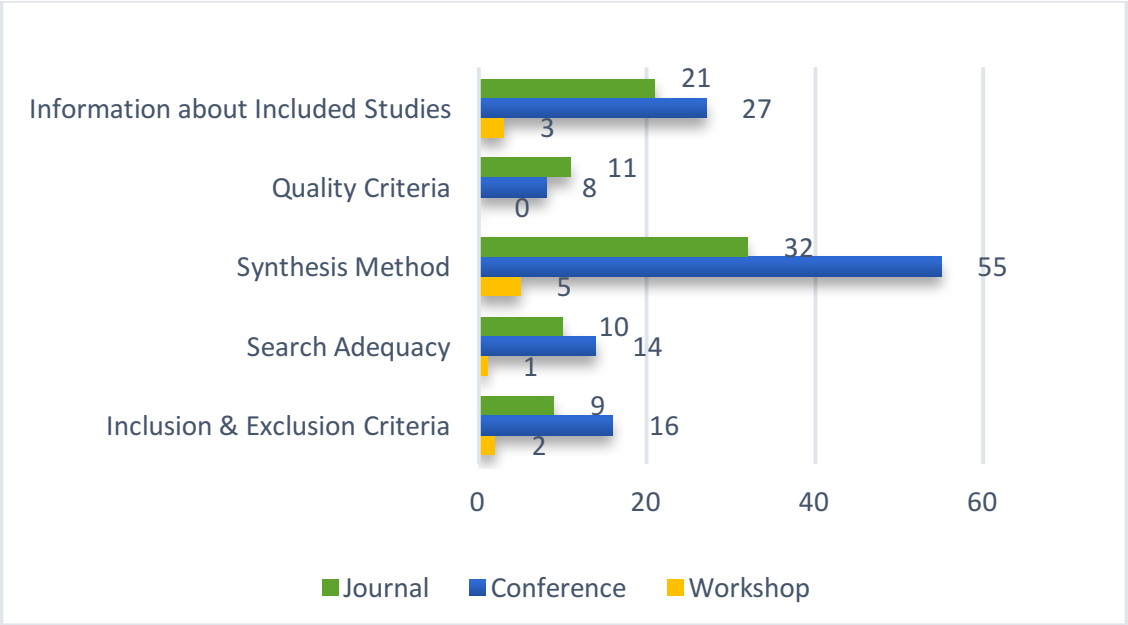

*Figure 10. Number of 0.5's in each quality criterion*

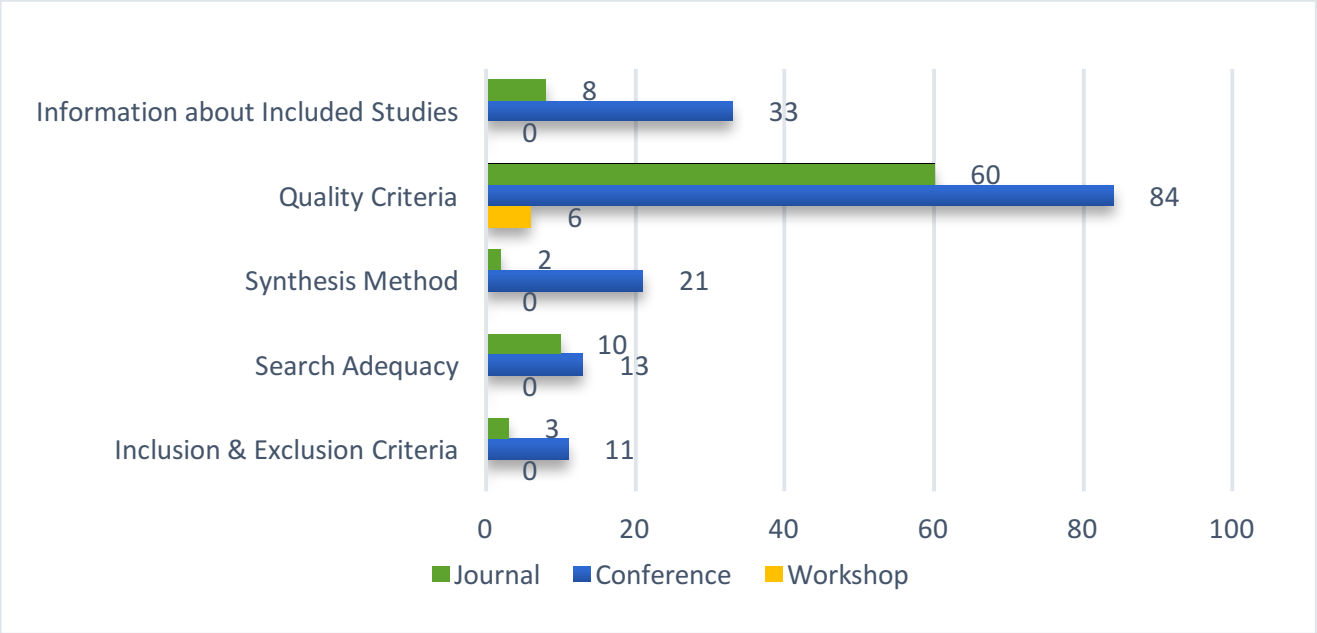

*Figure 11. Number of 0's in each quality criteria*

**RQ2.3 What are the strengths and weaknesses of published SMSs in terms of quality?**

To assess the strengths and weakness of the included SMSs, we analyze the quality scores of each of the 210 SMSs for the extracted quality assessment parameters corresponding to quality assessment questions suggested by DARE criteria. Individual analysis of each parameter provides insights into the strengths and weaknesses of published SMSs. We analyze the results separately for journals and conferences to gain further insights on how the responses vary between SMSs in conferences and journals. Recall that each attribute has three possible values: 1, 0.5, or 0. Figure 12 shows the quality score against each of the five identified parameters for the journal papers. A large number of SMSs published in journal papers do not provide quality assessment of included primary studies. Details of synthesis method is another area of concern as large number of papers only provide partial information on synthesizing results. Defining the IC and EC criteria, and the search process are strong points followed by inclusion of basic information about the included primary studies.

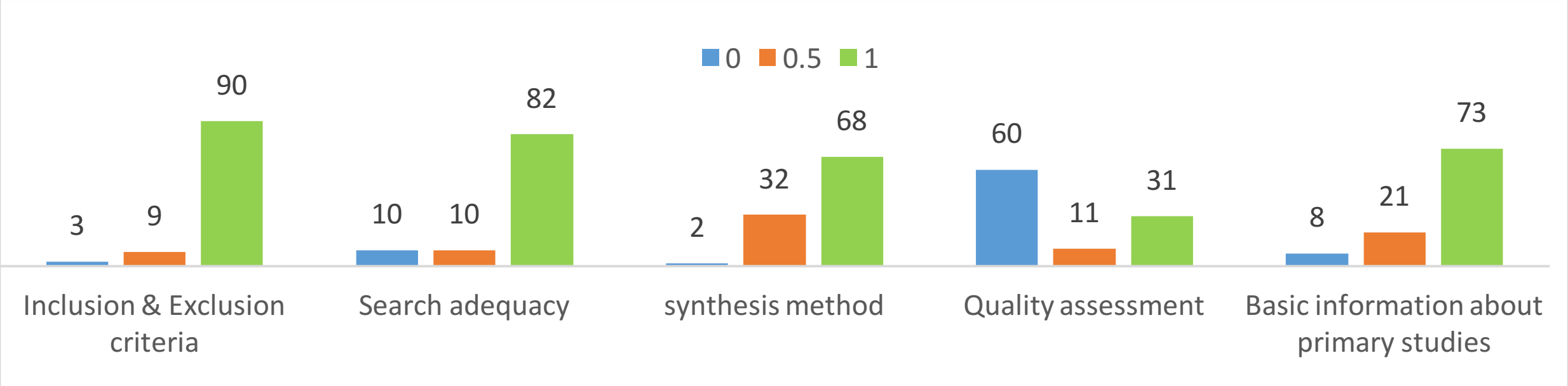

Figure 12 Aggregate scores for individual criteria for Journal papers

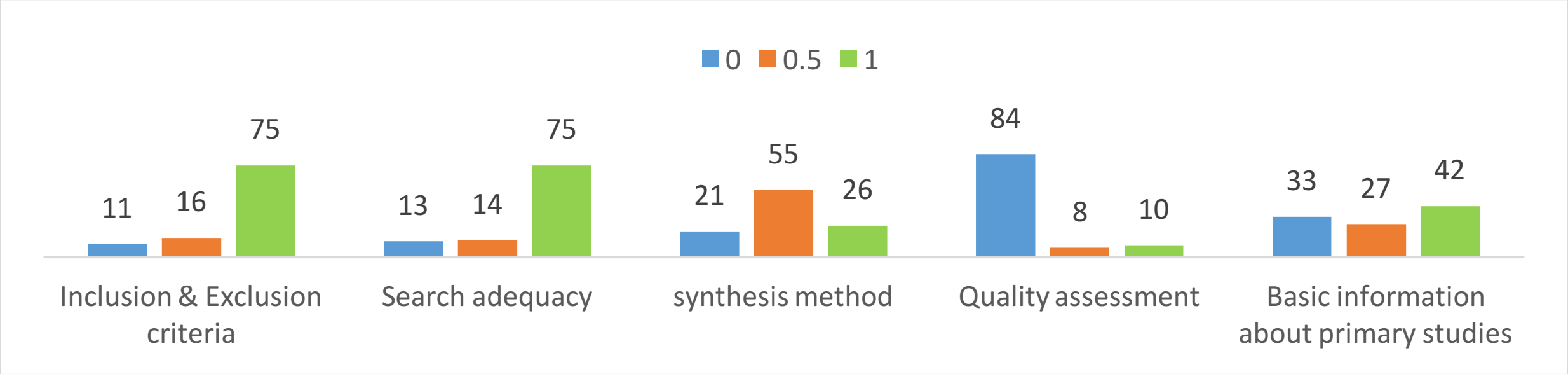

Figure 13 Aggregate scores for individual parameters for Conference Papers

Figure 13 shows the aggregates for individual quality parameters of SMSs published in conferences. SMSs in conferences have a lower mean quality score (mean quality score of 2.8) than journals. Quality assessment of included studies is again identified as a weak point of SMSs. However, the results are more alarming, as 84 of the 102 SMSs in conferences do not provide quality assessment of included studies. Additionally, a high number of papers either completely fail to provide information about the included primary studies or only provide partial information. A high number of conference papers do not discuss the synthesis method adequately. Surprisingly, we identified ten conference papers and two journal papers (S30, S32) that fail to provide the inclusion and exclusion criteria, which is a basic requirement of SMSs.

It can be argued that the lower quality scores for conferences are a result of space limitations. Conference papers tend to be shorter than journal papers and consequently some information cannot be presented in an adequate manner [40]. The quality score of SMSs in EASE conference which has the highest number of published SMSs in SE (for conferences) is higher than other conferences.

**RQ2.4 How does the quality of SMSs vary with publication venues?**

We identify the top publication venues (defined in terms of the total count of published SMSs) for both conferences and journals. To answer this RQ, we analyze the quality scores of SMSs published in top publication venues for both conferences and journals. As discussed previously, in total there are 63 distinct conferences and 36 journals that have published at least one SMS in SE. However, most of the SMSs are published in selected few venues. For example, out of 102 included SMSs from journals, 37 are published in the journal *Information and Software Technology (IST)* and 18 are published in *Journal of Systems and Software (JSS)*. For conferences, there are 102 identified SMSs in total and the most frequent venues are *Evaluation and Assessment in Software Engineering (EASE)* with 14 SMSs and *Empirical Software Engineering and Measurement (ESEM)* with ten published SMSs. No other venue has more than five published SMSs.

Table 8 Breakdown of quality score by top venues

| Venue | Type | No. of Papers | Mean Quality Score |
|---|---|---|---|
| **Information and Software Technology (IST)** | Journal | 37 | 3.97 |
| **Journal of Systems and Software (JSS)** | Journal | 18 | 3.97 |
| **Evaluation and Assessment in Software Engineering (EASE)** | Conference | 14 | 3.17 |
| **Empirical Software Engineering and Measurement (ESEM)** | Conference | 10 | 2.35 |
| **Other Journals (Excluding IST and JSS)** | Journal | 47 | 3.55 |
| **Other Conferences (Excluding EASE and ESEM)** | Conference | 78 | 2.82 |

When grouped together IST and JSS account for approximately half of all SMSs published in journals. Table 8 shows the distribution of the mean quality scores for top journals and conferences based on the number of papers. IST and JSS have the same mean quality score of 3.97. However,

Excluding IST and JSS, the mean quality scores still remain high (3.55) for all the remaining journals. For conferences, EASE has a mean quality score of 3.17 as compared to the mean quality score of 2.35 of ESEM, the second most frequent conference venue. Excluding EASE and ESEM, gives a mean quality score of 2.82 which is same as the overall mean for conferences.

To summarize, 55 out of 102 SMSs published in SE journals are published in IST and JSS, both of which have a high mean quality score for the published SMSs. The remaining 47 SMSs are published in 34 different journals. Only five SMSs in the journals fail to provide the necessary guidelines on conducting an SMS and have quality scores lower than 2. The SMSs published in conferences are more widely distributed. The EASE conference has published 14 SMSs and ESEM conference has published ten SMSs. The remaining 78 SMSs are published in 61 different conferences. Some of these conferences do not enforce the SMS guidelines (possibly due to page restrictions) resulting in 30 SMSs which have quality scores lower than or equal to 2.

**RQ2.5 What is the distribution of quality w.r.t different SE areas?**
We have shown quality distribution for both, keywords-based classification and SWEBOK-based knowledge areas. Figure 14 to 20 show the distribution of quality among different areas (based on our keywords-based classification). Overall we found that a significant number of SMSs do not meet the requirement of assessing the quality of included primary studies (QC4).

For example, Figure 14 shows the frequency of quality scores for each quality criterion in the area of software testing. There are 24 SMSs in software testing area. Out of 24 SMSs, 18 studies scored one in QC1 (inclusion and exclusion criteria). There are 22 SMSs scored one in QC2 (search adequacy). There are 13 SMSs that scored one in QC3 (synthesis method). QC4 refers to quality assessment of included primary studies. The results indicate that seven SMSs scored one in this criterion because they successfully report quality assessment of primary studies. Most of the SMSs in the area of software testing meet the four quality criteria except QC4 which represent quality assessment of included primary studies. QC4 has the highest number of zeroes, which indicates that majority of SMSs ignore quality assessment of included primary studies. Figure 15 shows the frequency distribution of quality scores for each quality criteria in the area of *Requirements Engineering*. There are a total of 18 SMSs in this area. It can be seen that there are 11 SMSs that have scored zero in QC4 (quality assessment of primary studies). Nine SMSs partially applied synthesis method (QC3) and six SMSs provide incomplete information about included primary studies (QC5) and therefore scored 0.5. We show the distribution of quality scores for individual knowledge areas based on SWEBOK, given from Figure 21 up to Figure 34. It can be noticed that majority of the SMSs from all the knowledge areas failed to meet QC4 (quality assessment) and partially fulfilled QC3 (synthesis method). However, most of the SMSs have met QC1 (inclusion and exclusion criterion) and QC2 (search adequacy criterion).

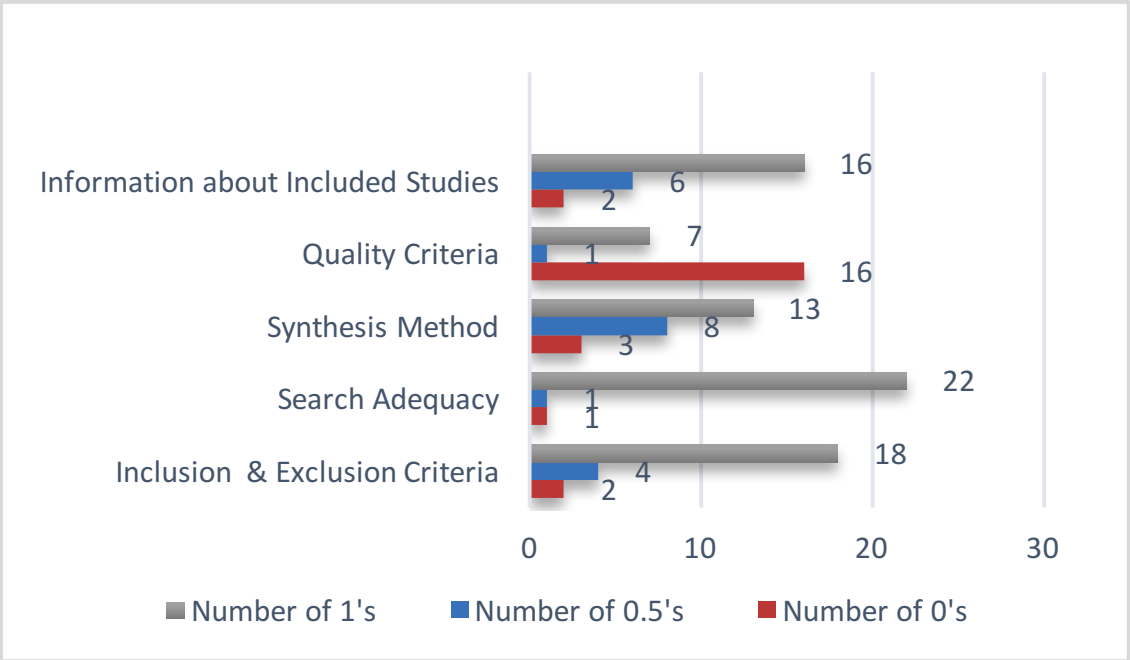

*Figure 14.Quality distribution of SMSs in the area of Software Testing*

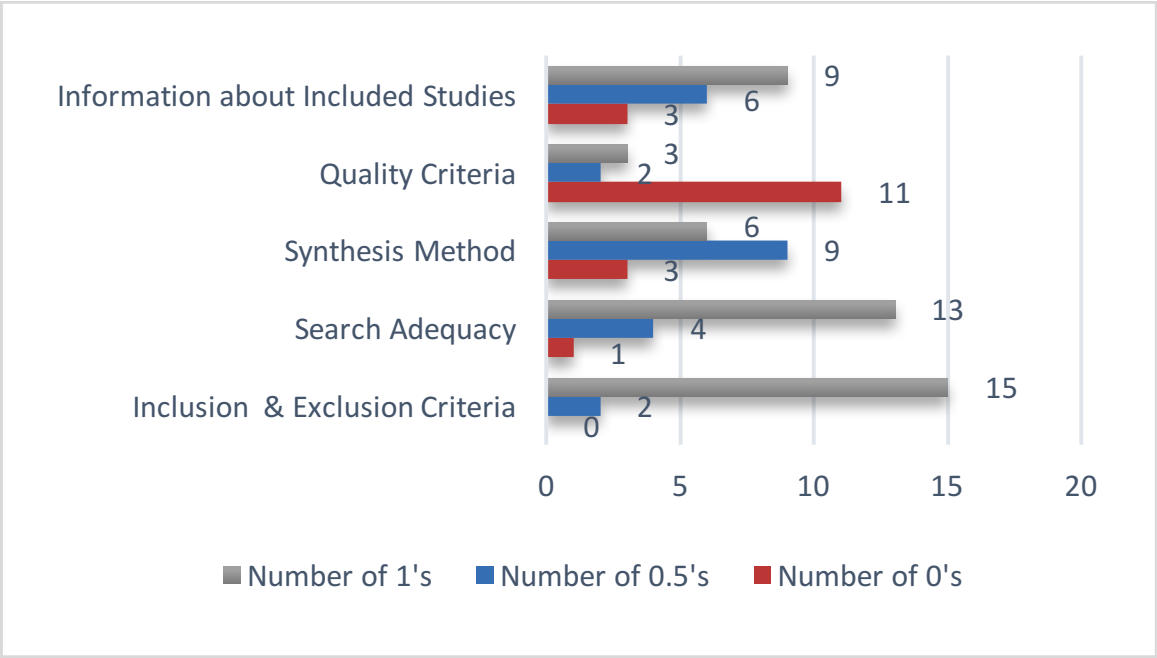

*Figure 15. Quality distribution of SMSs in the area of Requirement Engineering*

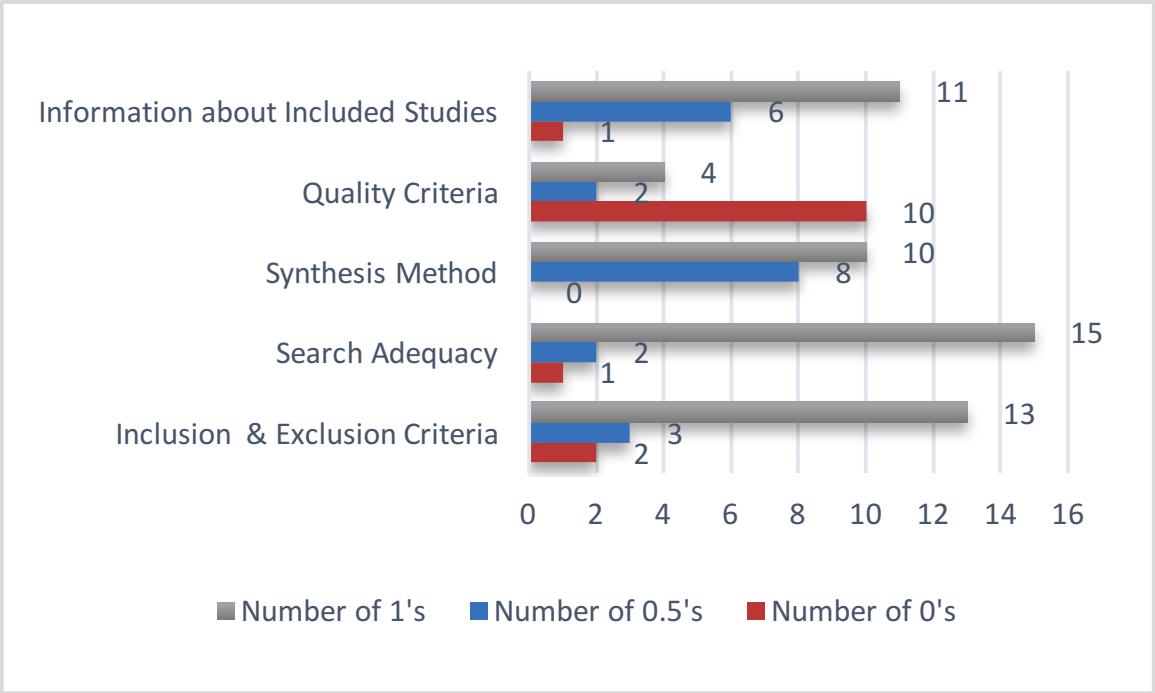

*Figure 16. Quality distribution of SMSs in the area of Product Line Engineering*

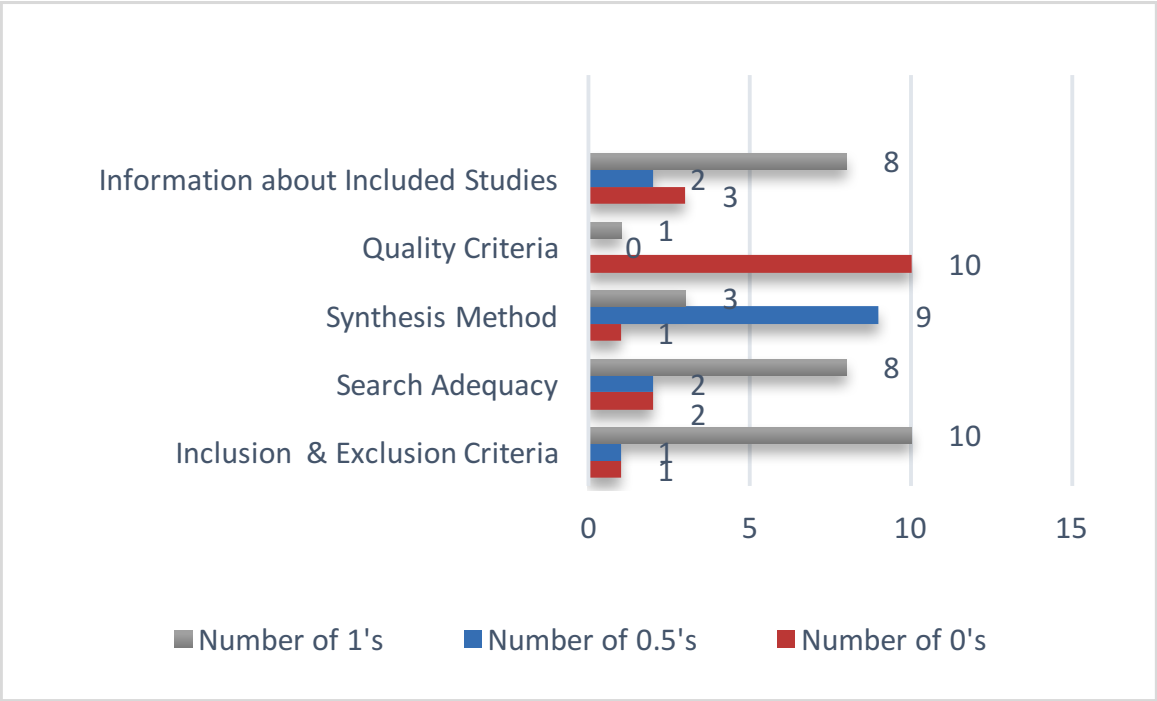

*Figure 17. Quality distribution of SMSs in the area of Model Driven Engineering*

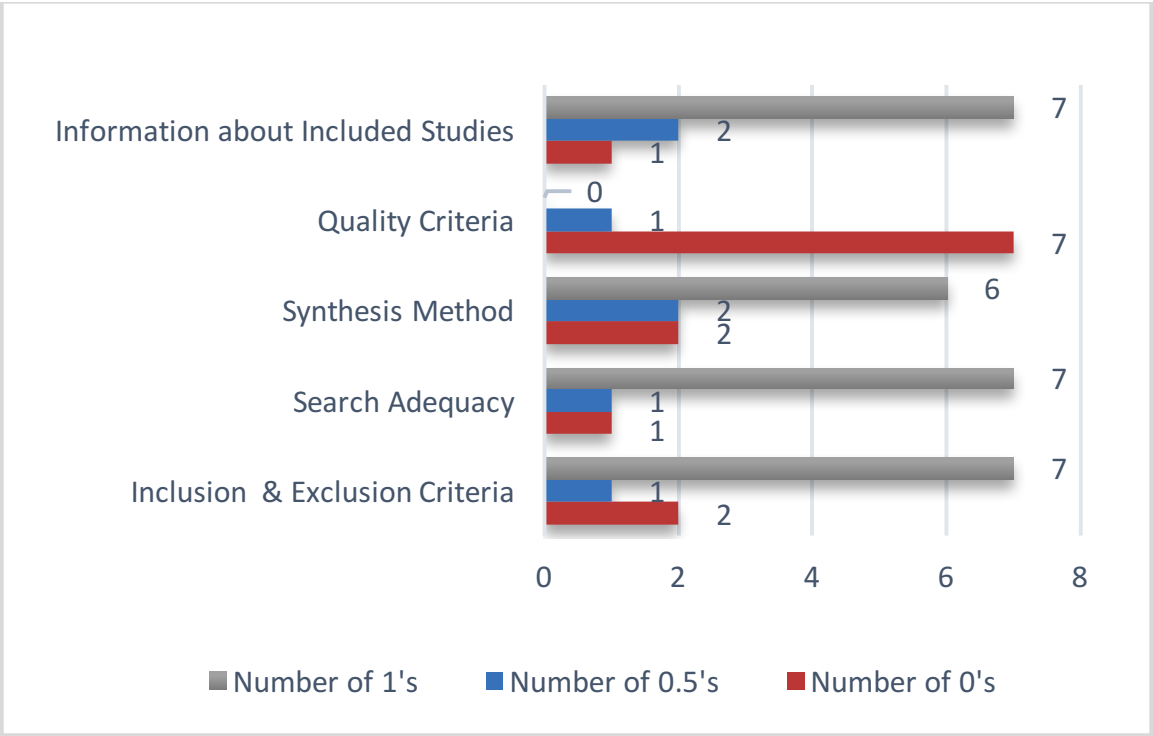

*Figure 18. Quality distribution of SMSs in the area of Agile Software Development*

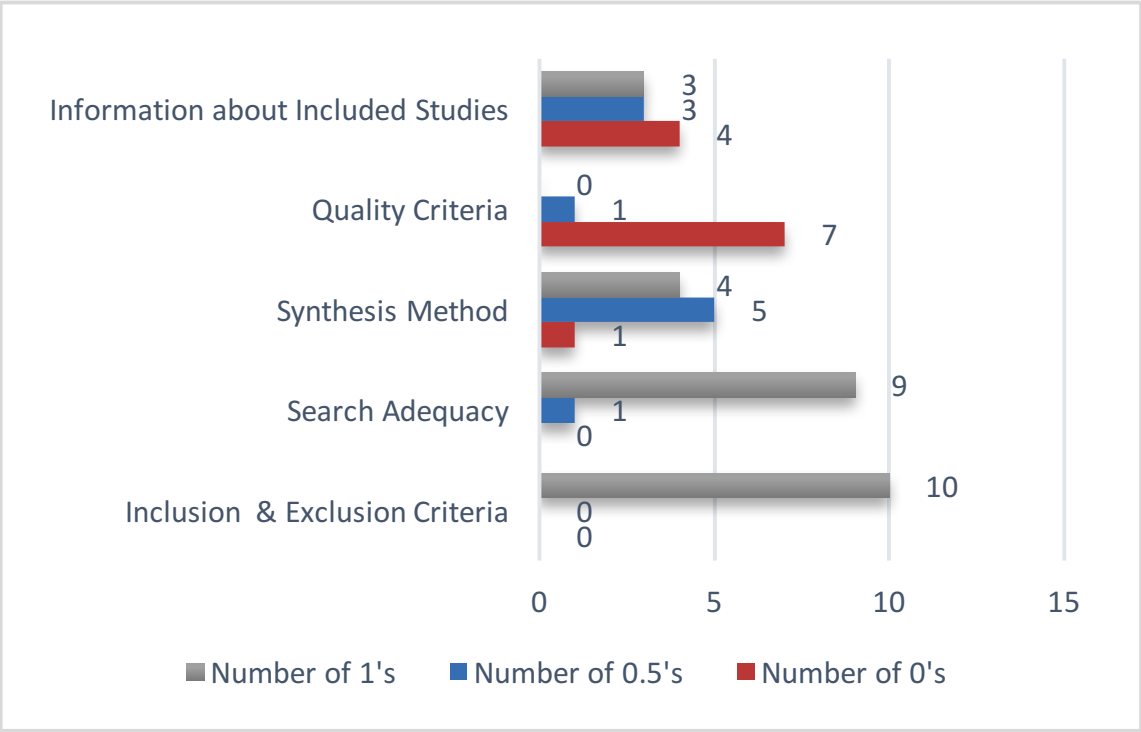

*Figure 19. Quality distribution of SMSs in the area of Meta Studies*

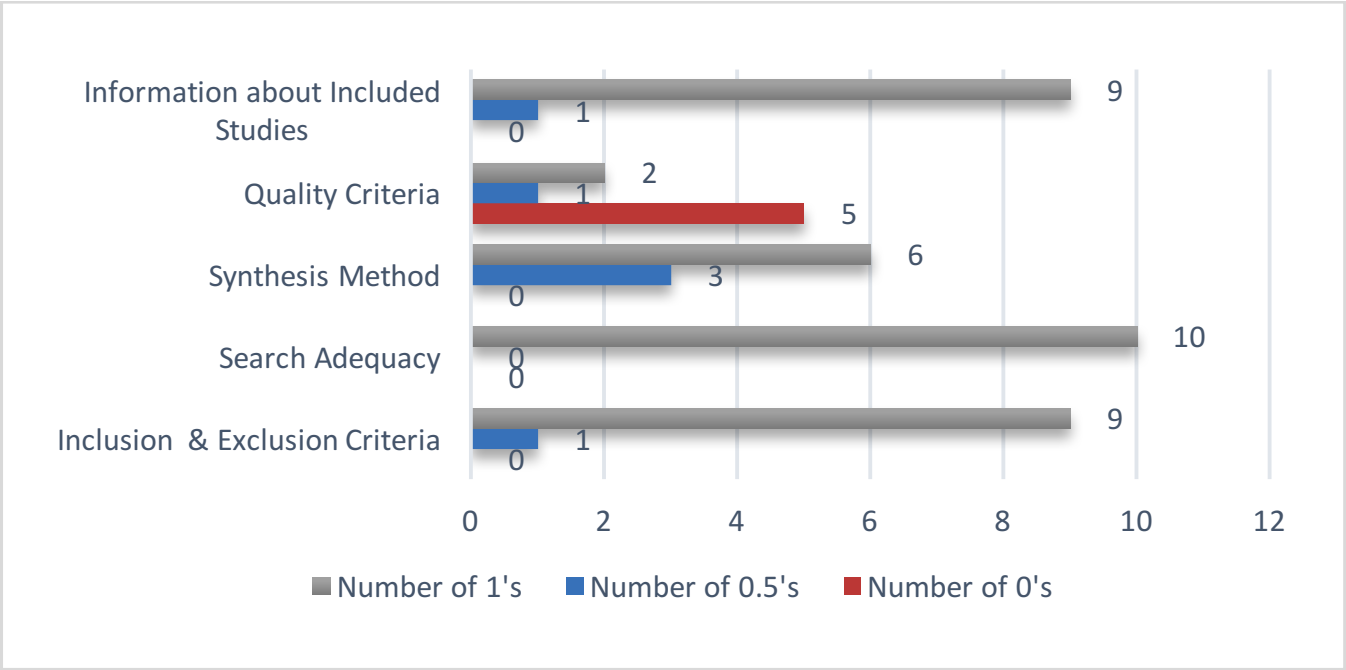

*Figure 20. Quality distribution of SMSs in the area of Global Software Development*

Figure 21 shows the quality distribution of SMSs in the area of requirements engineering. It can be seen that 14 out of 18 SMSs did not assess the quality of included primary papers (QC4). Three SMSs failed to include basic information regarding the included studies (QC5). Furthermore, three SMSs did not meet QC3, which is the use of synthesis method for analyzing the extracted data. One SMS did not mention any inclusion and exclusion criteria (QC1) and one SMS did not fulfil the search adequacy criterion (QC2). Also, out of 18 SMSs in requirements engineering area, ten studies partially perform synthesis method (QC3), 07 studies failed to report complete information about primary studies (QC5) and four studies do not follow an adequate search process (QC2). Two SMSs partially assess quality of primary studies (QC4) and two SMSs in *Requirement Engineering* failed to explicitly report inclusion and exclusion criteria (QC1).

Figure 22 shows the quality distribution of SMSs in the area of software design. There are 48 SMSs in software design area. Out of 48 SMSs, 33 studies did not evaluate the quality of included studies (QC4) and 13 did not meet the criterion of including information about primary studies (QC5). Eight SMSs failed to use appropriate synthesis method (QC3), four studies failed to report appropriate inclusion and exclusion criteria (QC1), and 4 studies did not perform adequate search process (QC2). We found 21 SMSs that document the synthesis method used to analyze the included studies but did not explicitly provide a reference to which synthesis method is used. We consider such papers as partially meeting the requirement of describing the synthesis method used (QC3). Nine SMSs do not explicitly report inclusion and exclusion criteria (QC1), seven studies provide incomplete information about primary studies (QC5). There are four studies in the area of software design that do not perform adequate search process (QC2). Finally, there are two SMSs that partially assess the quality of primary studies (QC4).

Figure 23 shows the quality distribution of SMSs in the area of *Software Construction*. There are 28 SMSs in the area of software construction. 75% of the total SMSs in this area did not evaluate the quality of included studies (QC4) and 16 SMSs partially fulfilled the criterion of synthesis method (QC3). In the area of software maintenance, eight out of 13 SMSs failed to fulfil quality assessment criteria (QC4) and five performed data synthesis (QC3) without providing any reference of the used method. Figure 24 shows quality distribution in software testing area. Most of the SMSs in this area meet four quality criteria except QC4 which indicates that majority of SMSs ignored quality assessment of included primary studies. Figure 25 presents the graph of quality distribution in the *Software Maintenance* area. Two studies partially meet quality assessment criteria (QC4), similarly, two studies partially report information about primary studies (QC5).

Figure 26 shows the distribution of quality scores in the area of *Software Configuration Management*. This area has a total of six SMSs. Four out of six studies in the area did not assess the quality of primary studies (QC5), whereas, only one study did not perform adequate search process (QC2). Consequently, four mapping studies in the partially synthesized the included literature (QC3). There are two studies that provide incomplete information about primary studies (QC5). Finally, only one study in this area did not explicitly reported inclusion and exclusion criteria (QC1).

Figure 27 presents the distribution of quality in the *Software Engineering Management* area. In this area, there are a total of 21 SMSs. Out of 21 SMSs, 14 failed to assess the quality of included studies (QC4), Nine SMSs partially applied synthesis method (QC3) and five studies incompletely reported information about primary studies (QC5).

Figure 28 shows the distribution of quality in the area of software engineering process. This area has a total of 21 SMSs out of which 15 did not asses the quality of primary studies (QC4). Nine SMSs partially applied synthesis method (QC3), whereas, four SMSs report insufficient information about primary studies (QC5).

Figure 29 shows the quality distribution of *Software Engineering Models and Methods*. There are a total of 31 SMSs in this area. 87 % of these SMSs did not evaluate the quality of included studies. Nearly half of the SMSs of this area partially applied synthesis method (QC3).

In the *Software Quality* area, 18 out 31 SMSs did not meet QC4 (quality criteria) and scored a zero, as shown in Figure 30. Additionally, seven SMSs failed to provide any information about primary studies (QC5). Five studies do not apply synthesis method (QC3). Three studies do not report inclusion and exclusion criteria (QC1) and two SMSs do not apply any adequate search process (QC2). All the studies that completely failed to fulfil a given quality criteria scored a zero. Nine out of 31 studies, partially fulfilled QC3 (synthesis method) and four SMSs partially fulfilled QC5 (primary studies information). There are three SMSs in software quality area that did not provide a through synthesis of literature (QC3). Four studies did not provide explicit inclusion and exclusion criteria (QC1). Finally, there are two SMSs in software quality area that failed to provide a quality assessment of included primary studies (QC4). All the SMSs that partially fulfilled a given quality criteria were assigned a score of 0.5.

Figure 31 shows the distribution of quality in the area of *Software Engineering Professional Practices*. There are 15 SMSs in this area. It can be seen from Figure 31 that most of the SMSs managed to score a 1 in all quality criteria except in QC4 (quality assessment of primary studies). 73% of the studies failed to assess the quality of included primary studies (QC4) hence scored 0 in QC4.

Figure 32 shows the distribution of quality scores in the area of *Software Engineering Economics*. There are a total of 20 SMSs in this area. 80% of the SMSs in this area failed to assess the quality of included primary studies (QC4), hence scored 0 in QC4. Similarly, five studies failed to provide information about primary studies (QC5) and 2 studies scored 0 in QC3 due to absence of synthesis method (QC3). 50% of the studies this area partially applied synthesis method (QC3).

Figure 33 shows the quality distribution in the area of computing foundations. Out of 21 SMSs, 14 studies failed to evaluate the quality of included studies (QC4), 04 failed to use any synthesis method (QC3) and include basic information regarding primary studies (QC5).

Figure 34 shows the distribution of quality in the area of engineering foundations. There is a total of 18 SMSs in this area, out of which 13 SMSs did not perform quality assessment of included primary studies (QC4) and four SMSs failed to provide information about primary studies (QC5). Eight studies scored 0.5 in QC3 (synthesis method) and five studies scored 0.5 in QC5 (primary studies information).

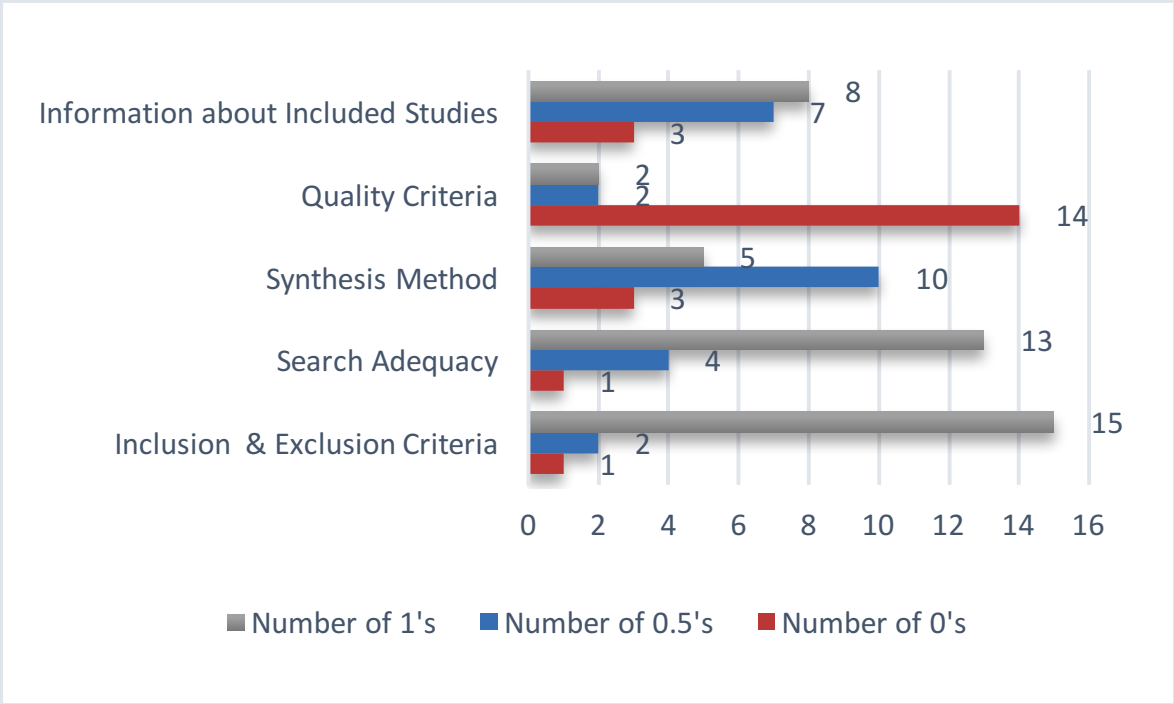

*Figure 21. Quality distribution of SMSs in the area of Requirement Engineering (SWEBOK)*

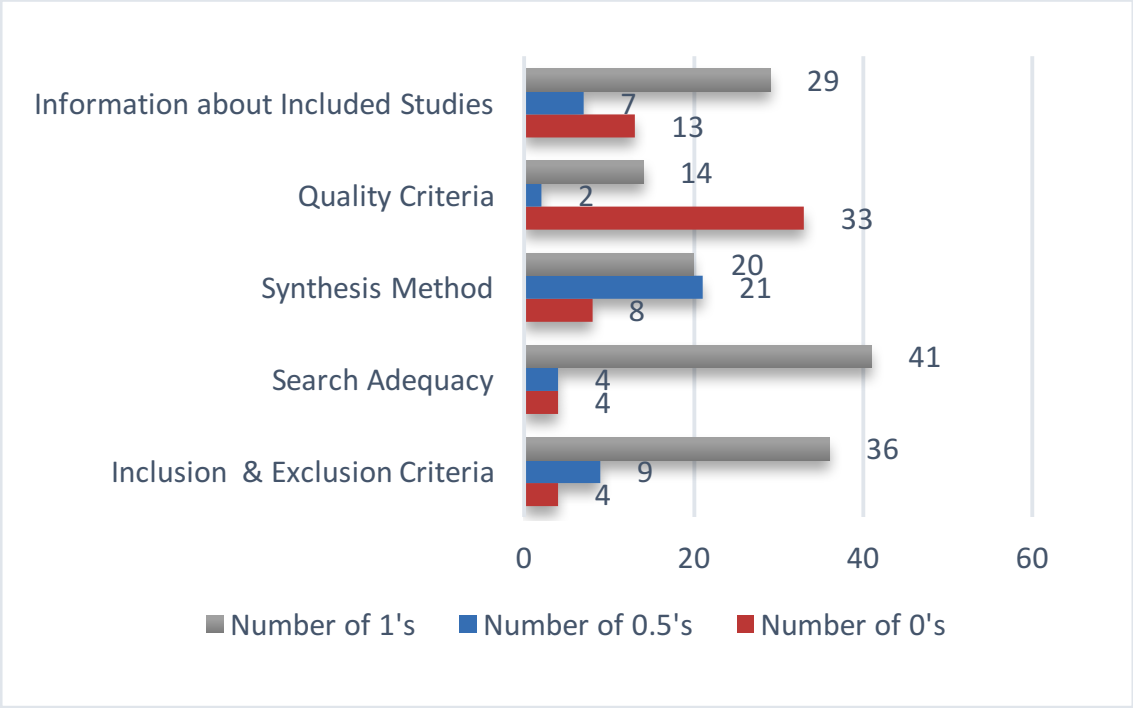

*Figure 22. Quality distribution of SMSs in the area of Software Design (SWEBOK)*

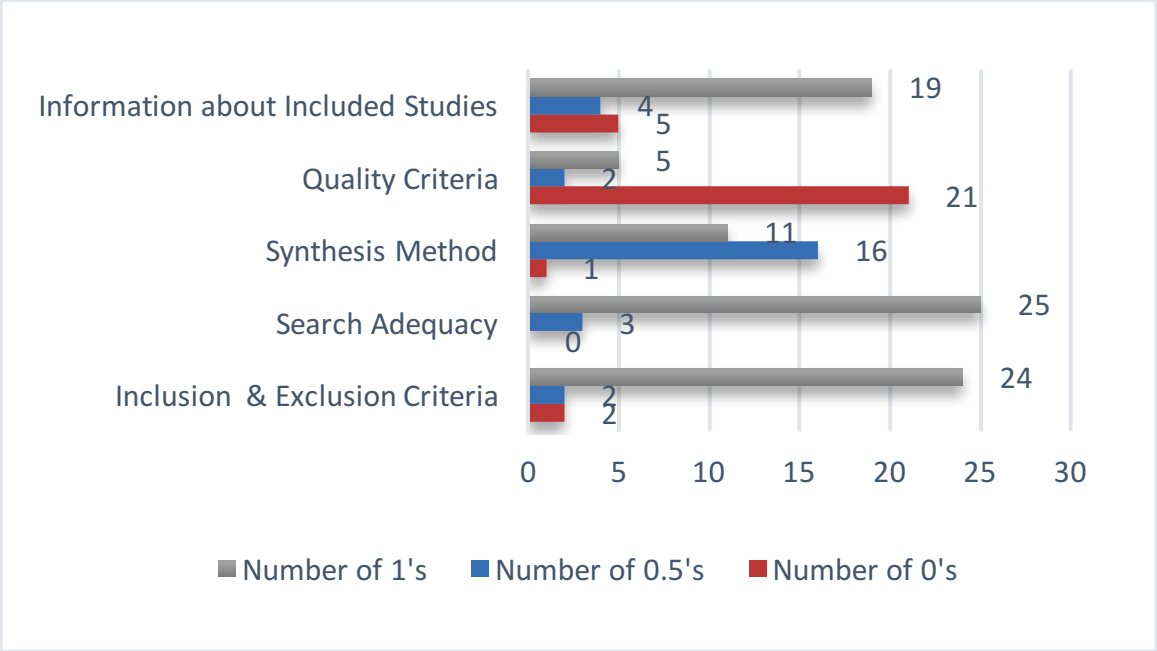

*Figure 23. Quality distribution of SMSs in the area of Software Construction (SWEBOK)*

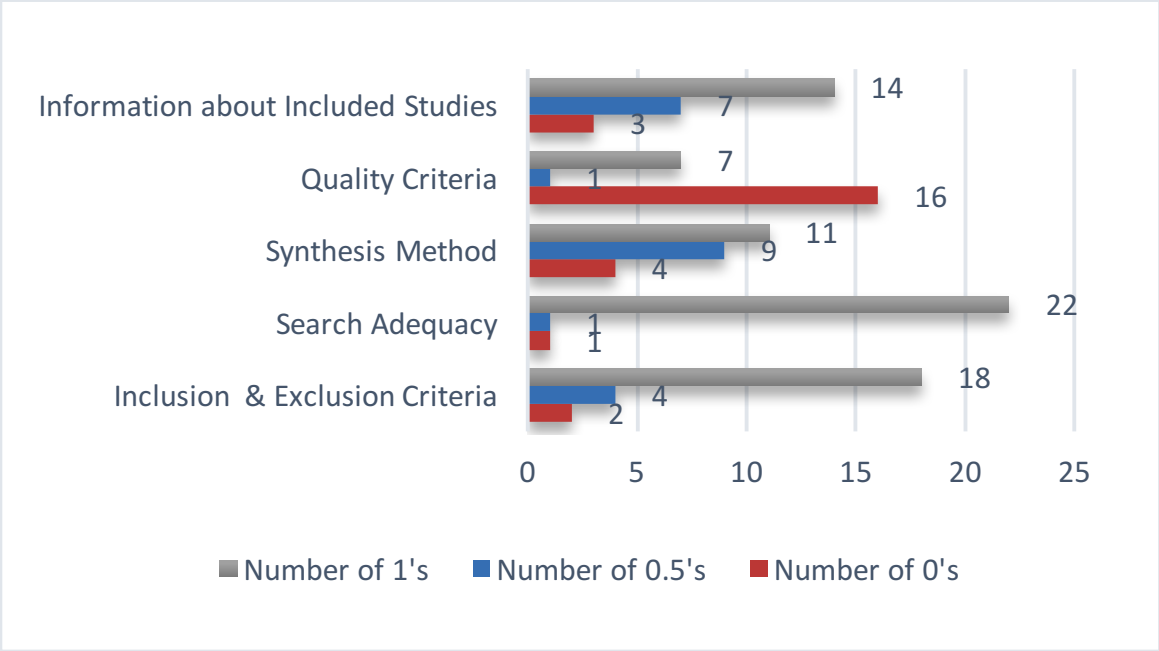

*Figure 24. Quality distribution of SMSs in the area of Software Testing (SWEBOK)*

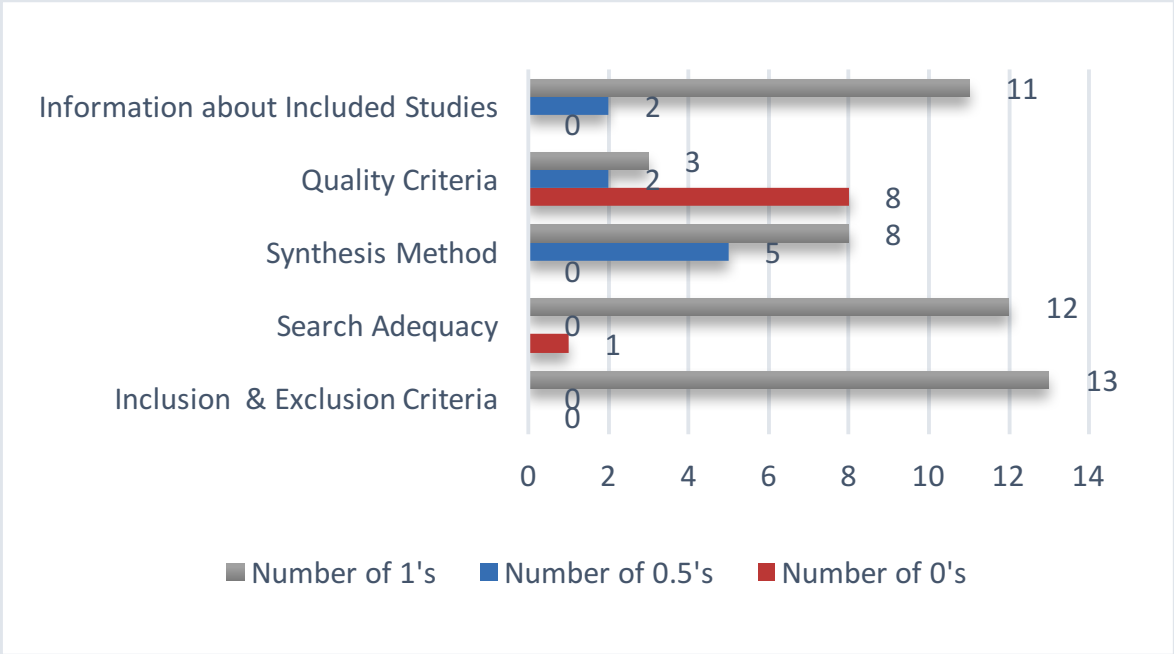

*Figure 25. Quality distribution of SMSs in the area of Software Maintenance (SWEBOK)*

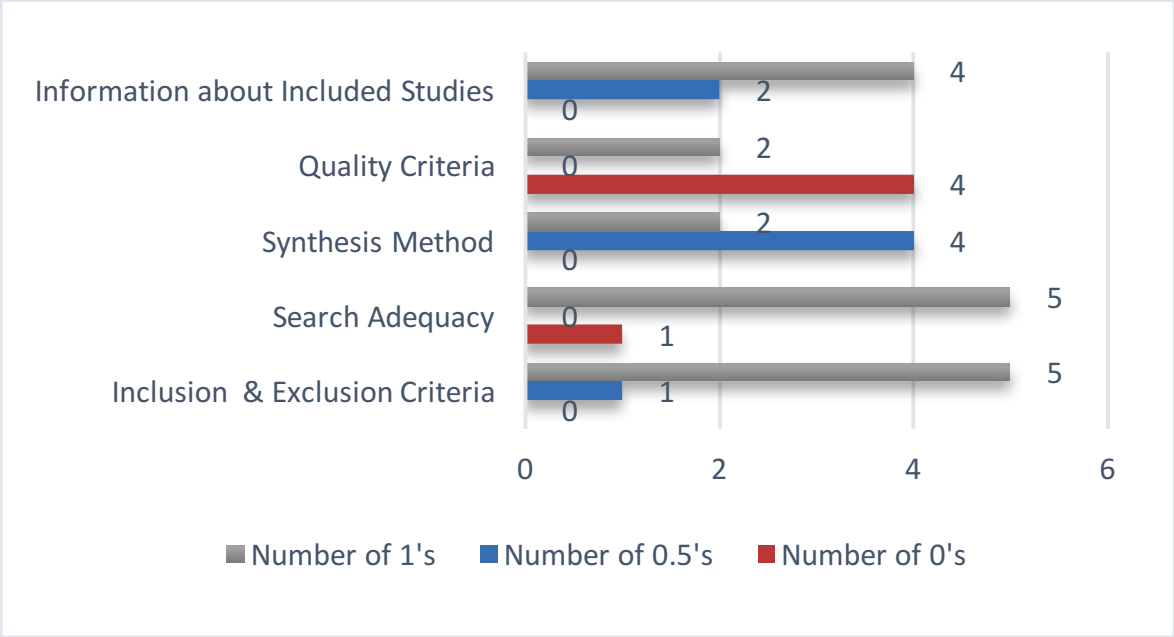

*Figure 26. Quality distribution of SMSs in the area of Software Configuration Management (SWEBOK)*

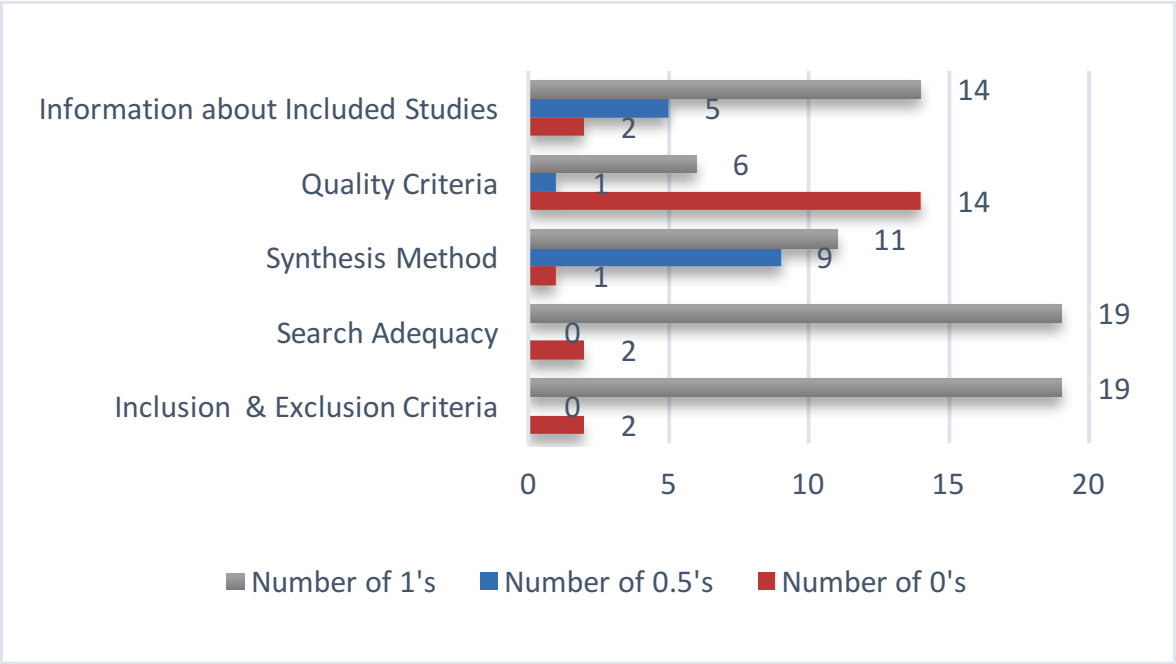

Figure 27. Quality distribution of SMSs in the area of Software Engineering Management (SWEBOK)

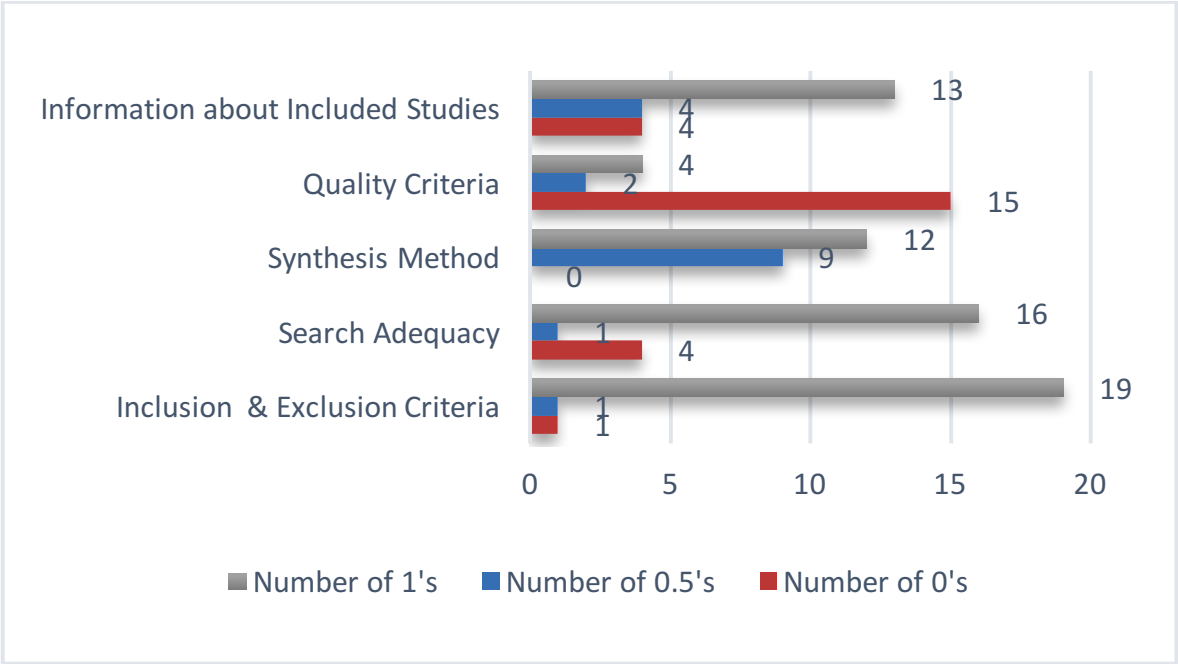

Figure 28. Quality distribution of SMSs in the area Software Engineering Process (SWEBOK)

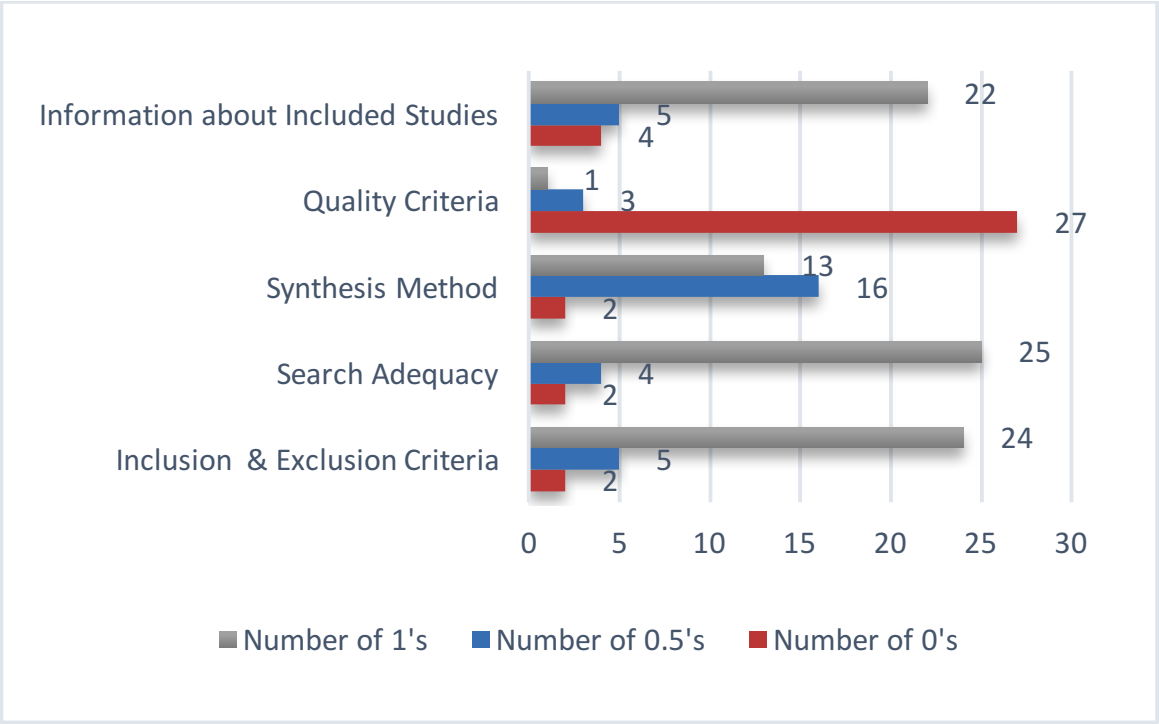

Figure 29. Quality distribution of SMSs in the area of Software Engineering Models & Methods (SWEBOK)

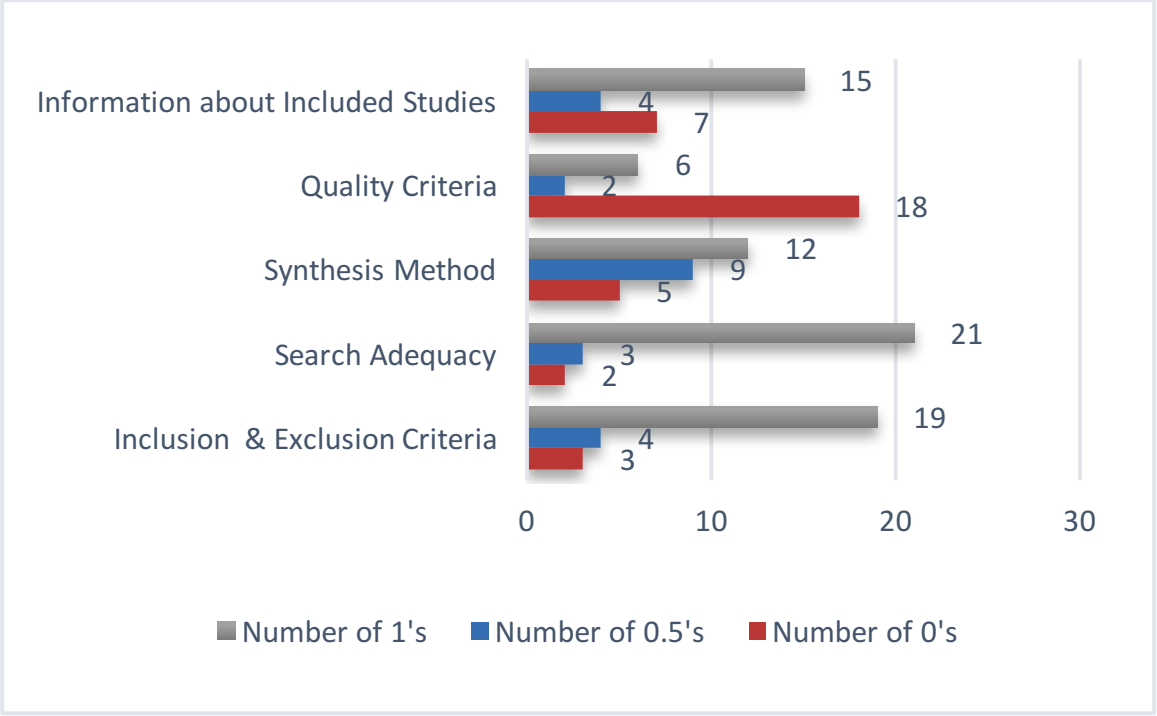

Figure 30. Quality distribution of SMSs in the area of Software Quality (SWEBOK)

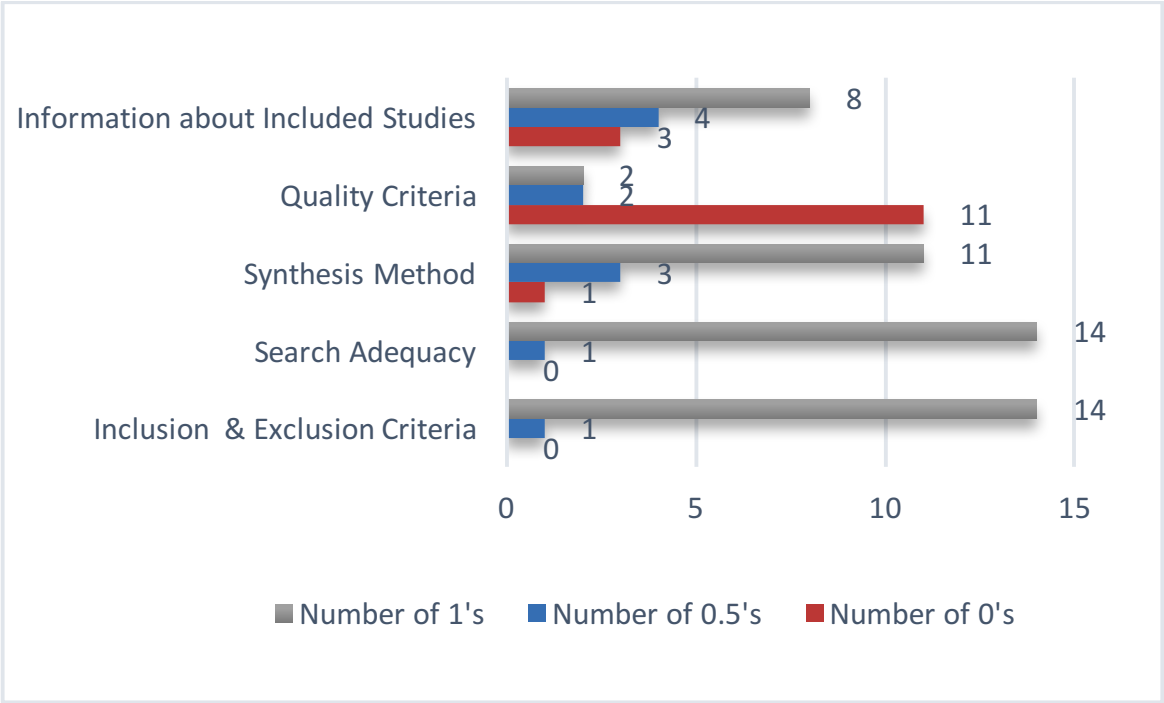

Figure 31. Quality distribution of SMSs in the area of Software Engineering Professional Practices (SWEBOK)

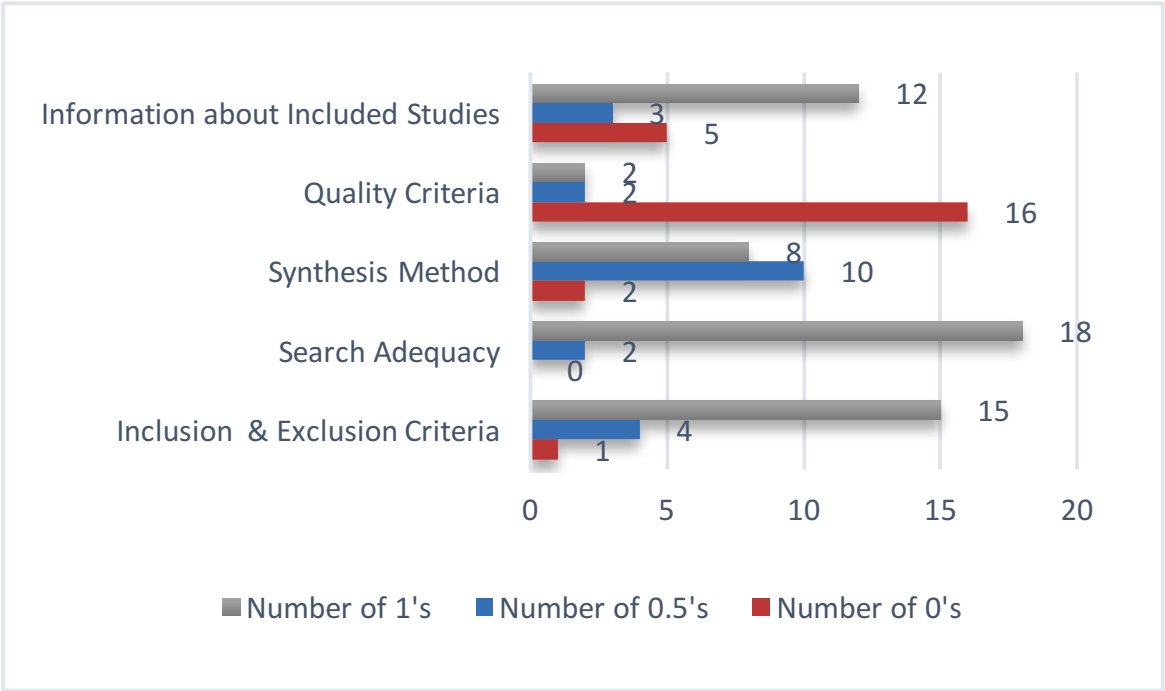

Figure 32. Quality distribution of SMSs in the area of Software Engineering Economics (SWEBOK)

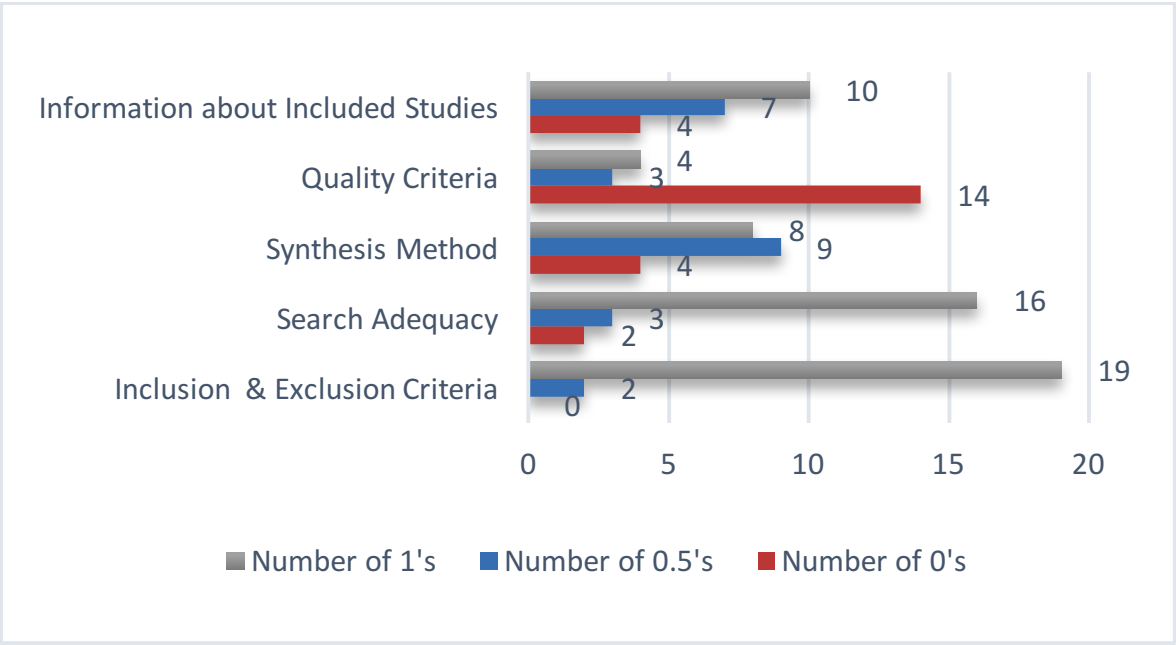

Figure 33. Quality distribution of SMSs in the area of Computing Foundations (SWEBOK)

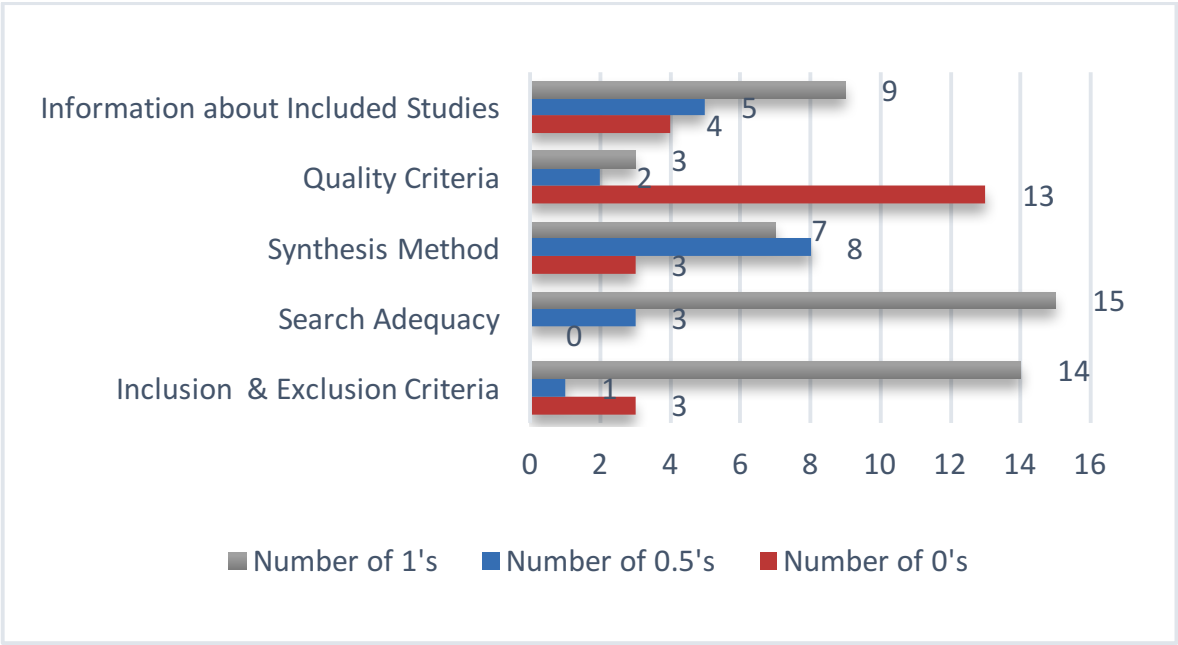

Figure 34. Quality distribution of SMSs in the area of Engineering Foundations (SWEBOK)

To answer *RQ 2.5*, we present and discuss a number of graphs (Figure 14-Figure 34) which shows the quality distribution of included SMSs. To summarize, we can say that most of the SMSs in different SE areas did not assess the quality of included primary studies (QC4) whereas most of the SMSs partially fulfilled the QC3 (Synthesis Method). Most of the SMSs do meet QC1 (inclusion and exclusion criterion) and QC2 (search adequacy criterion).

**RQ3 what are the current trends in SMSs relating to guidelines, data sources, types of questions and number of included studies?**

This question summarizes the information regarding various trends of SMSs in SE. These include information such as the most frequently followed guidelines, information on the most cited papers, and the most active researchers publishing SMSs.

**RQ3.1 Which guidelines are most frequently cited for conducting SMSs?**

We identified 15 different papers that contained guidelines for performing secondary studies (shown in Table A1). Figure 35 shows number of papers following each of the identified guidelines. We found 79 papers that followed multiple guideline. For example S186, cites both the guidelines proposed by Peterson *et al.* [1] and the guidelines proposed by Kitchenham *et al.* [2]. In such cases we credit both the papers. The guidelines by Peterson *et al.* [1] appear to be most widely used and we identified 119 SMSs that explicitly claim to follow these guidelines (not exclusively). We found six SMSs citing the guidelines proposed by Peterson *et al.* in [1]. Guidelines by Kitchenham *et al.* [2] are referred by 73 SMSs. Another 73 SMSs claim to follow the guidelines presented by Kitchenham *et al.* [3, 4, 6, 8] [5]. There are 17 SMSs that claim to follow the guidelines proposed in Budgen *et al.* [7]. We identified 12 papers that did not cite any guidelines and we could not infer the information about the followed guidelines. This is represented by *N/A* in Figure 35. We can conclude that works by Peterson *et al.* and Kitchenham *et al.* have the most influence on how the SMSs are conducted in SE and the guidelines proposed by the two works have a high level of acceptance in EBSE community.

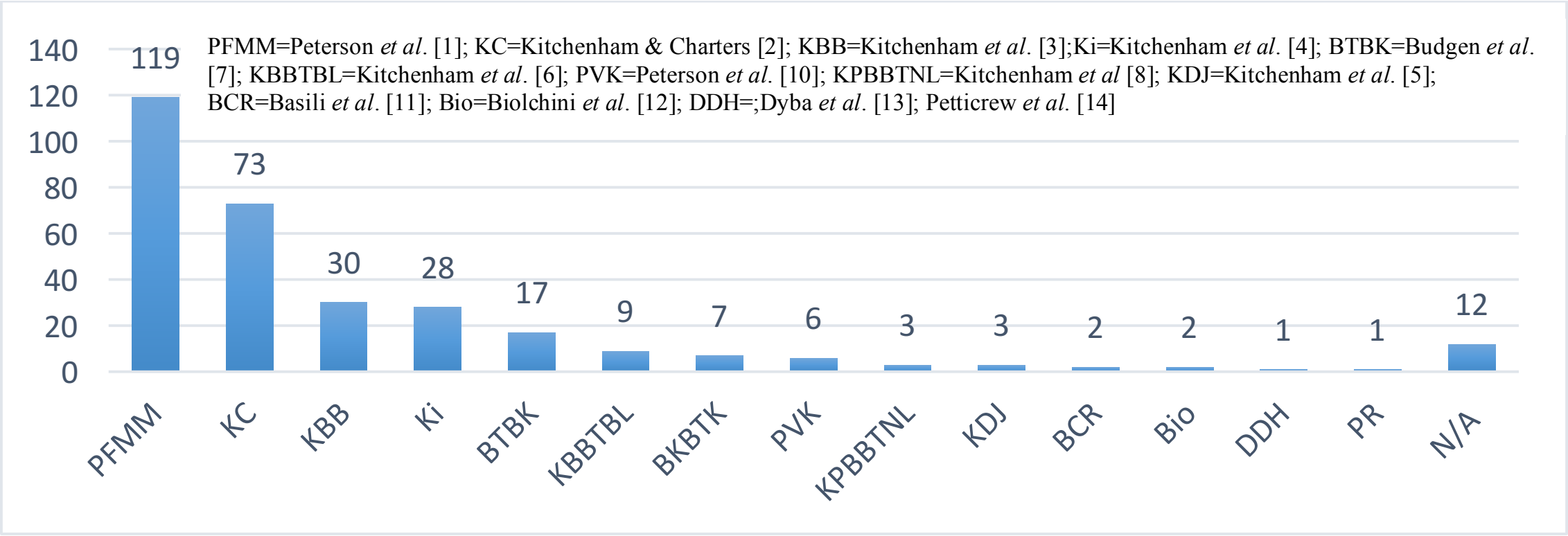

*Figure 35. Number of SMSs using each of the identified guidelines*

**RQ3.2 Which online databases are most frequently included as sources?**

We identified 28 different online databases used by the included 210 SMSs. Table A*2* in the Appendix shows the usage statistics for all of the online databases and Figure 36 shows the 15 mostly frequently used online databases. *IEEEXplore* is the most frequently cited database and is mentioned as a data source in 183 of 210 included SMSs. *ACM* is the second most frequently included source and is included in 177 SMSs. *Science Direct* is included in 117 SMSs while *Springer Link* is included in 98 SMSs. Other important data sources include *Scopus* (95), *Web of science* (56), *Google scholar* (51), *Compendex* (44) and *Inspec* (36). Whereas, *IET Digital Library, ArXiv*, *Inderscience*, *Kluwer online*, *EMBASE* and *sciVerse* are the least used online databases.

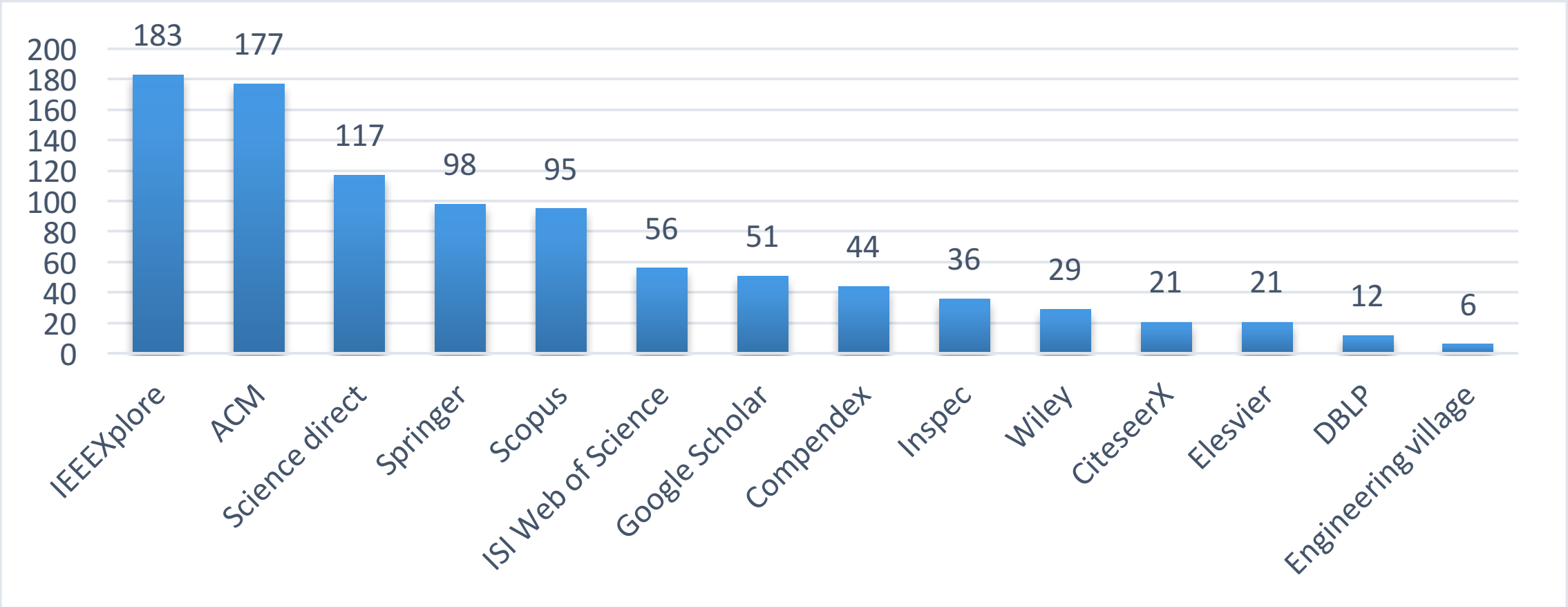

Figure 36. Online databases used by SMSs

**RQ3.3 What type of research questions are most frequently addressed by SMSs?**

Formulation of research questions is an essential part of conducting an SMS. An SMS provide insights on the included primary studies through answers to the posed research questions. In *RQ3.3*, we analyze the kind of questions being posed by SMSs in SE. The answer will allow us to analyze the usefulness of the information provided by SMSs.

Following our established protocol, two authors independently classified the type of questions addressed by the SMSs into four distinct categories, i.e., *current trends*, *demographics*, *research gaps* and *quality assessment*. Both authors analyzed every question posed by 210 SMSs and placed it into one of the above-mentioned categories. Any conflicts between the two researchers were resolved through meetings and discussions where all authors participated.

Table 9 presents the categories with the examples of questions included in these categories. *Current trends* include the questions regarding the existing primary work and identify the nature of existing work in the area. *Demographics* include questions that determine the characteristics of papers in the area, for example, top cited papers, top researchers and top venues, etc. *Research gaps* contain questions that identify gaps in the area and provide future directions. *Quality assessment* comprises of questions that assess the quality of the included primary studies.

Table 9. Categories of research questions with examples

| Category | Sample questions |
|---|---|
| **Current trends** | • What are the existing techniques in the area?<br>• What are the active areas of research in the area?<br>• What is the automation level of techniques?<br>• What is the type of contribution made by study?<br>• What is the type of research made by study. |
| **Demographics** | • Which are the most cited papers in the area?<br>• What is the annual distribution of papers?<br>• What are the top venues?<br>• Who are the top researchers?<br>• Which countries are most active in publishing relevant research papers? |
| **Research gaps** | • What are the limitations of the existing techniques?<br>• Which areas of research needs the attention of researchers?<br>• What are the potential gaps in the area?<br>• What are directions for new research? |
| **Quality assessment** | • What is the quality of papers published in the area?<br>• Is the quality of papers improving over time?<br>• What are the quality papers in the area?<br>• How to improve the quality of papers in the area?<br>• What is the reporting quality of papers? |

Figure 37 shows the distribution of SMSs containing each type of questions. Around 72% of the papers (152 out of 210) directly pose questions relating to the demographics in the area. Almost every SMS (207 out of 210) contained questions related to current trends. A closer analysis of the three remaining SMSs (S30, 145, S148) show some related information, but there were no research questions that explicitly identified current trends. Most SMSs aim to build a systematic map of the domain and identify research gaps in literature and we found 169 SMSs that explicitly defined such research questions. The remaining SMSs contained some discussion on research gaps but no research questions were explicitly posed to identify gaps. Quality assessment of included studies is a weak point as only 47 out of 210 included SMSs assess the quality of included studies. Quality assessment is important to ensure that the results of SMSs are not being biased by including very low-quality research papers.

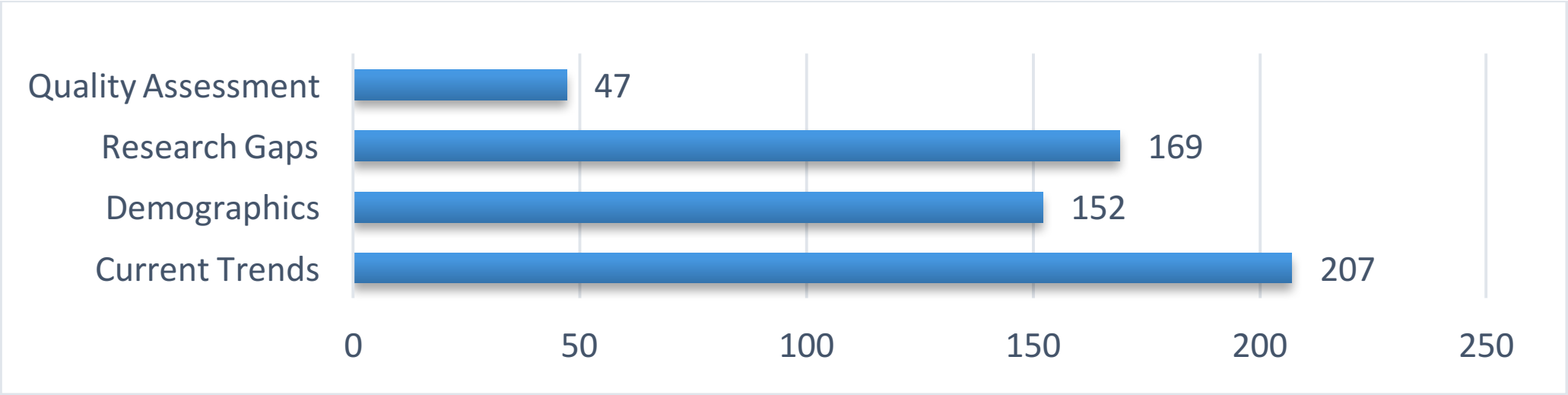

Figure 37. Types of questions addressed by SMSs

**RQ3.4 How many primary studies are included on average in published SMSs?**

Each included SMS contains a varying number of included studies. The included SMSs contain approximately 100 primary studies on average, with a high standard deviation of 148.28. A total of 20,466 primary studies are included in the published SMSs (includes repetitions and overlap between studies). S28 aggregates the least number of primary studies at 9, while S161 contains the highest number of included primary studies (1440); both the papers are conference papers. For journal papers, S2 contains the highest number of primary studies where 679 primary studies are included. Figure 38 (b) shows a scatter plot of number of primary studies included in each SMS for each year since 2007 while Figure 38 (a) shows the boxplot of included primary studies. One reason for such high number of included primary studies is the broadness of area. For example, S161 reports on a wide-ranging area of domain specific languages and S115 aggregates the mechanisms to perform empirical studies, both of which contain extensive works. The number of primary studies included in each SMS is given in Table A *1*.

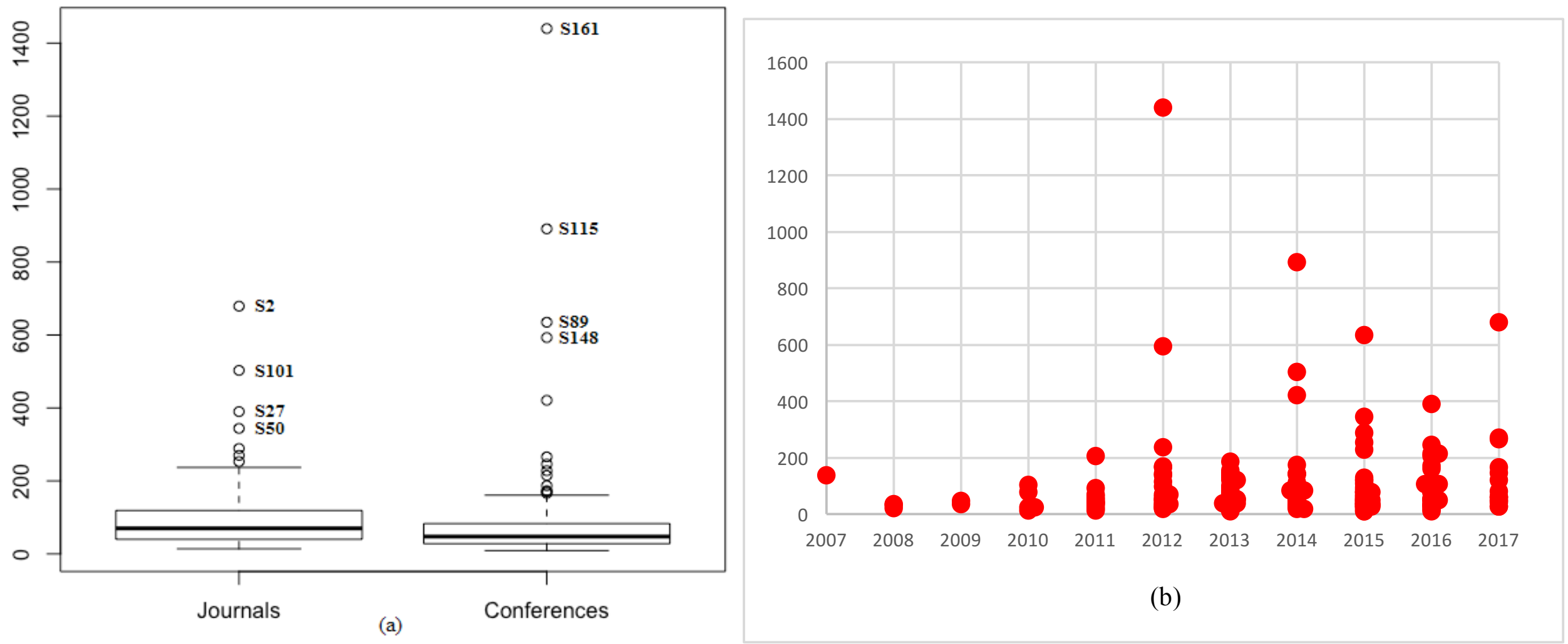

*Figure 38 Number of included primary studies (a) box plot (b) scatter plot*

**RQ3.5 Is there an increase in the number of included studies over the years?**

An SMS aggregates primary studies published in a given area. Number of primary studies included in SMSs in a given year can be taken as a rough indicator of how active the area is and how many primary studies are published each year. We use spearman's rank order correlation to analyze the relationship between the number of included studies over time. As more and more primary studies are published, we expect to see an increase in the number of included primary studies in recent years. The resultant values of $r_s = 0.13$ and p-value $= 0.04$ show a weak correlation but statistically significant results, indicating that the number of primary studies included in published SMSs is increasing.

**RQ4 What are the demographics of published SMSs in SE?**

A common contribution of secondary and tertiary studies is to provide demographic information on the area being studied. In the following we briefly answer various sub-questions regarding the demographics of SMSs in SE.

**RQ4.1 How many SMSs are published annually? What is the publication trend?**

Distribution of SMSs over the years is shown in Figure 39. It can be observed that since the publication of first SMS in 2007, there is a steady increase in number of SMSs published annually. In particular since 2010 there is a rapid increase in the number of published SMSs. Out of 210 papers, 102 were published in journals, 102 were published in conferences and only 6 were published in workshops. The year 2015 has the highest frequency of published SMSs at 52, whereas the following year has only 30 published SMSs. The data for 2017 is not complete as we only include papers till august 2017, however the data from first eight months (15 SMSs) suggest that the number of SMSs would remain lower than 2015. A possible reason can be publications of large number of SMSs aggregating various SE fields and consequently there are fewer gaps that remain to be covered by secondary studies.

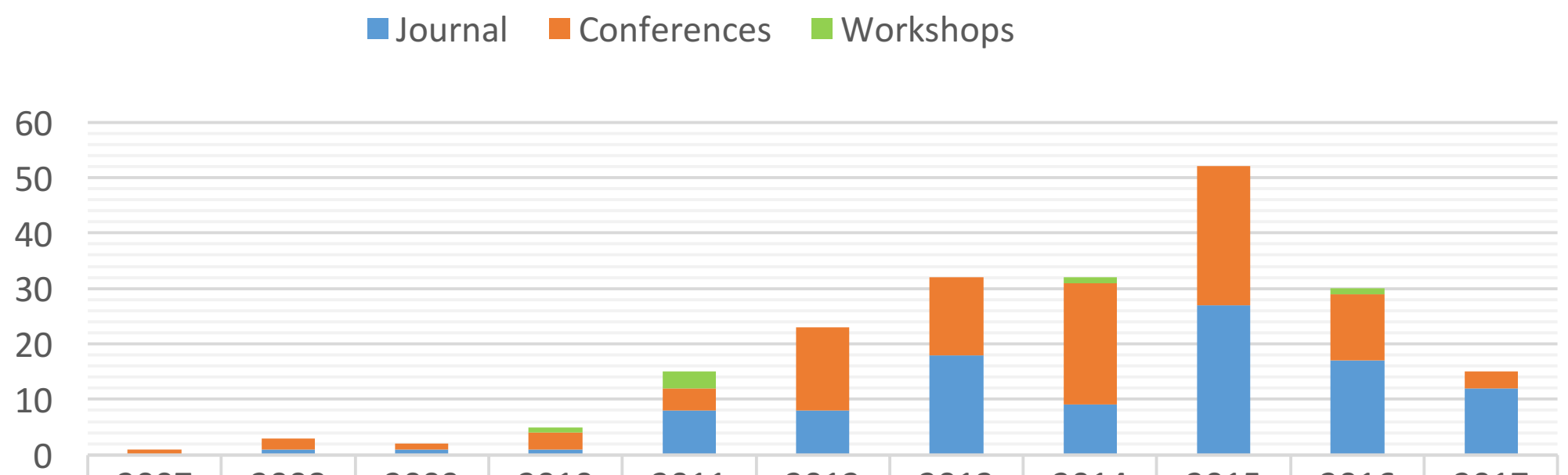

| | 2007 | 2008 | 2009 | 2010 | 2011 | 2012 | 2013 | 2014 | 2015 | 2016 | 2017 |
|---|---|---|---|---|---|---|---|---|---|---|---|
| Workshops | 0 | 0 | 0 | 1 | 3 | 0 | 0 | 1 | 0 | 1 | 0 |
| Conferences | 1 | 2 | 1 | 3 | 4 | 15 | 14 | 22 | 25 | 12 | 3 |
| Journal | 0 | 1 | 1 | 1 | 8 | 8 | 18 | 9 | 27 | 17 | 12 |

Figure 39. Annual distribution of papers

**RQ4.2 Which venues publish SMSs most frequently?**

Almost half of all SMSs published in journals are published in Information and Software Technology (IST) and Journal of Systems & Software's (JSS). Evaluation and Assessment in Software Engineering (EASE) and Empirical Software Engineering and Measurement (ESEM) are the most frequent conference venues for SMSs. These four publication venues combine for approximately 38% of all published SMSs in SE. Similar results have been observed in other tertiary and secondary studies [10, 29, 30]. Table 10 shows the publication venues that have more than three SMSs published. A full list appears in Table A *1* in the Appendix.

Table 10. Top venues

| S.no | Venue type | Venue name | Acronym | Count |
|---|---|---|---|---|
| **1** | Journal | Information and Software Technology | IST | 37 |
| **2** | Journal | Journal of Systems & Software's | JSS | 18 |
| **3** | Journal | Journal of Software's: Evolution and Process | JSEP | 4 |
| **4** | Journal | Empirical Software Engineering | ESE | 3 |
| **5** | Conference | Evaluation and Assessment in Software Engineering | EASE | 14 |
| **6** | Conference | Empirical Software Engineering and Measurement | ESEM | 10 |
| **7** | Conference | Software Engineering and Advanced Applications | SEAA | 4 |
| **8** | Conference | Asia-Pacific Software Engineering | APSE | 4 |

### RQ4.3 Which are the most cited papers in the area?

Citation count can loosely be taken as an indicator of the impact of a given research paper. The citation data included in this paper is extracted from Google Scholar on 11th September 2017. The Google Scholar citation counts are not the most reliable way of measuring actual citations as they do consider citations from technical reports, etc., in the count. Nevertheless these are a useful measure for identifying how much a paper has been cited. Our included SMSs are cited a combined total of 4,741 times with an average of nearly 22.57 citations per paper and a standard deviation of 37.87. Figure 40 (a) shows the number of citations for each included SMS. As expected, citations are highly influenced by publication date (among other factors), and papers published earlier tend to have more citations. Also journal papers tend to have higher citations as compared to conference papers.

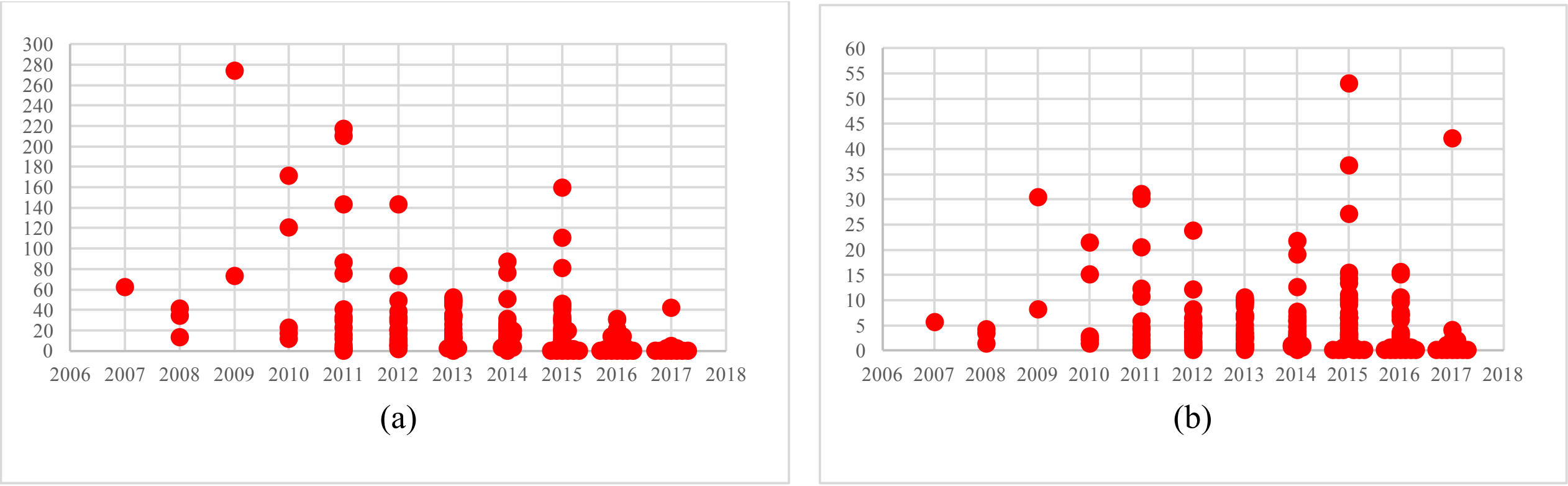

*Figure 40. (a) Total citations over the years (b) Normalized citations over the years*

Since these papers have different publication year, it would be appropriate to take into account the age (calculated as *current year – publication year*) of each SMS in evaluating the citation count. Figure 40 (b) shows the normalized number of citation for each SMS and is calculated as:

$$Normalized\ number\ of\ citations = \frac{Total\ Citations}{(Current\ year - publication\ year\ )}$$

Table 11 and Table 12 lists the top cited journal and conference papers using total citation count and normalized citation counts respectively. S181 is the highest cited SMS based on total citations while S60 is the highest cited SMS based on normalized citations.

Table 11. Top cited Papers based on number of citations (extracted on 11 September 2017)

| Ref | Paper Title | Venue | Year | # of citations | Normalized citations |
|---|---|---|---|---|---|
| **Top 5 cited SMSs published in Journals** | | | | | |
| **S181** | **A systematic review of search-based testing for non-functional system properties** | **IST** | **2009** | **274** | **30.44** |
| S168 | Software product line testing - A systematic mapping study | IST | 2011 | 217 | 31 |
| S165 | Usability evaluation methods for the web: A systematic mapping study | IST | 2011 | 210 | 30 |
| S177 | What's up with software metrics? - A preliminary mapping study | JSS | 2010 | 171 | 21.37 |
| S60 | Gamification in Education: A Systematic Mapping Study | ETS | 2015 | 159 | 53 |
| **Top 5 cited SMSs published in Conferences** | | | | | |
| **S179** | **Agile practices in global software engineering - A systematic map** | **ICGSE** | **2010** | **120** | **15** |
| S104 | A systematic mapping on gamification applied to education | SAC | 2014 | 87 | 21.75 |
| S182 | A systematic mapping study on empirical evaluation of software requirements specifications techniques | ESEM | 2009 | 73 | 8.11 |
| S185 | How software designs decay: A pilot study of pattern evolution | ESEM | 2007 | 62 | 5.63 |
| S172 | Alignment of Requirements Specification and Testing: A Systematic Mapping Study | ICSTW | 2011 | 40 | 5.71 |

Table 12. Top cited SMSs based on Normalized citation (extracted on 11 September 2017)

| Ref | Paper Title | Venue | Year | # of citations | Normalized citations |
|---|---|---|---|---|---|
| **Top 5 most cited Journal papers based on Normalized citations** | | | | | |
| **S60** | **Gamification in Education: A Systematic Mapping Study** | **ETS** | **2015** | **159** | **53** |
| S14 | Continuous deployment of software intensive products and services: A systematic mapping study | JSS | 2017 | 42 | 42 |
| S69 | A systematic mapping study on technical debt and its management | JSS | 2015 | 110 | 36.66 |
| S168 | Software product line testing - A systematic mapping study | IST | 2011 | 217 | 31 |
| S181 | A systematic review of search-based testing for non-functional system properties | IST | 2009 | 274 | 30.44 |
| **Top 5 most cited Conference papers based on Normalized citations** | | | | | |
| **S104** | **A systematic mapping on gamification applied to education** | **SAC** | **2014** | **87** | **21.75** |
| S179 | Agile practices in global software engineering - A systematic map | ICGSE | 2010 | 120 | 15 |
| S182 | A systematic mapping study on empirical evaluation of software requirements specifications techniques | ESEM | 2009 | 73 | 8.11 |
| S123 | Tools to support systematic literature reviews in software engineering: A mapping study | ESEM | 2013 | 34 | 6.8 |
| S97 | Integrating agile and user-centered design: A systematic mapping and review of evaluation and validation studies of agile-UX | AGILE | 2014 | 27 | 6.75 |

From Table 11 and Table 12 we can observe that most cited papers are published in the most frequent publication venues for journals. i.e., IST and JSS. S60 which is published in ETS is an exception. ETS has a wider audience in the education domain and the SMS covers an active area of research of utilizing game concepts to improve education.

### RQ4.4 Are SMSs published in journals cited more often than SMSs published in conferences?

Journal papers tend to receive higher number of citation as compared to conference. The 102 included SMSs published in journals have on average 33.19 citations per paper. The average number of citations for SMSs published in conferences is much lower at 12.0 citations per paper. The top 5 cited SMSs in SE are all published in journals. The highest cited conference paper has only 120 total citations compared to the highest cited journal paper that has 274 citations. There are 11 conference papers with no citations compared to 12 journal papers with no citations. There are 27 journal papers that have between 1-10 citations as compared to 52 conference papers in the same range, as shown in Table *13*. We found only 9 SMSs (out of 102) published in conference and 37 (out of 102) journal papers that have more than 30 citations. Compared to eight journal papers, there is only one SMS published in a conference that has more than 100 citations. Figure 41 (a) depicts the number of citations *vs.* the number of papers published in journals while Figure 41 (b) shows the number of citations *vs.* the number of papers published in conferences.

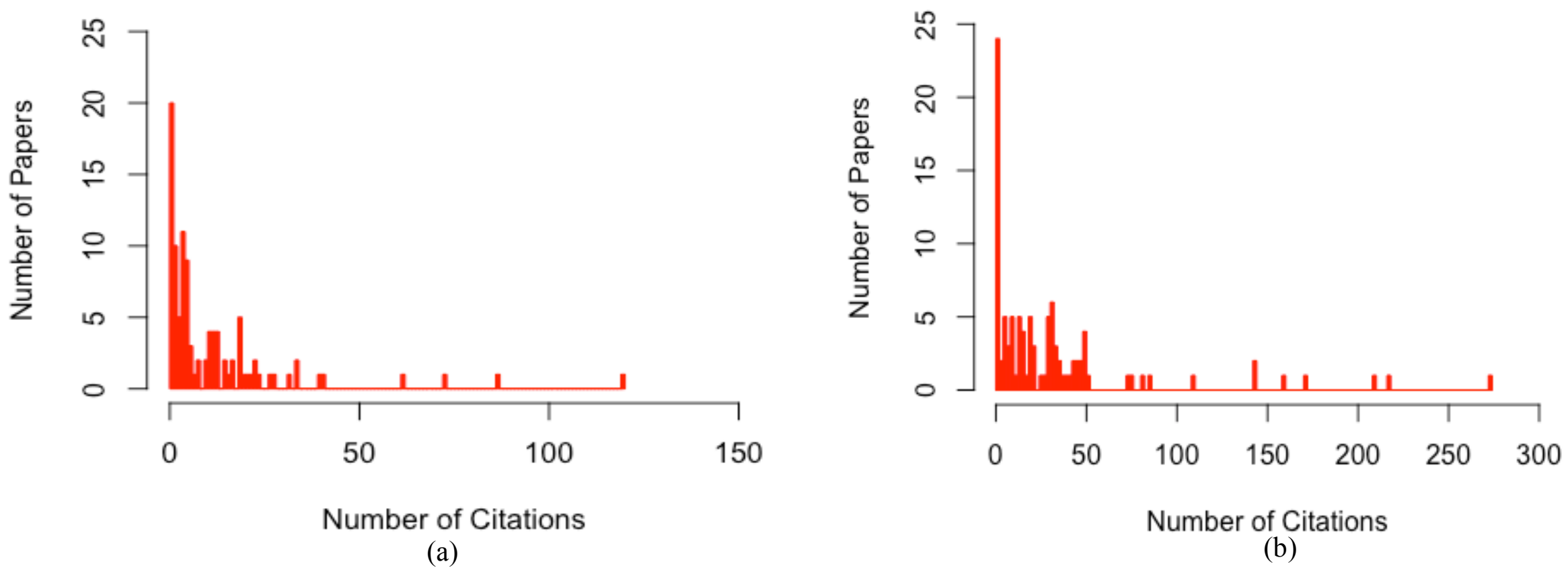

(a) (b)

*Figure 41 Number of papers vs citation count (a) Conferences, (b) Journals*

Table 13 Distribution of citation counts of SMSs

| Ranges (# of citations) | # Citations (journals) | # Citations (Conferences) | # Citations (Workshops) | # of Papers |
|---|---|---|---|---|
| 0 | 12 | 11 | 0 | 23 |
| 1 – 10 | 27 | 52 | 2 | 81 |
| 11 – 30 | 26 | 30 | 3 | 59 |
| 31 – 60 | 25 | 5 | 0 | 30 |
| 61 – 100 | 4 | 3 | 1 | 8 |
| 101 – 200 | 5 | 1 | 0 | 6 |
| 201 – 300 | 3 | 0 | 0 | 3 |

The results of one tailed Wilcoxon rank sum test [41] show that the journal papers have significantly higher citation counts as compared to conference papers ($n_1$ = 102, $n_2$ = 102, w = 5908, p = 0.00002). The magnitude of the difference is given by Cohen D = 0.57, which indicates a medium sized effect on citation counts due to the nature of publication.

**RQ4.5 Who are the active researchers with most published SMSs?**

To show the highest publishing authors in the area we have followed the same approach, followed in other studies, like [42], by counting the number of published SMSs by each author. For simplicity, we have shown the top authors who have at least published four SMSs in our included studies, as given in Table 14. The ranking of the authors, based solely on the number of publication is as follow: Fabio Q B da Silva has published 7 SMSs and is the author with most published SMSs. Ali Idri, Mario Piattini and Jose Luise Fernandez Aleman have published 6 SMSs each. They are followed by Vahid Garousi, Paria Avgeriou, Marcela Genero and Ambrosio Toval with 5 published SMSs. Consequently, Wasif Afzal and Robert Feldt have 04 publications each.

*Table 14. Key researchers*

| S.No | Author | References |
|---|---|---|
| 1 | Fabio Q B da Silva | S32, S47, S51, S109, S169, S171, S199 |
| 2 | Ali Idri | S25, S65, S66, S85, S141, S209 |
| 3 | Mario Piattini | S53, S81, S91, S97, S105, S159 |
| 4 | Jose Luis Fernandez Aleman | S25, S50, S85, S141, S165, S209 |
| 5 | Vahid Garousi | S55, S88, S124, S136, S201 |
| 6 | Paris Avgeriou | S9, S43, S69, S111, S122 |
| 7 | Marcela Genero | S91, S97, S105, S117, S143 |
| 8 | Ambrosio Toval | S25, S50, S85, S141, S209 |
| 9 | Wasif Afzal | S5, S59, S181, S194 |
| 10 | Robert Feldt | S172, S181, S183, S194 |

## 5. Discussion: Interpretation and Implications

In this tertiary study, we reviewed 210 secondary studies published in SE till August 2017. We identified 102 SMSs published in journals, 102 SMSs published in conferences and 6 SMSs published in workshops. In this section, we reflect on implications of our tertiary study.

### 5.1. Reflection on SM in SE

Systematic mapping studies have gradually become popular in SE with more than 60% of the SMSs published in the past four years: 2014 to 2017. We found that not all areas of SE are well covered by SMSs. Some of the SE areas, such as *software testing* and *requirements engineering* have significantly more SMSs than areas such as *software maintenance* and *software configuration management*. We did not find any SMS on SE tool usage and evaluations or on mathematical and computational foundations in SE. Tools play an important role in managing various SE activities and we consider a lack of focus on SE tools to be an important omission.

We found that while the overall quality of published SMSs is improving in recent years, there is a significant difference in the quality of SMSs published in journals as compared to SMSs published in conferences. More than half of the SMSs in journals are published in IST and JSS. The SMSs published in both these journals have a higher aggregate quality score as compared to the SMSs published in other journals. Most of the published SMSs focus on identifying current trends and research gaps in the existing literature. A significant number of SMSs do not perform any quality assessment of primary studies. This is an important factor to consider when conducting secondary studies, because a quality assessment may have a significant impact on the results drawn from an SMS. For example, lack of quality assessment criteria may result in an inaccurate demographic information drawn from an SMS.

With the passage of time, the SMSs are including more primary studies, as discussed in detail in *RQ3.5*. This can be attributed to the increasing amount of primary studies being published in various areas of SE. However, this also poses a challenge for the authors of SMSs. One key reason for conducting an SMS rather than an SLR is that they require relatively less effort [1, 4]. With the increasing number of primary studies the required effort for conducting an SMS is also significantly increasing. Consequently, we have identified some SMSs that exclude a number of mandatory steps provided in the various guidelines to reduce the effort. For example, the studies S2, S89 and S101 do cover a large number of primary studies, however they do not provide synthesis on the extracted data and basic information about the primary studies that are considered mandatory for conducting SMSs [27, 40].

Citations are considered as an indirect measure to evaluate to impact of a published work. Our study indicates that the SMSs published in journals have higher citation counts as compared to SMSs published in conferences. This can be attributed to a general perception of journal papers being more prestigious.

### 5.2. Implications for Software Engineering Education

Software Engineering Book of Knowledge (SWEBOK) [17] plays an important role in SE and is referenced by a number of curriculum guidelines, including ACM graduate curriculum guidelines [43]. We map the identified SMSs in the area of software engineering to the various knowledge areas included in SWEBOK. Such a mapping allows us to (i) identify areas that currently lack research focus, (ii) highlight areas that are more popular in terms of research, and (iii) identify any area that is attracting attention in terms of research publications but is not covered by SWEBOK. SMSs aggregate existing primary studies and a lack of SMSs in an area can be considered as an (weak) indicator of less papers being published in the area [19, 29].

The mapping of the SMSs on SWEBOK is shown in Table 6. The table shows a number of areas that lack SMSs, especially in certain 'sub areas' of software engineering (mentioned in the "sub areas" column of the table). We identify a distinct lack of SMSs covering SE tools ranging from requirement management tools to software design, analysis, and testing tools. In our opinion, there is, in general, a lack of research papers on empirical evaluation of existing commercial and open source tools. Most papers tend to focus on in-built research prototype tools that are not available for professional use. Also, at times, the publication venues can be harsh on papers that present tool comparisons. This is typically because such empirical evaluations mostly do not cover the entire spectrum of available tools and are considered as incomplete studies or such evaluations are considered to not provide significant 'research' contributions. Further efforts should be made to increase awareness about the tools. Surprisingly, important SE areas such as *configuration management* also lack sufficient coverage by SMSs. Configuration management plays an important role in software industry, therefore more efforts should be focused on research and evaluation of configuration management approaches.

Though the SWEBOK classification allows a mapping of existing SMSs in various SE knowledge areas in a consistent manner, there are some SMSs that are difficult to map onto an existing class. We identify a total of 18 papers that either only partially map to existing knowledge areas of SWEBOK or cannot be mapped at all. *Table 15* provides a mapping of such studies. Three new areas (not covered in SWEBOK) are identified; Software Engineering Education, Gamification in Software Engineering, and Social Computing. These three areas are attracting attention recently and in our opinion, require an explicit placement in SWEBOK knowledge areas. Seven studies that could only be partially mapped, but are from diverse domains are placed in the miscellaneous category.

Table 15. New proposed classes for SWEBOK

| S.no | Proposed Knowledge Area | SMSs References |
|---|---|---|
| 1. | Software Engineering Education | S49, S83, S60, S85, S90, S104, S195, S206 |
| 2. | Gamification in Software Engineering | S81, S60, S104, S105 |
| 3. | Social Computing | S191, S114 |
| 4. | Miscellaneous | S81, S96, S93, S68, S47, S109, S197 |

## 5.3. Implications for research community

Our study can benefit researchers in various ways. The classification of existing SMSs in SE can help new researchers in targeting research areas that currently lack secondary studies. Another way of identifying interesting areas could be to combine most popular areas across different phases of software development lifecycle as proposed by Garousi *et al.* [19]. Alternatively, researchers can focus on less explored areas by combining least popular areas or a hybrid combination of areas to identify topics not yet covered by existing SMSs. Thus building on the examples provided in [19], an interesting topic for a future SMS can be "*Empirical evaluation of tool support for Software Architecture Description Languages*" that combines the popularity of software architecture and empirical evaluations with low level focus on software engineering tools. Other gaps can also be directly identified as noted by a large number of sub-areas that have less or no published SMS, i.e., popular sub-areas of *Software Testin*g: *Software Testing Tools, Test Levels, Test Measures*, *Software Configuration Management* and *Mathematical Foundations (discussed in detail in section 4)*. Our study can also be useful for researchers new to systematic mapping studies by providing meta-information from the existing published SMSs. New researchers can use the identified information and trends as a set of recommendations to follow. For example, following are some of the recommendations that can be extracted from our study: (i) The guidelines for SMSs that are most widely accepted by editors and reviewers, (ii) The databases that are expected to be included when searching for source studies, (iii) The type of questions that should be posed (iv) The suitable venues for publishing an SMS, and (v) The type of publications (journal/conference) that have a higher impact. Finally, the presence of a significant number of papers that cite multiple guidelines that are being followed indicate that the current set of guidelines need further refinement and may be consolidated, particularly in the view of increasing number of primary studies.

### 5.4. Implications for practitioners

Most of the tertiary studies focus on providing information to researchers as well as practitioners. Our study can be useful for practitioners as it serves as an index of existing studies in SE domain. Practitioners can identify the maturity of secondary studies in various sub-areas, loosely based on the quality of the published secondary works in each of the SE sub areas. This allows the practitioners to adopt the practices proposed by researchers with greater confidence and helps in locating the relevant evidence to support their industry usage, and in convincing the higher management to adopt certain practices. The keyword based thematic analysis identify the more active areas of research. For example, the high number of published works in software design indicates a more mature sub-area as compared to software configuration management, where relatively fewer works are published. Similarly, the high maturity of software testing indicates that this is an active area of research with new findings being published regularly.

## 6. Comparison of results with previous tertiary studies

We conduct a tertiary study that aims to provide a map of all SMSs published in SE. We evaluate the quality of the published SMSs and provide implications for academics, researchers and industry practitioners. As discussed in literature review, a number of previous tertiary studies exist in the domain with different focus and research questions. In this section, we compare the results of our study with related published tertiary studies, indicating where our results reinforce the previously published results and where they diverge. Table 16 provides a detailed comparison of our results with the existing tertiary studies.

Similar to previous three studies [8, 10, 18], we also identify that the quality assessment is still a weak area for SMSs. Our study includes recent papers published till August 2017 and finds that the claim of inadequate quality assessment is still valid. In general, 71% percent of the SMSs fail to adequately perform the quality assessment of the included primary studies. Similarly, our results on the most often used databases are partially similar to the results presented in [29]. Our findings also reinforce the findings in [19] (covering SMSs in software testing) that most of the SMSs in software testing focus on software testing techniques and that there is lack of secondary studies on test levels, measures and tools. Our findings on the guidelines used by SMSs are in line with Peterson *et al.* [10] that most of the papers used the guidelines provided in [1] and [2] for conducting an SMS. We also confirm their claim that software testing and software design have most number of SMSs. We also identified “software engineering education” as an area to be included in SWEBOK classification along with four additional areas (given in Table 15). Moreover, our findings regarding the top venues are also in agreement with other tertiary studies [10, 29]. However, our findings are not in line with the claim made by [24] which states that few studies perform research synthesis. According to our findings, there are only 10% studies which failed to perform synthesis, whereas most of the studies performed synthesis in one way or another. Our results also differ from [10] which claims that number of secondary studies published in conferences are greater than journals. According to our findings, number of SMSs published in journals are greater than conference publications.

Table 16. Comparison of our results with existing tertiary studies

| Ref | Area | Points | Inline/Dissimilar |
|---|---|---|---|
| [19] | Software Testing | No SMS is found on test automation tools | Inline |
| | | Test management, organization and monitoring has no SMS | Inline |
| | | No SMS found on test planning and estimation | Inline |
| | | SMSs are of poor quality due to not assessing the quality of included papers | Inline |
| [29] | Agile Development | Majority of SMSs are following standard guidelines | Inline |
| | | The number of SMSs in the area of agile is on rise | Inline |
| | | IEEEXplore and ACM are the most used databases | Inline |
| | | Majority of the studies have good quality | Inline |
| | | Studies published in Journals have good quality than conferences | Inline |
| [28] | Agile Development | Most of the papers in the area of agile are of low or medium quality | Inline |
| | | Most of the papers in the area fail to assess the quality of primary studies | Inline |
| [24] | Meta study | Few studies include research synthesis | Dissimilar |
| **[9]** | GSD | Most of the systematic reviews in GSD have explicitly mentioned the number of included primary studies | Inline |
| | | Quality assessment of primary studies is missing from most of the systematic reviews published in the area of GSD | Inline |
| [18] | MDE | Most of the papers failed to assess the quality of included studies | Inline |
| **[10]** | Meta study | Overall quality of SMSs is improving | Inline |
| | | Most of the SMSs failed to assess the quality of primary studies | Inline |
| | | The number of SMSs published in conferences are more than published in journals | Dissimilar |
| | | IST and EASE are top publishing venues | Inline |
| | | Suggested addition of new areas for SWEBOK e.g. Software Engineering Education etc. | Inline |
| | | IEEEXplore and ACM are the top most used databases | Inline |

## 7. Threats to validity

In this section, we discuss the various potential threats to the validity of our results.

*Internal validity:* An internal threat to our study can be the incomplete selection of studies. To reduce the threat, we have rigorously conducted our search process by identifying the search keywords, formation of search strings and execution of search strings according to the instructions given by each online repository. Furthermore, we have applied our inclusion and exclusion criteria for final selection of papers. Any conflicts in the final selection process were resolved through discussion and voting process to reduce the biasness of individual author.

*External validity:* we conducted this tertiary study in the domain of software engineering. Therefore, the results and findings are only valid in the given area. We believe that this study is repeatable because of the systematic protocol followed to search, gather and evaluate the data.
*Construct validity:* The construct validity is related to the correctness of constructs studied in the paper. In our study, selected papers were systematically mapped to our classification schemes after the selection of papers. At each stage, two authors independently extracted data from papers according to the identified attributes. Then the results of mappings are finalized in different review and group meetings where all authors participated.
*Conclusion validity:* In order to reduce this threat, the conclusion must be drawn correctly and should be reproducible for other researchers. In this study most of the graphs, charts and tables, which discusses the trends and observations, are directly generated from the data and therefore our results are directly traceable to the data. The data from secondary studies are extracted carefully and systematically, which is explained in section 3. This ensures high degree of reliability that the conclusion drawn in the study is directly traceable to the data and hence can be reproduced by other researchers. Another threat is the scoring scheme used for summarizing the overall quality of SMSs. We use DARE criteria for assessing the quality of the included studies. The assessment of quality depends on presence and absence or partial presence of information in the included studies. To categorize the studies broadly based on the quality criteria, we sum the obtained values as per the current practice followed by the recent SLR in top journals of the domain [6, 8, 29, 34]. There is a potential risk of arriving at wrong conclusion about the quality of the paper. However, we use the quality scores to place studies in broad categories and do not exclude studies based on quality scores calculated in this manner.

## 8. Conclusion

The goal of our tertiary study was to systematically map the published Systematic Mapping Studies (SMSs) in Software Engineering (SE) to identify various trends, including, the guidelines used, frequent publication venues, most cited papers, type of research questions answered with the goal of identifying gaps in literature, and to highlight future research directions. We provide quality assessment of the published SMSs and contrast the quality and differences in the impact (in terms of citations) of SMSs published in journals with the quality of SMSs published in conferences. We build a systematic map of SMSs in SE and identify active research areas through thematic analysis and discuss implications for software engineering education through mapping on SWEBOK knowledge areas.

Our study includes a total of 210 published SMSs till august 2017. While software engineering is a relatively new discipline, there is a huge body of published work. Identifying interesting works and developing an overall understanding of what is happening in software engineering is a daunting task for any new researcher and practitioner. Tertiary studies like ours play an important role by aggregating and indexing published secondary studies, which in turn aggregate the primary studies. This provides an easier and systematic approach to locate the most relevant information.

Our summarized findings in response to our posed research questions are as follows:
(i) *RQ1: which areas in SE are addressed by SMSs?* We classified the included studies using thematic analysis to determine the most active areas of research. Thematic analysis does not conform to a uniform classification but is useful in determining the research areas gaining a lot of attention. To obtain a more uniform view of the landscape of SMSs in SE, we map the identified SMSs on SWEBOK knowledge areas. Our results show that not all areas in SE are equally covered or active. Areas such as software testing, software design, requirements engineering, product line engineering contain highest count of SMSs. On the other hand, some important areas such as software maintenance and configuration management only have a few SMSs. Results also show a lack of SMSs on tools used in various stages of software development lifecycle. Given the importance of tools for practitioners we recommend secondary (and primary) studies to aggregate the available literature.
(ii) *RQ2: what are the trends relating to quality of published SMSs?* We evaluated the quality of the published SMSs using the widely accepted DARE criteria in the EBSE community. We found the overall quality of SMSs in SE to be adequate. Journal papers have significantly higher quality scores than conference papers. In our opinion, while available space limits do effect the quality, the results also show that the journal reviewers tend to be more familiar with SMSs and hence only papers with proper reporting are accepted. This is corroborated by high concentration of SMSs published in journals in IST and JSS. Almost half of all published SMSs in journals appear in these two prestigious venues. SMSs published in conferences are more widely spread, with only 24 out of 102 included SMSs from conferences appearing in top two conferences. Even the papers in the second most frequent conference venue ESEM tend to have low quality scores. One important weakness of SMSs is the lack of quality assessment of included studies. Relatively few SMSs include such assessment and its importance needs to be emphasized.
(iii) *RQ3:* What are the current trends in SMSs relating to guidelines, data sources, types of questions and number of included studies? We extract the information on the most frequently cited guidelines in SMSs, the frequently used databases, and the types of research questions answered in SMSs. We found that 79 of 210 papers cited multiple guidelines. While most papers cite the guidelines by Kitchenham and Peterson, the exact citation differs. There are a number of papers containing guidelines and revisions to those guidelines. This fragmentation of citations is a potential source of confusion for new researchers and we recommend that the guidelines should be revised and consolidated. The revisions could also take into account the large body of work that is now available and recommendations can be made to reduce or eliminate or update some steps from the existing guidelines. The importance of quality assessment questions should be highlighted as they are frequently missing from the list of questions posed by SMSs. We found that 2015 was the peak year in terms of frequency of SMSs published. Since 2015, a decreasing trend in the number of published SMSs is observed. On the other hand, we see a significant increase in the number of primary studies included in published SMSs, indicating a fast increase in the body of available primary studies in SE.

(iv) *RQ4:* What are the demographics of published SMSs in SE? Demographic information is a frequent outcome of all secondary and tertiary studies. We found an increasing trend in number of SMSs being published. However, we note a drop in frequency of published SMSs since the peak year of 2015. We identify IST and JSS as the journals with highest number of published SMSs while EASE and ESEM are the most frequent conferences. We found that the journal papers have significantly more impact than conferences papers in terms of citations. On the other hand, we found a large number of papers with very few (less than 10) citations. Even accounting for SMSs published in 2017 (and hence not having enough time to accumulate citations), few SMSs appear to have high impact. There is no SMS in conferences that have more than 100 citations, while the highest cited SMS published in IST has 274 citations. Of the most cited SMSs, 3 are published in IST, 1 in JSS and 1 in ETS. Fabia Q B da Silva is the researcher with most published SMSs.

Additionally, we reflect on the current trends in SE and discuss the implications of our study on software engineering education and for new researchers. We point out the hot areas in SE that are not covered by SWEBOK classification and recommend including them. SWEBOK classification is used as reference when developing SE curriculum in a number of universities, and therefore it is important to include areas that are of interest to researchers and practitioners. We identify quality assessment and information on primary studies as the two main factors contributing to low quality scores. Our study results can also be used by new researchers as concrete recommendations on which type of questions to pose, which databases to use, where to publish, etc., by new researchers. The results of our study partially re-confirm results of previous studies that found in-adequate quality assessment in published secondary studies. Moreover, our findings regarding the top venues are in agreement with other tertiary studies [10, 29]. However, our findings differ with the claim made by [24] which states that few studies perform research synthesis. We found that only approximately 10% of the studies failed to perform synthesis. Our results also differ from [10] which claims that number of secondary studies published in conferences are greater than journals. We found more SMSs published in journals than in conferences. Based on our results, we recommend more emphasis be placed on neglected areas such as tool support for software engineering practices, software configuration management and mathematical foundations for software engineering. In future, our study can be used as basis of performing multi-vocal study on areas that are not well covered by SMSs.

## Included Systematic Mapping Studies

| | |
|---|---|
| S1 | Fernández, D. M., Ognawala, S., Wagner, S., & Daneva, M. (2017). Where do we stand in requirements engineering improvement today? First results from a mapping study. In *Proceedings of the 8th ACM/IEEE International Symposium on Empirical Software Engineering and Measurement*. https://doi.org/10.1145/2652524.2652555 |
| S2 | Haghighatkhah, A., Banijamali, A., Pakanen, O.-P., Oivo, M., & Kuvaja, P. (2017). Automotive software engineering: A systematic mapping study. *Journal of Systems and Software*, *128*(June), 25–55. https://doi.org/10.1016/j.jss.2017.03.005 |
| S3 | Assuncao, W. K. G., Lopez-Herrejon, R. E., Linsbauer, L., Vergilio, S. R., & Egyed, A. (2017). Reengineering legacy applications into software product lines: a systematic mapping. *Empirical Software Engineering*, 1–45. https://doi.org/10.1007/s10664-017-9499-z |
| S4 | Caballero-Hernández, J. A., Palomo-Duarte, M., Dodero, J. M., Caballero Hernández, J. A., Duarte, M. P., & Dodero Beardo, J. M. (2017). Skill assessment in learning experiences based on serious games: A Systematic Mapping Study. *Computers & Education*, *113*, 42–60. https://doi.org/https://doi.org/10.1016/j.compedu.2017.05.008 |
| S5 | Abbaspour Asadollah, S., Sundmark, D., Eldh, S., Hansson, H., & Afzal, W. (2017). 10 Years of research on debugging concurrent and multicore software: a systematic mapping study. *Software Quality Journal*, *25*(1), 49–82. https://doi.org/10.1007/s11219-015-9301-7 |
| S6 | Enríquez, J. G., Domínguez-Mayo, F. J., Escalona, M. J., Ross, M., & Staples, G. (2017). Entity reconciliation in big data sources: A systematic mapping study. *Expert Systems with Applications*, *80*, 14–27. https://doi.org/10.1016/j.eswa.2017.03.010 |
| S7 | Ferreira, T. N., Vergilio, S. R., & de Souza, J. T. (2017). Incorporating user preferences in search-based software engineering: A systematic mapping study. *Information and Software Technology*, *90*, 55–69. https://doi.org/10.1016/j.infsof.2017.05.003 |
| S8 | Usman, M., Britto, R., Börstler, J., & Mendes, E. (2017). Taxonomies in software engineering: A Systematic mapping study and a revised taxonomy development method. *Information and Software Technology*, *85*(January), 43–59. https://doi.org/10.1016/j.infsof.2017.01.006 |
| S9 | Garcés, L., Ampatzoglou, A., Avgeriou, P., & Nakagawa, E. Y. (2017). Quality attributes and quality models for ambient assisted living software systems: A systematic mapping. *Information and Software Technology*, *82*, 121–138. https://doi.org/10.1016/j.infsof.2016.10.005 |
| S10 | Marimuthu, C., & Chandrasekaran, K. (2017). Software Engineering Aspects of Green and Sustainable Software : A Systematic Mapping Study. In *In Proceedings of the 10th Innovations in Software Engineering Conference* (pp. 34–44). ACM. |
| S11 | Meldrum, S., Licorish, S. A., Tony, B., & Savarimuthu, R. (2017). Crowdsourced Knowledge on Stack Overflow : A Systematic Mapping Study. In *Proceedings of the 21st International Conference on Evaluation and Assessment in Software Engineering* (pp. 180–185). ACM. |
| S12 | Nguyen, P. H., Ali, S., & Yue, T. (2017). Model-based security engineering for cyber-physical systems: A systematic mapping study. *Information and Software Technology*, *83*(November), 116–135. https://doi.org/10.1016/j.infsof.2016.11.004 |
| S13 | Nascimento, A. M., & Silveira, D. S. da. (2017). A systematic mapping study on using social media for business process improvement. *Computers in Human Behavior*, *73*, 670–675. https://doi.org/10.1016/j.chb.2016.10.016 |
| S14 | Rodríguez, P., Haghighatkhah, A., Lwakatare, L. E., Teppola, S., Suomalainen, T., Eskeli, J., … Oivo, M. (2017). Continuous deployment of software intensive products and services: A systematic mapping study. *Journal of Systems and Software*, *123*, 263–291. https://doi.org/10.1016/j.jss.2015.12.015 |
| S15 | Vale, T., de Almeida, E. S., Alves, V., Kulesza, U., Niu, N., & de Lima, R. (2017). Software product lines traceability: A systematic mapping study. *Information and Software Technology*, *84*, 1–18. https://doi.org/10.1016/j.infsof.2016.12.004 |
| S16 | Dusse, F., Júnior, P. S., Alves, A. T., Novais, R., Vieira, V., & Mendonça, M. (2016). Information visualization for emergency management: A systematic mapping study. *Expert Systems with Applications*, *45*(October 2015), 424–437. https://doi.org/10.1016/j.eswa.2015.10.007 |
| S17 | Cavalcante, E., Pereira, J., Alves, M. P., Maia, P., Moura, R., Batista, T., … Pires, P. F. (2016). On the interplay of Internet of Things and Cloud Computing: A systematic mapping study. *Computer Communications*, *89–90*, 17–33. https://doi.org/10.1016/j.comcom.2016.03.012 |
| S18 | González-Ladrón-de-Guevara, F., Fernández-Diego, M., & Lokan, C. (2016). The usage of ISBSG data fields in software effort estimation: A systematic mapping study. *Journal of Systems and Software*, *113*(December), 188–215. https://doi.org/10.1016/j.jss.2015.11.040 |
| S19 | Nunes, F., & Albuquerque, A. B. (2016). A Proposal for Systematic Mapping Study of Software Security Testing, Verification and Validation. *International Journal of Computer, Electrical, Automation, Control and Information Engineering*, *10*(6), 1004–1010. |
| S20 | Zein, S., Salleh, N., & Grundy, J. (2016). *Journal of Systems and Software*, *117*(July), 334–356. https://doi.org/10.1016/j.jss.2016.03.065 |
| S21 | Abdallah, A., Lapalme, J., & Abran, A. (2016). Enterprise Architecture Measurement: A Systematic Mapping Study. *2016 4th International Conference on Enterprise Systems (ES)*, 13–20. https://doi.org/10.1109/ES.2016.8 |
| S22 | Alves, N. S. R., Mendes, T. S., de Mendonça, M. G., Spínola, R. O., Shull, F., & Seaman, C. (2016). Identification and management of technical debt: A systematic mapping study. *Information and Software Technology*, *70*, 100–121. https://doi.org/10.1016/j.infsof.2015.10.008 |
| S23 | Axelsson, J., & Skoglund, M. (2016). Quality assurance in software ecosystems: A systematic literature mapping and research agenda. *Journal of Systems and Software*, *114*, 69–81. https://doi.org/10.1016/j.jss.2015.12.020 |
| S24 | Barn, B. S., Clark, T., Ali, A., & Arif, R. (2016). A Systematic Mapping Study of the Current Practice of Indian Software Engineering. In *Proceedings of the 9th India Software Engineering Conference* (pp. 89–98). ACM. |
| S25 | Chadli, S. Y., Idri, A., Ros, J. N., Fernández-Alemán, J. L., de Gea, J. M. C., & Toval, A. (2016). *Software project management tools in global software development: a systematic mapping study*. *SpringerPlus* (Vol. 5). Springer International Publishing. https://doi.org/10.1186/s40064-016-3670-7 |
| S26 | Pahl, C., & Jamshidi, P. (2016). Microservices: A Systematic Mapping Study. *Proceedings of the 6th International Conference on Cloud Computing and Services Science*, (October), 137–146. https://doi.org/10.5220/0005785501370146 |
| S27 | Kosar, T., Bohra, S., & Mernik, M. (2016). Domain-Specific Languages: A Systematic Mapping Study. *Information and Software Technology*, *71*(November), 77–91. https://doi.org/10.1016/j.infsof.2015.11.001 |
| S28 | Eleutério, J. D. A. S., Gaia, F. N., Bondavalli, A., Lollini, P., Rodrigues, G. N., & Rubira, C. M. F. (2016). On the Dependability for Dynamic Software Product Lines A Comparative Systematic Mapping Study. *Proc. of SEAA*, 323–330. https://doi.org/10.1109/SEAA.2016.40 |
| S29 | de F. Farias, M. A., Novais, R., Júnior, M. C., da Silva Carvalho, L. P., Mendonça, M., & Spínola, R. O. (2016). A systematic mapping study on mining software repositories. *Proceedings of the 31st Annual ACM Symposium on Applied Computing - SAC '16*, (August), 1472–1479. https://doi.org/10.1145/2851613.2851786 |
| S30 | Gabriel, A., Monticolo, D., Camargo, M., & Bourgault, M. (2016). Creativity support systems: A systematic mapping study. *Thinking Skills and Creativity*, *21*, 109–122. https://doi.org/10.1016/j.tsc.2016.05.009 |
| S31 | Jabbari, R., bin Ali, N., Petersen, K., & Tanveer, B. (2016). What is DevOps? A Systematic Mapping Study on Definitions and Practices. *Proceedings of the Scientific Workshop Proceedings of XP2016 on - XP '16 Workshops*, 1–11. https://doi.org/10.1145/2962695.2962707 |
| S32 | Airton, F., Germano, S., Quesado, E., Dornelas, M., Silva, B., & Maciel, P. (2016). Benchmark applications used in mobile cloud computing research: a systematic mapping study. *The Journal of Supercomputing*, *72*(4), 1431–1452. https://doi.org/10.1007/s11227-016-1674-2 |
| S33 | Khan, N. F., & Ikram, N. (2016). Security Requirements Engineering: A Systematic Mapping (2010-2015). *2016 International Conference on Software Security and Assurance (ICSSA)*, 31–36. https://doi.org/10.1109/ICSSA.2016.13 |
| S34 | Lopez-Herrejon, R. E., Illescas, S., & Egyed, A. (2016). Visualization for Software Product Lines: A Systematic Mapping Study. *4th IEEE Working Conference on Software Visualization*, 26–35. https://doi.org/10.1109/VISSOFT.2016.11 |
| S35 | Magues, D. A., Castro, J. W., & Acuna, S. T. (2016). Usability in agile development: A systematic mapping study. *2016 XLII Latin American Computing Conference (CLEI)*, 1–8. https://doi.org/10.1109/CLEI.2016.7833347 |
| S36 | Montalvillo, L., & Díaz, O. (2016). Requirement-driven evolution in software product lines: A systematic mapping study. *Journal of Systems and Software*, *122*, 110–143. https://doi.org/10.1016/j.jss.2016.08.053 |
| S37 | Alshuqayran Nuha, A. N. E. R. (2016). A Systematic Mapping Study in Microservice Architecture. *IEEE 9th International Conference on Service Oriented Computing and Application*, 8. https://doi.org/10.1109/SOCA.2016.15 |
| S38 | Horkoff, J., Aydemir, F. B., Cardoso, E., Li, T., Mate, A., Paja, E., … Giorgini, P. (2016). Goal-Oriented Requirements Engineering: A Systematic Literature Map. *2016 IEEE 24th International Requirements Engineering Conference (RE)*, 106–115. https://doi.org/10.1109/RE.2016.41 |
| S39 | Saint-Louis, P., & Lapalme, J. (2016). Investigation of the Lack of Common Understanding in the Discipline of Enterprise Architecture : A Systematic Mapping Study. *Proceedings - IEEE International Enterprise Distributed Object Computing Workshop, EDOCW*, *2016–September*, 74–82. https://doi.org/10.1109/EDOCW.2016.7584364 |
| S40 | Aguileta, A. A., & Gómez, O. S. (2016). Software engineering research in mexico: A systematic mapping study. *International Journal of Software Engineering and Its Applications*, *10*(12), 75–92. https://doi.org/10.14257/ijseia.2016.10.12.07 |
| S41 | Souag, A., Mazo, R., Salinesi, C., & Comyn-Wattiau, I. (2016). Reusable knowledge in security requirements engineering: a systematic mapping study. *Requirements Engineering*, *21*(2), 251–283. https://doi.org/10.1007/s00766-015-0220-8 |
| S42 | Lindström, B., Ali, S. M., & Blom, M. (2016). Testability and Software Performance : A Systematic Mapping Study. *ACM Symposium on Applied Computing*, 1566–1569. |
| S43 | Yang, C., Liang, P., & Avgeriou, P. (2016). A systematic mapping study on the combination of software architecture and agile development. *Journal of Systems and Software*, *111*(October 2015), 157–184. https://doi.org/10.1016/j.jss.2015.09.028 |
| S44 | Batot, E., Sahraoui, H., Syriani, E., Molins, P., & Sboui, W. (2016). Systematic Mapping Study of Model Transformations for Concrete Problems. In *n Model-Driven Engineering and Software Development (MODELSWARD), 2016 4th International Conference on* (pp. 176–183). IEEE. |
| S45 | Salehi, S., Selamat, A., & Fujita, H. (2015). Systematic mapping study on granular computing. *Knowledge-Based Systems*, *80*(May), 78–97. https://doi.org/10.1016/j.knosys.2015.02.018 |
| S46 | Feloni, D. F. G., & Braga, R. T. V. (2015). Methodologies for Evaluation and Improvement of Software Processes in the Context of Quality and Maturity Models : a Systematic Mapping. *18th Ibero American Conference on Software Engineering (CIbSE)*, 1–14. |
| S47 | Cruz, S., Da Silva, F. Q. B., & Capretz, L. F. (2015). Forty years of research on personality in software engineering: A mapping study. *Computers in Human Behavior*, *46*, 94–113. https://doi.org/10.1016/j.chb.2014.12.008 |
| S48 | Dalal, D., Gill, N. S., & Singh, L. (2015). Built-in testing in component-based software - a mapping study. *Computing for Sustainable Global Development (INDIACom), 2015 2nd International Conference on*, 159–168. |
| S49 | Connor, R. V, Mitasiunas, A., & Ross, M. (2015). A Systematic Mapping Study on Software Process Education. *1st International Workshop on Software Process Education, Training and Professionalism (SPETP 2015)*. https://doi.org/urn:nbn:de:0074-1368-8 |
| S50 | Cruz, Z., Fernández-Alemán, J., & Toval, A. (2015). Security in cloud computing: A mapping study. *Computer Science and Information Systems*, *12*(1), 161–184. https://doi.org/10.2298/CSIS140205086C |
| S51 | De Magalhães, C. V. C., Da Silva, F. Q. B., Santos, R. E. S., & Suassuna, M. (2015). Investigations about replication of empirical studies in software engineering: A systematic mapping study. *Information and Software Technology*, *64*(July), 76–101. https://doi.org/10.1016/j.infsof.2015.02.001 |
| S52 | Abelein, U., & Paech, B. (2015). Understanding the Influence of User Participation and Involvement on System Success -- A Systematic Mapping Study. *Empirical Software Engineering*, *20*(1), 28–81. https://doi.org/10.1007/s10664-013-9278-4 |
| S53 | García-Mireles, G. A., Moraga, M. Á., García, F., & Piattini, M. (2015). Approaches to promote product quality within software process improvement initiatives: A mapping study. *Journal of Systems and Software*, *103*, 150–166. https://doi.org/10.1016/j.jss.2015.01.057 |
| S54 | Paiva, S. L. D. C., & Simao, A. D. S. (2015). A Systematic Mapping Study on Test Generation from Input/Output Transition Systems. *Proceedings - 41st Euromicro Conference on Software Engineering and Advanced Applications, SEAA 2015*, (August), 333–340. https://doi.org/10.1109/SEAA.2015.66 |
| S55 | Zhi, J., Garousi-Yusifoılu, V., Sun, B., Garousi, G., Shahnewaz, S., & Ruhe, G. (2015). Cost, benefits and quality of software development documentation: A systematic mapping. *Journal of Systems and Software*, *99*(January), 175–198. https://doi.org/10.1016/j.jss.2014.09.042 |
| S56 | Wakil, K., & Jawawi, D. N. A. (2015). Model driven web engineering: A systematic mapping study. *E-Informatica Software Engineering Journal*, *9*(1), 87–122. https://doi.org/10.5277/E-INF150106 |
| S57 | Carroll, C., Falessi, D., Forney, V., Frances, A., Izurieta, C., & Seaman, C. (2015). A Mapping Study of Software Causal Factors for Improving Maintenance. *International Symposium on Empirical Software Engineering and Measurement*, *2015–November*(805), 235–238. https://doi.org/10.1109/ESEM.2015.7321183 |
| S58 | Arruda, M. F., Ciˆ, G., & Bulc, R. D. F. (2015). Requirements Specification of Context-Aware Systems : A Systematic Mapping. *Revista de Sistemas de Informaçao Da FSMA*, *16*(1), 41–51. |
| S59 | Flemstr, D., Sundmark, D., & Afzal, W. (2015). Vertical Test Reuse for Embedded Systems : A Systematic Mapping Study. *Software Engineering and Advanced Applications (SEAA), 2015 41st Euromicro Conference on*. |
| S60 | Dicheva, D., Dichev, C., Agre, G., & Angelova, G. (2015). Gamification in Education: A Systematic Mapping Study. *Educational Technology & Society*, *18*(3), 75–88. https://doi.org/10.1109/EDUCON.2014.6826129. |
| S61 | Gbaffonou, B. A. D., Lapalme, J., & Champagne, R. (2015). Service-Oriented Architecture: A Mapping Study. *2015 International Conference on Enterprise Systems (ES)*, 33–42. https://doi.org/10.1109/ES.2015.11 |
| S62 | Guedes, G., Silva, C., Soares, M., & Castro, J. (2015). Variability Management in Dynamic Software Product Lines: A Systematic Mapping. *2015 IX Brazilian Symposium on Components, Architectures and Reuse Software*, 90–99. https://doi.org/10.1109/SBCARS.2015.20 |
| S63 | Heikkila, V. T., Damian, D., Lassenius, C., & Paasivaara, M. (2015). A Mapping Study on Requirements Engineering in Agile Software Development. *Proceedings - 41st Euromicro Conference on Software Engineering and Advanced Applications, SEAA 2015*, 199–207. https://doi.org/10.1109/SEAA.2015.70 |
| S64 | Heredia, A., Colomo-Palacios, R., & De Amescua, A. (2015). Software Business Models from a Distribution Perspective: A Systematic Mapping Study. *Procedia Computer Science*, *64*(1877), 395–402. https://doi.org/10.1016/j.procs.2015.08.516 |
| S65 | Idri, A., Amazal, F. A., & Abran, A. (2015). Analogy-based software development effort estimation: A systematic mapping and review. *Information and Software Technology*, *58*, 206–230. https://doi.org/10.1016/j.infsof.2014.07.013 |
| S66 | Idri, A., Abnane, I., & Abran, A. (2015). Systematic Mapping Study of Missing Values Techniques in Software Engineering Data. *International Conference on Software Engineering, Artificial Intelligence, Networking and Parallel/Distributed Computing (SNPD), 2015 16th IEEE/ACIS*, 1–8. |
| S67 | Gonçales, L. J., Farias, K., Scholl, M., Roberto Veronez, M., & de Oliveira, T. C. (2015). Comparison of Design Models: A Systematic Mapping Study. *International Journal of Software Engineering and Knowledge Engineering*, *25*(09n10), 1765–1769. https://doi.org/10.1142/S021819401571014X |
| S68 | Riaz, M., Breaux, T., & Williams, L. (2015). How have we evaluated software pattern application? A systematic mapping study of research design practices. *Information and Software Technology*, *65*, 14–38. https://doi.org/10.1016/j.infsof.2015.04.002 |
| S69 | Li, Z., Avgeriou, P., & Liang, P. (2015). A systematic mapping study on technical debt and its management. *Journal of Systems and Software*, *101*, 193–220. https://doi.org/10.1016/j.jss.2014.12.027 |
| S70 | Kabbedijk, J., Bezemer, C.-P., Jansen, S., & Zaidman, A. (2015). Defining multi-tenancy: A systematic mapping study on the academic and the industrial perspective. *Journal of Systems and Software*, *100*, 139–148. https://doi.org/10.1016/j.jss.2014.10.034 |
| S71 | Souza, E. F., Falbo, R. D., & Vijaykumar, N. L. (2015). Knowledge management initiatives in software testing: A mapping study. *Information and Software Technology*, *57*(1), 378–391. https://doi.org/10.1016/j.infsof.2014.05.016 |
| S72 | Lacerda, A., Queiroz, R. De, & Barbosa, M. (2015). A Systematic Mapping on Security Threats in Mobile Devices. *Internet Technologies and Applications (ITA), 2015.*, 286–291. https://doi.org/10.1109/ITechA.2015.7317411 |
| S73 | Lopes, A. M. Z., Pedro, L. Z., Isotani, S., & Bittencourt, I. I. (2015). Quality Evaluation of Web-Based Educational Software: A Systematic Mapping. *2015 IEEE 15th International Conference on Advanced Learning Technologies*, 250–252. https://doi.org/10.1109/ICALT.2015.88 |
| S74 | Lopez-Herrejon, R. E., Linsbauer, L., & Egyed, A. (2015). A systematic mapping study of search-based software engineering for software product lines. *Information and Software Technology*, *61*, 33–51. https://doi.org/10.1016/j.infsof.2015.01.008 |
| S75 | Lopez-Herrejon, R. E., Fischer, S., Ramler, R., & Egyed, A. (2015). A first systematic mapping study on combinatorial interaction testing for software product lines. *2015 IEEE Eighth International Conference on Software Testing, Verification and Validation Workshops (ICSTW)*, (Iwct), 1–10. https://doi.org/10.1109/ICSTW.2015.7107435 |
| S76 | Maplesden, D., Tempero, E., Hosking, J., & Grundy, J. C. (2015). Performance Analysis for Object-Oriented Software: A Systematic Mapping. *IEEE Transactions on Software Engineering*, *41*(7), 691–710. https://doi.org/10.1109/TSE.2015.2396514 |
| S77 | Ameller, D., Burgus, X., Collell, O., Costal, D., Franch, X., & Papazoglou, M. P. (2015). Development of service-oriented architectures using model-driven development: A mapping study. *Information and Software Technology*, *62*(1), 42–66. https://doi.org/10.1016/j.infsof.2015.02.006 |
| S78 | Ingibergsson, J. T. M., Schultz, U. P., & Kuhrmann, M. (2015). On the use of safety certification practices in autonomous field robot software development: A systematic mapping study. *Lecture Notes in Computer Science (Including Subseries Lecture Notes in Artificial Intelligence and Lecture Notes in Bioinformatics)*, *9459*(1), 335–352. https://doi.org/10.1007/978-3-319-26844-6_25 |
| S79 | Stevanetic, S., & Zdun, U. (2015). Software metrics for measuring the understandability of architectural structures - A systematic mapping study. *ACM International Conference Proceeding Series*, *27–29–Apri*, Nanjing University-. https://doi.org/10.1145/2745802.2745822 |
| S80 | Paz, F. (2015). Usability Evaluation Methods for Software Development: A Systematic Mapping Review. *2015 8th International Conference on Advanced Software Engineering & Its Applications (ASEA)*, 1–4. https://doi.org/10.1109/ASEA.2015.8 |
| S81 | Pedreira, O., Garc??a, F., Brisaboa, N., & Piattini, M. (2015). Gamification in software engineering - A systematic mapping. *Information and Software Technology*, *57*(1), 157–168. https://doi.org/10.1016/j.infsof.2014.08.007 |
| S82 | Pitangueira, A. M., Maciel, R. S. P., & Barros, M. (2015). Software requirements selection and prioritization using SBSE approaches: A systematic review and mapping of the literature. *Journal of Systems and Software*, *103*(C), 267–280. https://doi.org/10.1016/jjss.2014.09.038 |
| S83 | Knutas, A., Ikonen, J., & Porras, J. (2015). Computer-Supported Collaborative Learning in Software Engineering Education : A Systematic Mapping Study, *7*(May 2016). |
| S84 | Abdelmaboud, A., Jawawi, D. N. A., Ghani, I., Elsafi, A., & Kitchenham, B. (2015). Quality of service approaches in cloud computing: A systematic mapping study. *Journal of Systems and Software*, *101*(March), 159–179. https://doi.org/10.1016/j.jss.2014.12.015 |
| S85 | Ouhbi, S., Idri, A., Fernández-Alemán, J. L., & Toval, A. (2015). Requirements engineering education: a systematic mapping study. *Requirements Engineering*, *20*(2), 119–138. https://doi.org/10.1007/s00766-013-0192-5 |
| S86 | Santos, A. R., de Oliveira, R. P., & de Almeida, E. S. (2015). Strategies for consistency checking on software product lines. *EASE - International Conference on Evaluation and Assessment in Software Engineering*, (Section 3), 1–14. https://doi.org/10.1145/2745802.2745806 |
| S87 | Fonseca, Vinícius Soares; Barcellos, Monalessa Perini; Falbo, R. de A. (2015). Integration of Software Measurement Supporting Tools : A Mapping Study. *Seke*, 6. https://doi.org/10.18293/SEKE2015-058 |
| S88 | Garousi Yusifoɫlu, V., Amannejad, Y., & Betin Can, A. (2015). Software test-code engineering: A systematic mapping. *Information and Software Technology*, *58*(July), 123–147. https://doi.org/10.1016/j.infsof.2014.06.009 |
| S89 | Kuhrmann, M., Konopka, C., Nellemann, P., Diebold, P., & Münch, J. (2015). Software process improvement: Where is the evidence?: Initial findings from a systematic mapping study. *International Conference on Software and Systems Process, ICSSP 2015*, *24–26–Augu*, 107–116. https://doi.org/10.1145/2785592.2785600 |
| S90 | Nascimento, D. M. C., Bittencourt, R. A., & Chavez, C. (2015). Open source projects in software engineering education: a mapping study. *Computer Science Education*, *25*(1), 67–114. https://doi.org/10.1080/08993408.2015.1033159 |
| S91 | Vargas-Enriquez, J., Garcia-Mundo, L., Genero, M., & Piattini, M. (2015). A Systematic Mapping Study on Gamified Software Quality. *2015 7th International Conference on Games and Virtual Worlds for Serious Applications (VS-Games)*, 1–8. https://doi.org/10.1109/VS-GAMES.2015.7295763 |
| S92 | Rafique, S., Humayun, M., Hamid, B., Abbas, A., Akhtar, M., & Iqbal, K. (2015). Web application security vulnerabilities detection approaches: A systematic mapping study. *2015 IEEE/ACIS 16th International Conference on Software Engineering, Artificial Intelligence, Networking and Parallel/Distributed Computing (SNPD)*, 1–6. https://doi.org/10.1109/SNPD.2015.7176244 |
| S93 | Hussan Munir, Krzysztof Wnuk, Per Runeson. (2015). Open Innovation in Software Engineering: A Systematic Mapping Study. *Empirical Software Engineering*, *21*(2), 684–723. |
| S94 | Cavalcanti, Y. C., Da Mota Silveira Neto, P. A., Machado, I. D. C., Vale, T. F., De Almeida, E. S., & Meira, S. R. D. L. (2014). Challenges and opportunities for software change request repositories: A systematic mapping study. *Journal of Software: Evolution and Process*, *26*(7), 620–653. https://doi.org/10.1002/smr.1639 |
| S95 | Ardini, A., Hosseini, M., Alrobai, A., Shahri, A., Phalp, K., & Ali, R. (2014). Survey: Social computing for software engineering: A mapping study. *Computer Science Review*, *13–14*, 75–93. Retrieved from http://10.0.3.248/j.cosrev.2014.08.002%5Cnhttps://login.e.bibl.liu.se/login?url=https://search.ebscohost.com/login.aspx?direct=true&db=edselp&AN=S1574013714000124&site=eds-live&scope=site |
| S96 | Castelluccia, D., & Boffoli, N. (2014). Service-oriented product lines. *ACM SIGSOFT Software Engineering Notes*, *39*(2), 1–6. https://doi.org/10.1145/2579281.2579294 |
| S97 | Jurca, G., Hellmann, T. D., & Maurer, F. (2014). Integrating agile and user-centered design: A systematic mapping and review of evaluation and validation studies of agile-UX. *Proceedings - 2014 Agile Conference, AGILE 2014*, 24–32. https://doi.org/10.1109/AGILE.2014.17 |
| S98 | Qiu, D., Li, B., Ji, S., & Leung, H. (2014). Regression Testing of Web Service: A Systematic Mapping Study. *ACM Computing Surveys*, *47*(2), 1–46. https://doi.org/10.1145/2631685 |
| S99 | González, J. E., Juristo, N., & Vegas, S. (2014). A systematic mapping study on testing technique experiments: Has the situation Changed since 2000? *Proceedings of the 8th ACM/IEEE International Symposium on Empirical Software Engineering and Measurement - ESEM '14*, 1–4. https://doi.org/10.1145/2652524.2652569 |
| S100 | Seriai, A., Benomar, O., Cerat, B., & Sahraoui, H. (2014). Validation of Software Visualization Tools: A Systematic Mapping Study. *2014 Second IEEE Working Conference on Software Visualization*, 60–69. https://doi.org/10.1109/VISSOFT.2014.19 |
| S101 | Febrero, F., Calero, C., & Moraga, M. Ngeles. (2014). A Systematic Mapping Study of Software Reliability Modeling. *Information and Software Technology*, *56*(8), 839–849. https://doi.org/10.1016/j.infsof.2014.03.006 |
| S102 | Diebold, P., & Dahlem, M. (2014). Agile Practices in Practice: A Mapping Study. *Proceedings of the 18th International Conference on Evaluation and Assessment in Software Engineering (EASE '14)*, (May 2014), 30:1--30:10. https://doi.org/10.1145/2601248.2601254 |
| S103 | Assunção, W. K. G., & Vergilio, S. R. (2014). Feature location for software product line migration. *Proceedings of the 18th International Software Product Line Conference on Companion Volume for Workshops, Demonstrations and Tools - SPLC '14*, 52–59. https://doi.org/10.1145/2647908.2655967 |
| S104 | de Sousa Borges, S., Durelli, V. H. S., Macedo Reis, H., & Isotani, S. (2014). A systematic mapping on gamification applied to education. *Proceedings of the 29th Annual ACM Symposium on Applied Computing - SAC '14*, (Icmc), 216–222. https://doi.org/10.1145/2554850.2554956 |
| S105 | Vargas, J. A., García-Mundo, L., Genero, M., & Piattini, M. (2014). A systematic mapping study on serious game quality. *CEUR Workshop Proceedings*, *1196*, 52–57. https://doi.org/10.1145/2601248.2601261 |
| S106 | Dasanayake, S., Markkula, J., & Oivo, M. (2014). Concerns in Software Development : A Systematic Mapping Study. *Proceesings of 18th International Conference on Evaluation and Assessment in Software Engineering*, 14–17. |
| S107 | Penzenstadler, B., Raturi, A., Richardson, D., Calero, C., Femmer, H., & Franch, X. (2014). Systematic Mapping Study on Software Engineering for Sustainability (SE4S). *18th International Conference on Evaluation and Assessment in Software Engineering, EASE*, (November), 1–14. https://doi.org/10.1145/2601248.2601256 |
| S108 | Souza, F., Papadakis, M., Durelli, V., & Delamaro, M. E. (2014). Test Data Generation Techniques for Mutation Testing: A Systematic Mapping. *Pages.Cs.Aueb.Gr*. https://doi.org/10.13140/RG.2.1.3699.9209 |
| S109 | Santos, R. E. S., de Magalh~aes, C. V. C., & da Silva, F. Q. B. (2014). The Use of Systematic Reviews in Evidence Based Software Engineering: A Systematic Mapping Study. *Proceedings of the 8th ACM/IEEE International Symposium on Empirical Software Engineering and Measurement*, (September), 53:1--53:4. https://doi.org/10.1145/2652524.2652553 |
| S110 | Paternoster, N., Giardino, C., Unterkalmsteiner, M., & Gorschek, T. (2014). Software Development in Startup Companies: A Systematic Mapping Study. *Information and Software Technology*, *56*(10), 1200–1218. |
| S111 | Tofan, D., Galster, M., Avgeriou, P., & Schuitema, W. (2014). Past and future of software architectural decisions - A systematic mapping study. *Information and Software Technology*, *56*(8), 850–872. https://doi.org/10.1016/j.infsof.2014.03.009 |
| S112 | Vlietland, J., & van Vliet, H. (2014). Alignment issues in chains of scrum teams. *Lecture Notes in Business Information Processing*, *182 LNBIP*(June), 301–306. https://doi.org/10.1007/978-3-319-08738-2 |
| S128 | Rosli, M. M., Tempero, E., & Luxton-Reilly, A. (2013). Can We Trust Our Results? A Mapping Study on Data Quality. *Software Engineering Conference (APSEC, 2013 20th Asia-Pacific*, *1*, 116–123. https://doi.org/10.1109/APSEC.2013.26 |
| S129 | Catal, C., & Mishra, D. (2013). Test case prioritization: A systematic mapping study. *Software Quality Journal*, *21*(3), 445–478. https://doi.org/10.1007/s11219-012-9181-z |
| S130 | Catal, C., & Atalay, M. (2013). A systematic mapping study on architectural analysis. *Proceedings of the 2013 10th International Conference on Information Technology: New Generations, ITNG 2013*, 661–664. https://doi.org/10.1109/ITNG.2013.102 |
| S131 | Cedillo, P., Fernandez, a, Insfran, E., & Abrahão, S. (2013). Quality of Web Mashups: A Systematic Mapping Study. *4th International Workshop on Quality in Web Engineering (QWE 2013), Co-Located with the 13th International Conference on Web Engineering (ICWE), Aalborg, Denmark.* Retrieved from http://users.dsic.upv.es/~afernandez/files/publications/pdf/Cedillo-et-al-QWE13.pdf |
| S132 | Costa, C., & Murta, L. (2013). Version Control in Distributed Software Development: A Systematic Mapping Study. *Proceedings of IEEE 8th International Conference on Global Software Engineering - ICGSE'13*, 90–99. https://doi.org/10.1109/ICGSE.2013.19 |
| S133 | Pergher, M., & Rossi, B. (2013). Requirements prioritization in software engineering: A systematic mapping study. *2013 3rd International Workshop on Empirical Requirements Engineering, EmpiRE 2013 - Proceedings*, (July 2013), 40–44. https://doi.org/10.1109/EmpiRE.2013.6615215 |
| S134 | Mehmood, A., & Jawawi, D. N. A. (2013). Aspect-oriented model-driven code generation: A systematic mapping study. *Information and Software Technology*, *55*(2), 395–411. https://doi.org/10.1016/j.infsof.2012.09.003 |
| S135 | Mohabbati, B., Asadi, M., Gašević, D., Hatala, M., & Müller, H. A. (2013). Combining service-orientation and software product line engineering: A systematic mapping study. *Information and Software Technology*, *55*(11), 1845–1859. https://doi.org/10.1016/j.infsof.2013.05.006 |
| S136 | Banerjee, I., Nguyen, B., Garousi, V., & Memon, A. (2013). Graphical user interface (GUI) testing: Systematic mapping and repository. *Information and Software Technology*, *55*(10), 1679–1694. https://doi.org/10.1016/j.infsof.2013.03.004 |
| S137 | Kaisti, M., Rantala, V., Mujunen, T., Hyrynsalmi, S., Könnölä, K., Mäkilä, T., & Lehtonen, T. (2013). Agile methods for embedded systems development - a literature review and a mapping study. *EURASIP Journal on Embedded Systems*, *2013*(1), 15. https://doi.org/10.1186/1687-3963-2013-15 |
| S138 | Laguna, M. A., & Crespo, Y. (2013). A systematic mapping study on software product line evolution: From legacy system reengineering to product line refactoring. *Science of Computer Programming*, *78*(8), 1010–1034. https://doi.org/10.1016/j.scico.2012.05.003 |
| S139 | VANHANEN, J., & MÄNTYLÄ, M. V. (2013). a Systematic Mapping Study of Empirical Studies on the Use of Pair Programming in Industry. *International Journal of Software Engineering and Knowledge Engineering*, *23*(9), 1221–1267. https://doi.org/10.1142/S0218194013500381 |
| S140 | Mehmood, A., & Jawawi, D. N. A. (2013). Aspect-oriented code generation for integration of aspect orientation and model-driven engineering. *Information and Software Technology 55, no. 2 (2013): 395-411*. https://doi.org/10.1109/ICOSST.2012.6472845 |
| S141 | Ouhbi, S., Idri, A., Fernández-alemán, J. L., & Toval, A. (2013). Software Quality Requirements: a systematic mapping study. *Asia-Pacific Software Engineering Conference*, 231–238. https://doi.org/10.1109/APSEC.2013.40 |
| S142 | de Mendonca, V. R. L., Rodrigues, C. L., Soares, F. A. a De, & Vincenzi, A. M. R. (2013). Static Analysis Techniques and Tools: A Systematic Mapping Study. *ICSEA 2013, The Eighth International Conference on Software Engineering Advances*, (c), 72–78. Retrieved from http://www.thinkmind.org/index.php?view=article&articleid=icsea_2013_3_30_10407 |
| S143 | Fernández-Sáez, A. M., Genero, M., & Chaudron, M. R. V. (2013). Empirical studies concerning the maintenance of UML diagrams and their use in the maintenance of code: A systematic mapping study. *Information and Software Technology*, *55*(7), 1119–1142. https://doi.org/10.1016/j.infsof.2012.12.006 |
| S144 | Nik Daud, N. M., & Kadir, W. M. N. W. (2012). Systematic mapping study of quality attributes measurement in service oriented architecture. In *Information Science and Digital Content Technology (ICIDT), 2012 8th International Conference on* (Vol. 3, pp. 626–631). |
| S145 | Hellmann, T. D., Sharma, A., Ferreira, J., & Maurer, F. (2012). Agile testing: Past, present, and future - Charting a systematic map of testing in agile software development. *Proceedings - 2012 Agile Conference, Agile 2012*, 55–63. https://doi.org/10.1109/Agile.2012.8 |
| S146 | Elberzhager, F., Münch, J., & Nha, V. T. N. (2012). A systematic mapping study on the combination of static and dynamic quality assurance techniques. *Information and Software Technology*, *54*(1), 1–15. https://doi.org/10.1016/j.infsof.2011.06.003 |
| S147 | Lemos, J., Alves, C., Duboc, L., & Rodrigues, G. N. (2012). A Systematic Mapping Study on Creativity in Requirements Engineering. *Proceedings of the 27th Annual ACM Symposium on Applied Computing (SAC '12)*, (January 2012), 1083–1088. https://doi.org/10.1145/2245276.2231945 |
| S148 | Christian Manteuffel, S. I. (2012). A systematic mapping study on sustainable software engineering: A research preview. *9th SC@ RUG 2011-2012*, 35–39. |
| S149 | Malik, B., & Zafar, S. (2012). A Systematic Mapping Study on Software Engineering Education. *The Seventh International Conference on Software Engineering Advances (ICSEA 2012)*, *6*(January 2016), 2061–2071. Retrieved from http://waset.org/publications/8396/a-systematic-mapping-study-on-software-engineering-education |
| S150 | Saraiva, J., Barreiros, E., Almeida, A., Lima, F., Alencar, A., Lima, G., … Castor, F. (2012). Aspect-oriented software maintenance metrics: a systematic mapping study. *16th International Conference on Evaluation & Assessment in Software Engineering (EASE 2012)*, *2012*(1), 253–262. https://doi.org/10.1049/ic.2012.0033 |
| S151 | Burns, C., Ferreira, J., Hellmann, T. D., & Maurer, F. (2012). Usable results from the field of API usability: A systematic mapping and further analysis. *Proceedings of IEEE Symposium on Visual Languages and Human-Centric Computing, VL/HCC*, 179–182. https://doi.org/10.1109/VLHCC.2012.6344511 |
| S152 | Anjum, M., & Budgen, D. (2012). A Mapping Study of the Definitions used for Service Oriented Architecture. *Proceedings of the 16th International Conference on Evaluation and Assessment in Software Engineering, EASE 2012*, (October), 57–61. https://doi.org/10.1049/ic.2012.0008 |
| S153 | Elberzhager, F., Rosbach, A., Münch, J., & Eschbach, R. (2012). Reducing test effort: A systematic mapping study on existing approaches. *Information and Software Technology*, *54*(10), 1092–1106. https://doi.org/10.1016/j.infsof.2012.04.007 |
| S154 | Castro, O., Espinoza, A., & Martínez-Martínez, A. (2012). Findings based on a systematic mapping study on software product value estimation. *Tendencias En Investigacion E Inovacion En Ingenieria de Software: Un Enfoque Practico*, (April), 57–64. |
| S155 | Acuna, S. T., Castro, J. W. J. W., Dieste, O., & Juristo, N. (2012). A systematic mapping study on the open source software development process. *16th International Conference on Evaluation & Assessment in Software Engineering (EASE 2012)*, 42–46. https://doi.org/10.1049/ic.2012.0005 |
| S156 | Barney, S., Petersen, K., Svahnberg, M., Aurum, A., & Barney, H. (2012). Software quality trade-offs: A systematic map. *Information and Software Technology*, *54*(7), 651–662. https://doi.org/10.1016/j.infsof.2012.01.008 |
| S157 | Mohebzada, J. G., Ruhe, G., & Eberlein, A. (2012). Systematic mapping of recommendation systems for requirements engineering. *2012 International Conference on Software and System Process (ICSSP)*, 200–209. https://doi.org/10.1109/ICSSP.2012.6225965 |
| S158 | Kusumo, D. S., Staples, M., Zhu, L., Zhang, H., & Jeffery, R. (2012). Risks of off-the-shelf-based software acquisition and development: A systematic mapping study and a survey. *Proceedings of the 16th International Conference on Evaluation and Assessment in Software Engineering*, 233–242. https://doi.org/10.1049/ic.2012.0031 |
| S159 | Portillo-Rodríguez, J., Vizcaíno, A., Piattini, M., & Beecham, S. (2012). Tools used in GlobaAgile practices in global software engineering-A systematic mapl Software Engineering: A systematic mapping review. *Information and Software Technology*, *54*(7), 663–685. https://doi.org/10.1016/j.infsof.2012.02.006 |
| S160 | Shah, S. M. A., Morisio, M., & Torchiano, M. (2012). An Overview of Software Defect Density: A Scoping Study. *2012 19th Asia-Pacific Software Engineering Conference*, *1*, 406–415. https://doi.org/10.1109/APSEC.2012.93 |
| S161 | Nascimento, L. M. do, Viana, D. L., Neto, P. A. M. S., Martins, D. A. O., Garcia, V. C., & Meira, S. R. L. (2012). A Systematic Mapping Study on Domain-Specific Languages. *Seventh International Conference on Software Engineering and Advances (ICSEA 2012)*, (c), 179–187. |
| S162 | Tahir, A., & MacDonell, S. G. (2012). A systematic mapping study on dynamic metrics and software quality. *2012 28th IEEE International Conference on Software Maintenance (ICSM)*, 326–335. https://doi.org/10.1109/ICSM.2012.6405289 |
| S163 | Wendler, R. (2012). The maturity of maturity model research: A systematic mapping study. *Information and Software Technology*, *54*(12), 1317–1339. https://doi.org/10.1016/j.infsof.2012.07.007 |
| S164 | Noppen, J., Noppen, J., van Den Broek, P., van Den Broek, P., Aksit, M., & Aksit, M. (2007). Requirements Engineering: Foundation for Software Quality. *Lecture Notes on Computer Science*, *4542*(January), 247–261. https://doi.org/10.1007/978-3-540-73031-6 |
| S165 | Fernandez, A., Insfran, E., & Abrahão, S. (2011). Usability evaluation methods for the web: A systematic mapping study. *Information and Software Technology*, *53*(8), 789–817. https://doi.org/10.1016/j.infsof.2011.02.007 |
| S166 | Bilal, M. A. Q., Ahmed, S., & Shahzad, F. (2011). a Systematic Mapping on Selection of Open Source Software License : Economic and Social Perspective. *Journal of Theoretical and Applied Information Technology*, *25*(2). |
| S167 | Da Mota Silveira Neto, P. A., Carmo MacHado, I. Do, McGregor, J. D., De Almeida, E. S., & De Lemos Meira, S. R. (2011). A systematic mapping study of software product lines testing. *Information and Software Technology*, *53*(5), 407–423. https://doi.org/10.1016/j.infsof.2010.12.003 |
| S168 | Engström, E., & Runeson, P. (2011). Software product line testing - A systematic mapping study. *Information and Software Technology*, *53*(1), 2–13. https://doi.org/10.1016/j.infsof.2010.05.011 |
| S169 | da Silva, F. Q. B., Prikladnicki, R., França, A. C. C., Monteiro, C. V. F., Costa, C., & Rocha, R. (2011). An evidence-based model of distributed software development project management: results from a systematic mapping study. *Journal of Software Maintenance and Evolution: Research and Practice (JSME)*, (Special Issue), n/a-n/a. https://doi.org/10.1002/smr.563 |
| S170 | Arshad, A., & Usman, M. (2011). Security at software architecture level: A systematic mapping study. *Evaluation & Assessment in Software Engineering (EASE 2011), 15th Annual Conference on*, (May), 164–168. https://doi.org/10.1049/ic.2011.0020 |
| S171 | da Silva, F. Q. B., Suassuna, M., Lopes, R. F., Gouveia, T. B., Franca, A. C. A., de Oliveira, J. P. N., … Santos, A. L. M. (2011). Replication of Empirical Studies in Software Engineering: Preliminary Findings from a Systematic Mapping Study. *Replication in Empirical Software Engineering Research (RESER), 2011 Second International Workshop on*, 61–70. https://doi.org/10.1109/RESER.2011.14 |
| S172 | Barmi, Z. A., & Feldt, R. (2011).Alignment of Requirements Specification and Testing: A Systematic Mapping Study. In *n Software Testing, Verification and Validation Workshops (ICSTW)* (pp. 476–485). IEEE. https://doi.org/10.1109/ICSTW.2011.58 |
| S173 | Kusumo, D., Zhu, L., Staples, M., & Zhang, H. (2011). A Systematic Mapping Study on Off-The-Shelf-based Software Acquisition. *Australasian Conference on …*. |
| S174 | Petersen, K. (2011). Measuring and predicting software productivity: A systematic map and review. *Information and Software Technology*, *53*(4), 317–343. https://doi.org/10.1016/j.infsof.2010.12.001 |
| S175 | Maglyas, A., Nikula, U., & Smolander, K. (2011). What do we know about software product management? - a systematic mapping study. *Software Product Management (IWSPM)*, (August), 26–35. https://doi.org/10.1109/IWSPM.2011.6046201 |
| S176 | Barreiros, E., Almeida, A., Saraiva, J., & Soares, S. (2011). A Systematic Mapping Study on Software Engineering Testbeds. *2011 International Symposium on Empirical Software Engineering and Measurement*, *6*(11), 107–116. https://doi.org/10.1109/ESEM.2011.19 |
| S177 | Kitchenham, B. (2010). What's up with software metrics? - A preliminary mapping study. *Journal of Systems and Software*, *83*(1), 37–51. https://doi.org/10.1016/j.jss.2009.06.041 |
| S178 | Fauzi, S. S. M., Bannerman, P. L., & Staples, M. (2010). Software Configuration Management in Global Software Development: A Systematic Map. *2010 Asia Pacific Software Engineering Conference*, *3*(1), 404–413. https://doi.org/10.1109/APSEC.2010.53 |
| S179 | Jalali, S., & Wohlin, C. (2010). Agile practices in global software engineering - A systematic map. *Proceedings - 5th International Conference on Global Software Engineering, ICGSE 2010*, 45–54. https://doi.org/10.1109/ICGSE.2010.14 |
| S180 | Nakagawa, E. Y., Feitosa, D., & Felizardo, K. R. (2010). Using systematic mapping to explore software architecture knowledge. *ICSE Workshop on Sharing and Reusing Architectural Knowledge*, (January), 29–36. https://doi.org/10.1145/1833335.1833340 |
| S181 | Afzal, W., Torkar, R., & Feldt, R. (2009). A systematic review of search-based testing for non-functional system properties. *Information and Software Technology*, *51*(6), 957–976. https://doi.org/10.1016/j.infsof.2008.12.005 |
| S182 | Condori-Fernandez, N., Daneva, M., Sikkel, K., Wieringa, R., Dieste, O., & Pastor, O. (2009). A systematic mapping study on empirical evaluation of software requirements specifications techniques. *2009 3rd International Symposium on Empirical Software Engineering and Measurement*, (1), 502–505. https://doi.org/10.1109/ESEM.2009.5314232 |
| S183 | Mujtaba, S., Petersen, K., Feldt, R., & Mattsson, M. (2008). Software product line variability: A systematic mapping study. *School of Engineering, Blekinge Inst. of Technology*, (November 2015). |
| S184 | Pretorius, R., & Budgen, D. (2008). A mapping study on empirical evidence related to the models and forms used in the uml. *Proceedings of the Second ACM-IEEE International Symposium on Empirical Software Engineering and Measurement - ESEM '08*, 342. https://doi.org/10.1145/1414004.1414076 |
| S185 | Izurieta, C., & Bieman, J. M. (2007). How software designs decay: A pilot study of pattern evolution. *Proceedings - 1st International Symposium on Empirical Software Engineering and Measurement, ESEM 2007*, 449–451. https://doi.org/10.1109/ESEM.2007.58 |
| S186 | Elizabeth Bjarnason. "Distances between requirements engineering and later software development activities: a systematic map." In *International Working Conference on Requirements Engineering: Foundation for Software Quality*, pp. 292-307. Springer, Berlin, Heidelberg, 2013. |
| S187 | Feyh, M. and Petersen, K., 2013. Lean software development measures and indicators-a systematic mapping study. In *Lean Enterprise Software and Systems* (pp. 32-47). Springer, Berlin, Heidelberg. |
| S188 | Fotrousi, F., Fricker, S.A., Fiedler, M. and Le-Gall, F., 2014, June. Kpis for software ecosystems: A systematic mapping study. In *International Conference of Software Business* (pp. 194-211). Springer, Cham. |
| S189 | Kuhrmann, M., Méndez Fernández, D. and Tiessler, M., 2014. A mapping study on the feasibility of method engineering. *Journal of Software: Evolution and Process*, *26*(12), pp.1053-1073. |
| S190 | da Silva, T.S., Silveira, F.F., Silveira, M.S., Hellmann, T. and Maurer, F., 2015, June. A systematic mapping on agile UCD across the major agile and HCI conferences. In *International Conference on Computational Science and Its Applications* (pp. 86-100). Springer, Cham. |
| S191 | Giuffrida, R. and Dittrich, Y., 2013. Empirical studies on the use of social software in global software development–A systematic mapping study. *Information and Software Technology*, *55*(7), pp.1143-1164. |
| S192 | Mehmood, A. and Jawawi, D.N., 2012, December. A systematic map of integration of aspect orientation and model-driven engineering. In *Open Source Systems and Technologies (ICOSST), 2012 International Conference on* (pp. 1-6). IEEE. |
| S193 | Ruiz-Rube, I., Dodero, J.M., Palomo-Duarte, M., Ruiz, M. and Gawn, D., 2012. Uses and applications of spem process models. a systematic mapping study. *Journal of Software Maintenance and Evolution: Research and Practice*, *1*(32), pp.999-1025. |
| S194 | Afzal, Wasif, Richard Torkar, and Robert Feldt. "A Systematic Mapping Study on Non-Functional Search-based Software Testing." In SEKE, vol. 8, pp. 488-493. 2008. |
| S195 | Marques, Maíra R., Alcides Quispe, and Sergio F. Ochoa. "A systematic mapping study on practical approaches to teaching software engineering." In Frontiers in Education Conference (FIE), 2014 IEEE, pp. 1-8. IEEE, 2014. |
| S196 | Abdellatief, Majdi, Abu Bakar Md Sultan, Abdul Azim Abdul Ghani, and Marzanah A. Jabar. "A mapping study to investigate component-based software system metrics." Journal of systems and software 86, no. 3 (2013): 587-603. |
| S197 | Novais, Renato Lima, André Torres, Thiago Souto Mendes, Manoel Mendonça, and Nico Zazworka. "Software evolution visualization: A systematic mapping study." Information and Software Technology 55, no. 11 (2013): 1860-1883. |
| S198 | da Silva, Ivonei Freitas, Paulo Anselmo da Mota Silveira Neto, Pádraig O'Leary, Eduardo Santana de Almeida, and Silvio Romero de Lemos Meira. "Agile software product lines: a systematic mapping study." Software: Practice and Experience 41, no. 8 (2011): 899-920. |
| S199 | Silva, Fabio QB, Rafael Prikladnicki, A. César C. França, Cleviton VF Monteiro, Catarina Costa, and Rodrigo Rocha. "An evidence-based model of distributed software development project management: results from a systematic mapping study." Journal of software: Evolution and Process 24, no. 6 (2012): 625-642. |
| S200 | Hosseini, Mahmood, Alimohammad Shahri, Keith Phalp, Jacqui Taylor, and Raian Ali. "Crowdsourcing: A taxonomy and systematic mapping study." Computer Science Review 17 (2015): 43-69. |
| S201 | Farhoodi, Roshanak, Vahid Garousi, Dietmar Pfahl, and Jonathan Sillito. "Development of scientific software: A systematic mapping, a bibliometrics study, and a paper repository." International Journal of Software Engineering and Knowledge Engineering 23, no. 04 (2013): 463-506. |
| S202 | de Lima Salgado, André, and André Pimenta Freire. "Heuristic evaluation of mobile usability: A mapping study." In International Conference on Human-Computer Interaction, pp. 178-188. Springer, Cham, 2014. |
| S203 | Pekar, Viktor, Michael Felderer, and Ruth Breu. "Improvement methods for software requirement specifications: A mapping study." In Quality of Information and Communications Technology (QUATIC), 2014 9th International Conference on the, pp. 242-245. IEEE, 2014. |
| S204 | de Lima Fontão, Awdren, Rodrigo Pereira dos Santos, and Arilo Claudio Dias-Neto. "Mobile software ecosystem (mseco): a systematic mapping study." In Computer Software and Applications Conference (COMPSAC), 2015 IEEE 39th Annual, vol. 2, pp. 653-658. IEEE, 2015. |
| S205 | Fauzi, Shukor Sanim Mohd, Paul L. Bannerman, and Mark Staples. "Software configuration management in global software development: A systematic map." In Software Engineering Conference (APSEC), 2010 17th Asia Pacific, pp. 404-413. IEEE, 2010. |
| S206 | Nascimento, Debora Maria, Kenia Cox, Thiago Almeida, Wendell Sampaio, Roberto Almeida Bittencourt, Rodrigo Souza, and Christina Chavez. "Using Open Source Projects in software engineering |
| | education: A systematic mapping study." In Frontiers in Education Conference, 2013 IEEE, pp. 1837-1843. IEEE, 2013. |
| S207 | Paz, Freddy, and José Antonio Pow-Sang. "A systematic mapping review of usability evaluation methods for software development process." International Journal of Software Engineering and Its Applications 10, no. 1 (2016): 165-178. |
| S208 | Barbosa12, Olavo, and Carina Alves. "A systematic mapping study on software ecosystems." (2011). |
| S209 | Ouhbi, Sofia, Ali Idri, Jose Luis Fernández Alemán, and Ambrosio Toval. "Evaluating software product quality: A systematic mapping study." In Software Measurement and the International Conference on Software Process and Product Measurement (IWSM-MENSURA), 2014 Joint Conference of the International Workshop on, pp. 141-151. IEEE, 2014. |
| S210 | Ampatzoglou, Apostolos, Sofia Charalampidou, and Ioannis Stamelos. "Research state of the art on GoF design patterns: A mapping study." Journal of Systems and Software 86, no. 7 (2013): 1945-1964. |

# Appendix A

Table A 1. Key parameters extracted from SMSs

| S.no | Year | Type of Publication | Primary Studies | Guidelines Used* | Years covered | Quality Score | Research Area |
|---|---|---|---|---|---|---|---|
| **S1** | 2017 | Conference | 58 | PFMM | 1996-2012 | 3 | Requirement engineering |
| **S2** | 2017 | Journal | 679 | PFMM, KBB | 1990-2015 | 2.5 | Automated software engineering |
| **S3** | **2017** | **Journal** | **119** | **FMM, PVK** | **2001-2015** | **3.5** | **Software product line engineering** |
| **S4** | 2017 | Journal | 165 | PFMM, KC | 2010-2014 | 3 | Game based learning |
| **S5** | 2017 | Journal | 145 | PFMM | 2005-2015 | 2.5 | Software development |
| **S6** | 2017 | Journal | 61 | KPBBTNL, KC | 2010-2017 | 3 | Big data |
| **S7** | 2017 | Journal | 40 | PFMM | 2008-2015 | 3 | Search based software engineering |
| **S8** | 2017 | Journal | 270 | PFMM, Wi | 2000-2017 | 2.5 | Evidence-Based Software Engineering |
| **S9** | **2017** | **Journal** | **27** | **KC** | **2001-2017** | **5** | **Ambient Assisted Living, Model driven engineering** |
| **S10** | 2017 | Conference | 82 | PFMM, DD | 2010-2016 | 3 | Sustainable software development |
| **S11** | 2017 | Conference | 266 | PFMM, Wi | 2009-2017 | 3 | Crowdsourcing |
| **S12** | 2017 | Journal | 48 | KBB, Ki, PVK, PFMM | 2001-2016 | 3 | Model driven engineering |
| **S13** | 2017 | Journal | 28 | KBB, PFMM, PVK, | 2007-2015 | 3 | Social media, Business Process improvement |
| **S14** | **2017** | **Journal** | **50** | **KC, PFMM** | **2001-2014** | **5** | **Software configuration management** |
| **S15** | **2017** | **Journal** | **62** | **BKBTK** | **2001-2015** | **3.5** | **Software product line engineering** |
| **S16** | 2016 | Journal | 96 | PFMM, KC | 2002-2015 | 2.5 | Software visualization |
| **S17** | **2016** | **Journal** | **35** | **KBB, PFMM, DDH** | **2011-2014** | **4** | **Internet of Things** |
| **S18** | **2016** | **Journal** | **107** | **PVK, KC, PFMM, BTBK** | **2000-2013** | **3.5** | **software cost estimation** |
| **S19** | 2016 | Journal | 55 | Ki, KC, KBBTBL, PFMM, BTBK | 1998-2016 | 3 | Software testing |
| **S20** | **2016** | **Journal** | **79** | **PFMM, KC** | **2005-2015** | **4** | **Software Testing** |
| **S21** | 2016 | Conference | 16 | Ki, PFMM | 2004-2015 | 3 | Software architecture |
| **S22** | **2016** | **Journal** | **100** | **BTBK, KC** | **2010-2014** | **4** | **Software Maintenance** |
| **S23** | **2016** | **Journal** | **32** | **N/A** | **2011-2014** | **4.5** | **Software ecosystems** |
| **S24** | 2016 | Conference | 50 | PFMM | 2008-2015 | 1 | software engineering practices |
| **S25** | **2016** | **Journal** | **76** | **KC** | **1990-2014** | **5** | **Global Software Development** |
| **S26** | **2016** | **Conference** | **21** | **PFMM, KBBTBL** | **2013-2015** | **4** | **Micro-services** |
| **S27** | 2016 | Journal | 390 | KBB, PFMM | 2006-2012 | 2.5 | Domain-Specific Languages |
| **S28** | 2016 | Conference | 9 | BKBTK, PFMM | 2000-2016 | 2 | Software Product line |
| **S29** | 2016 | Conference | 107 | KBBTBL, PFMM, BKBTK | 2010-2014 | 2 | Software Repository Mining (MSR) |
| **S30** | 2016 | Journal | 49 | KC | 2007-2016 | 1.5 | Automated software engineering |
| **S31** | 2016 | Workshop | 49 | Keele, PVK | 2013-2015 | 3 | Software process development |
| **S32** | 2016 | Journal | 47 | PFMM | 2005-2015 | 2.5 | Mobile applications, Cloud computing |
| **S33** | 2016 | Conference | 215 | N/A | 2010-2015 | 2.5 | Requirement engineering |
| **S34** | **2016** | **Conference** | **32** | **PFMM, KC, KBB** | **2007-2015** | **3.5** | **Software product line engineering** |
| **S35** | 2016 | Conference | 161 | PFMM | 2002-2015 | 3 | Agile software development |
| **S36** | **2016** | **Journal** | **107** | **N/A** | **2010-2015** | **4.5** | **Software product line engineering, Requirement engineering** |
| **S37** | 2016 | Journal | 33 | KBBTBL, PFMM | 2014-2016 | 3 | Software architecture |
| **S38** | 2016 | Conference | 246 | PFMM | 1997-2015 | 3 | Requirement engineering |
| **S39** | 2016 | Conference | 171 | KBB | 1989-2015 | 2 | Software architecture |
| **S40** | 2016 | Journal | 206 | Ki, BTBK, KC, PVK | 2001-2015 | 2.5 | Software engineering practices |
| **S41** | **2016** | **Journal** | **95** | **PFMM** | **2000-2013** | **4.5** | **Requirement engineering** |
| **S42** | 2016 | Conference | 26 | PFMM | 1990-2014 | 2.5 | Software Testing |
| **S43** | **2016** | **Journal** | **54** | **Basili** | **2001-2014** | **4.5** | **Agile software development** |
| **S44** | 2016 | Conference | 82 | PFMM | 2005-2014 | 2 | Model driven engineering |
| **S45** | 2015 | Journal | 119 | PFMM, Wi | 2012-2014 | 2.5 | Software paradigms |
| **S46** | 2015 | Conference | 17 | KBBTBL, PFMM | 2001-2014 | 3 | Software process model |
| **S47** | **2015** | **Journal** | **90** | **KDJ, KC** | **1970-2010** | **4.5** | **Human factors in software engineering** |
| **S48** | **2015** | **Conference** | **26** | **N/A** | **1995-2014** | **4.5** | **Software testing** |
| **S49** | 2015 | Conference | 33 | PFMM | 2000-2014 | 2 | Professional practices |
| **S50** | 2015 | Journal | 344 | PFMM | 2009-2012 | 3 | Cloud computing |
| **S51** | **2015** | **Journal** | **37** | **PFMM** | **1996-2012** | **4.5** | **Evidence based software engineering** |
| **S52** | **2015** | **Journal** | **58** | **KC** | **1959-2011** | **4** | **Software Development** |
| **S53** | **2015** | **Journal** | **74** | **Bio** | **1997-2013** | **4.5** | **Software quality assurance** |
| **S54** | 2015 | Conference | 84 | PFMM | 1997-2012 | 3 | Software testing |
| **S55** | **2015** | **Journal** | **69** | **PFMM, KBBTBL, KBB, BTBK** | **1971-2011** | **4.5** | **Software Documentation** |
| **S56** | **2015** | **Journal** | **289** | **PFMM** | **2000-2013** | **3.5** | **Model driven engineering** |
| **S57** | 2015 | Conference | 78 | KC | 1985-2013 | 2.5 | Software maintenance |
| **S58** | 2015 | Conference | 49 | KC, KBBTBL | 2005-2014 | 3 | Requirement engineering |
| **S59** | **2015** | **Conference** | **11** | **KC, PFMM** | **2005-2015** | **5** | **Embedded systems, Software testing** |
| **S60** | 2015 | Journal | 34 | N/A | 2010-2014 | 3 | Game based learning |
| **S61** | 2015 | Conference | 105 | KBB, PFMM | 2000-2013 | 1.5 | Service-Oriented Architecture |
| **S62** | **2015** | **Conference** | **54** | **KC** | **2006-2015** | **5** | **Software product line engineering** |
| **S63** | 2015 | Conference | 28 | KC | 2000-2014 | 2 | Agile Software Development, Requirement engineering |
| **S64** | 2015 | Journal | 51 | PFMM | 1997-2014 | 3 | Model driven engineering |
| **S65** | **2015** | **Journal** | **65** | **KBB, KC** | **1990-2012** | **5** | **Software Cost estimation** |
| **S66** | **2015** | **Conference** | **35** | **PFMM** | **2000-2013** | **4** | **Data mining** |
| **S67** | **2015** | **Journal** | **40** | **BTBK** | **2003-2014** | **3.5** | **Model driven engineering** |
| **S68** | **2015** | **Journal** | **30** | **BTBK, KC, PFMM** | **1995-2015** | **5** | **Software architecture** |
| **S69** | **2015** | **Journal** | **94** | **KDJ, KC** | **1992-2013** | **5** | **Software management** |
| **S70** | 2015 | Journal | 48 | KBBTBL, Ki | 2000-2011 | 2 | Software as a service |
| **S71** | **2015** | **Journal** | **15** | **KBB, KC** | **2003-2013** | **5** | **Software Testing** |
| **S72** | 2015 | Conference | 10 | PFMM | 2012-2014 | 1.5 | Mobile computing |
| **S73** | 2015 | Conference | 78 | PFMM | 1996-2013 | 1.5 | Web application |
| **S74** | **2015** | **Journal** | **77** | **PFMM** | **2001-2014** | **4.5** | **Software Product line engineering** |
| **S75** | 2015 | Conference | 47 | PFMM | 2006-2015 | 2.5 | Software Product line engineering, software testing |
| **S76** | **2015** | **Journal** | **253** | **KPBBTNL, PFMM** | **2000-2013** | **4.5** | **Object oriented software** |
| **S77** | 2015 | Journal | 129 | KBB | 2003-2013 | 2.5 | Service oriented architecture, Model driven engineering |
| **S78** | **2015** | **Conference** | **49** | | **1987-2015** | **3.5** | **Robotics** |
| **S79** | **2015** | **Conference** | **25** | **KBB, KC, PFMM** | **1990-2014** | **4.5** | **Software architecture** |
| **S80** | 2015 | Conference | 228 | KC | 2000-2015 | 1 | Software development |
| **S81** | **2015** | **Journal** | **29** | **KBB, PFMM** | **2011-2014** | **5** | **Gamification in software engineering** |
| **S82** | **2015** | **Journal** | **39** | **KI, PFMM** | **2001-2013** | **4.5** | **Requirement engineering** |
| **S83** | **2015** | **Conference** | **121** | **KC, PFMM, Ki** | **2003-2013** | **3.5** | **Computer based learning** |
| **S84** | **2015** | **Journal** | **67** | **BTBK, PFMM, KBBTBL, KPBBTNL** | **2008-2012** | **3.5** | **Cloud computing** |
| **S85** | **2015** | **Journal** | **79** | **PFMM** | **1995-2012** | **5** | **Requirement engineering** |
| **S86** | 2015 | Conference | 24 | KC, PFMM | 2001-2013 | 3 | Software Product line engineering |
| **S87** | **2015** | **Conference** | **36** | **KC** | **1988-2014** | **3.5** | **Software metrics** |
| **S88** | **2015** | **Journal** | **60** | **KC, PFMM** | **1998-2012** | **5** | **Software testing** |
| **S89** | 2015 | Conference | 635 | Ki, PFMM | 1989-2013 | 2.5 | Software Process Improvement |
| **S90** | 2015 | Journal | 53 | KC, PFMM | 1994-2014 | 3 | Software engineering education |
| **S91** | 2015 | Conference | 35 | KC, PFMM, KBB | 2008-2014 | 2 | Software quality engineering |
| **S92** | 2015 | Conference | 41 | N/A | 1994-2014 | 3 | Security of web applications |
| **S93** | **2015** | **Journal** | **33** | **KBB, PFMM** | **2003-2013** | **5** | **Software engineering** |
| **S94** | **2014** | **Journal** | **142** | **PFMM** | **1990-2013** | **5** | **Software repositories** |
| **S95** | **2014** | **Journal** | **139** | **PFMM** | **2006-2013** | **4.5** | **Software engineering** |
| **S96** | 2014 | Conference | 71 | KBB, PFMM | 2008-2012 | 1.5 | Service-Oriented architecture, Software product line engineering |
| **S97** | 2014 | Conference | 76 | Wi | 2002-2013 | 2.5 | User centred design, Agile software development |
| **S98** | **2014** | **Journal** | **30** | **PFMM** | **2000-2013** | **5** | **Software testing** |
| **S99** | 2014 | Conference | 22 | PFMM | 2000-2013 | 2 | Software testing |
| **S100** | 2014 | Conference | 87 | PFMM | 2000-2012 | 2.5 | Software visualization |
| **S101** | 2014 | Journal | 503 | PFMM | 2003-2014 | 2.5 | Model driven engineering |
| **S102** | **2014** | **Conference** | **24** | **KBB, PFMM** | **2010-2013** | **4** | **Agile software development** |
| **S103** | **2014** | **Conference** | **63** | **PFMM** | **2011-2014** | **3.5** | **Software Product line engineering** |
| **S104** | **2014** | **Conference** | **26** | **PFMM** | **2011-2013** | **3.5** | **Gamification Applied to Education** |
| **S105** | 2014 | Conference | 112 | KC | 2014 | 1 | Quality assurance |
| **S106** | 2014 | Conference | 421 | KC, KBB, PFMM | N/A | 2.5 | Software Development |
| **S107** | **2014** | **Conference** | **83** | **KC** | **1999-2013** | **4** | **Sustainable software engineering** |
| **S108** | **2014** | **Conference** | **19** | **BTBK** | **1990-2014** | **3.5** | **Software Testing** |
| **S109** | 2014 | Conference | 52 | Ki, KC | 2004-2013 | 2.5 | Evidence based software engineering |
| **S110** | **2014** | **Journal** | **43** | **KC, PFMM** | **1994-2013** | **3.5** | **Software Development** |
| **S111** | **2014** | **Journal** | **144** | **PMM, BTBK, KC** | **2002-2012** | **4** | **Software architecture** |
| **S112** | 2014 | Conference | 34 | PMM | 2004-2013 | 3 | Software process improvement |
| **S113** | **2014** | **Journal** | **83** | **PMM** | **1985-2012** | **4** | **Data-intensive systems** |
| **S114** | **2014** | **Conference** | **48** | **KC, Wo** | **2000-2013** | **5** | **Model driven engineering** |
| **S115** | **2014** | **Conference** | **891** | **DybaKJ** | **1996-2013** | **3.5** | **Evidence based software engineering** |
| **S116** | 2014 | Conference | 85 | PMM | 2002-2013 | 3 | Software Testing |
| **S117** | **2014** | **Conference** | **94** | **PMM** | **2000-2012** | **4** | **Model driven engineering** |
| **S118** | **2014** | **Conference** | **30** | **PMM** | **2000-2013** | **4** | **Software architecture** |
| **S119** | **2014** | **Journal** | **79** | **Pe, PW** | **1999-2011** | **4.5** | **Software traceability** |
| **S120** | **2013** | **Conference** | **10** | **PMM** | **2001-2009** | **3.5** | **Formal Methods** |
| **S121** | 2013 | Journal | 14 | KC | 2007-2012 | 2.5 | Requirement engineering |
| **S122** | **2013** | **Journal** | **55** | **KC** | **2000-2011** | **5** | **Software architecture** |
| **S123** | 2013 | Conference | 14 | KC | 2004-2013 | 2 | Automated software engineering |
| **S124** | **2013** | **Journal** | **79** | **PMM** | **2000-2011** | **4** | **Software testing** |
| **S125** | **2013** | **Journal** | **91** | **KC, Pe** | **2006-2010** | **5** | **Global software development** |
| **S126** | 2013 | Conference | 30 | KBB | 2000-2013 | 3 | Software evolution |
| **S127** | **2013** | **Conference** | **38** | **PMM** | **2001-2011** | **3.5** | **Global Software Development** |
| **S128** | 2013 | Conference | 64 | N/A | 2008-2012 | 2.5 | Evidence based software engineering |
| **S129** | **2013** | **Journal** | **120** | **PMM, BTBK, KC** | **2000-2012** | **4** | **Software testing** |
| **S130** | 2013 | Conference | 63 | KDJ | 2002-2012 | 2 | Evidence-Based Software Engineering |
| **S131** | **2013** | **Conference** | **187** | **PFMM, KC** | **2006-2012** | **3.5** | **Quality assurance** |
| **S132** | 2013 | Conference | 29 | PFMM | 2002-2012 | 2.5 | Software Configuration Management |
| **S133** | **2013** | **Conference** | **39** | **BTBK, PFMM** | **2000-2009** | **3.5** | **Requirement engineering** |
| **S134** | **2013** | **Journal** | **65** | **PFMM** | **2002-2011** | **3.5** | **Model driven engineering** |
| **S135** | **2013** | **Journal** | **81** | **PFMM** | **2000-2011** | **5** | **Software product line engineering** |
| **S136** | **2013** | **Journal** | **136** | **Basili** | **1991-2011** | **4** | **Software testing** |
| **S137** | **2013** | **Journal** | **28** | **KC** | **2003-2012** | **3.5** | **Embedded systems, Agile software development** |
| **S138** | **2013** | **Journal** | **74** | **PFMM** | **1997-2012** | **4** | **Software product line engineering** |
| **S139** | **2013** | **Journal** | **154** | **KC** | **2001-2009** | **3.5** | **Evidence based software engineering** |
| **S140** | **2013** | **Journal** | **38** | **PFMM** | **2002-2009** | **3.5** | **Model driven engineering** |
| **S141** | **2013** | **Conference** | **51** | **BKDK** | **1990-2012** | **4** | **Requirement engineering** |
| **S142** | **2013** | **Conference** | **51** | **PFMM, KC** | **2003-2013** | **4** | **Quality assurance** |
| **S143** | **2013** | **Journal** | **38** | **KC** | **1997-2010** | **5** | **Software maintenance** |
| **S144** | **2012** | **Conference** | **53** | **Ki** | **2007-2011** | **4** | **Service oriented architecture** |
| **S145** | 2012 | Conference | 166 | PFMM | 2002-2011 | 1 | Agile software Development, Software testing |
| **S146** | **2012** | **Journal** | **51** | **PFMM, KC** | **1985-2010** | **3.5** | **Quality assurance** |
| **S147** | 2012 | Conference | 46 | Ki, PFMM | 2002-2011 | 3 | Requirement engineering |
| **S148** | 2012 | Journal | 593 | PFMM | 2005-2011 | 2 | Sustainable software engineering |
| **S149** | 2012 | Conference | 70 | Ki, PFMM | 2010 | 2 | software engineering |
| **S150** | **2012** | **Conference** | **138** | **Ki** | **1996-2011** | **3.5** | **Software maintenance** |
| **S151** | 2012 | Conference | 28 | PFMM, KC | 2004-2011 | 1.5 | Usability engineering |
| **S152** | 2012 | Conference | 98 | PR | 2000-2010 | 1.5 | Service oriented architecture |
| **S153** | **2012** | **Journal** | **144** | **PFMM** | **1991-2010** | **4** | **Software testing** |
| **S154** | 2012 | Conference | 36 | KC | 1998-2011 | 2 | Software Cost estimation |
| **S155** | **2012** | **Conference** | **29** | **PFMM** | **2000-2012** | **4** | **Software development process** |
| **S156** | **2012** | **Journal** | **168** | **Ki, Pe, DD** | **2005-2010** | **4** | **Quality Assurance** |
| **S157** | 2012 | Journal | 23 | PFMM | 2004-2009 | 3 | Requirement engineering |
| **S158** | **2012** | **Conference** | **36** | **BTBK, KC** | **1986-2010** | **4** | **OTS based software development** |
| **S159** | **2012** | **Journal** | **60** | **BTBK, PFMM** | **2000-2010** | **4** | **Global software development** |
| **S160** | 2012 | Conference | 19 | N/A | 2000-2011 | 2 | Defect density |
| **S161** | 2012 | Conference | 1440 | Ki, KC, PFMM | 1966-2011 | 1.5 | Domain-specific languages |
| **S162** | **2012** | **Conference** | **22** | **KC** | **2001-2011** | **3.5** | **Software metrics** |
| **S163** | 2012 | Journal | 237 | KC, Ki | 1993-2010 | 1.5 | CMMI |
| **S164** | **2011** | **Journal** | **53** | **PFMM** | **1997-2012** | **3.5** | **Requirement engineering** |
| **S165** | **2011** | **Journal** | **206** | **BTBK, Ki, PFMM** | **2003-2009** | **5** | **Usability engineering** |
| **S166** | 2011 | Journal | 18 | N/A | 2001-2010 | 2.5 | Software economics |
| **S167** | **2011** | **Journal** | **45** | **PFMM** | **1993-2009** | **4.5** | **Software Product line engineering, Software testing** |
| **S168** | 2011 | Journal | 64 | PFMM | 2001-2008 | 3 | Software product line engineering, Software testing |
| **S169** | **2011** | **Journal** | **70** | **Ki** | **1997-2009** | **3.5** | **Global Software Development** |
| **S170** | 2011 | Conference | 40 | PFMM | 1998-2011 | 2 | Software architecture |
| **S171** | **2011** | **Workshop** | **93** | **KC** | **1994-2010** | **3.5** | **Evidence based software engineering** |
| **S172** | 2011 | Conference | 36 | PFMM | 2002-2010 | 3 | Software testing, Requirement engineering |
| **S173** | 2011 | Conference | 56 | PFMM, KC | 1987-2009 | 2 | OTS based software development |
| **S174** | **2011** | **Journal** | **38** | **PFMM, DD** | **1991-2007** | **5** | **software productivity** |
| **S175** | 2011 | Workshop | 25 | KC | 1997-2010 | 3 | Software product management |
| **S176** | **2011** | **Conference** | **13** | **Ki** | **1989-2009** | **4** | **Software testing** |
| **S177** | **2010** | **Journal** | **103** | **KC, Ki** | **2000-2005** | **4** | **software metrics** |
| **S178** | 2010 | Conference | 24 | PFMM | 1999-2010 | 2.5 | Global software development, Configuration management |
| **S179** | **2010** | **Conference** | **77** | **KC** | **1999-2009** | **4** | **Global software development, Agile software development** |
| **S180** | 2010 | Workshop | 13 | PFMM | 1999-2009 | 3 | Software architecture |
| **S181** | **2009** | **Journal** | **35** | **Ki** | **1996-2007** | **5** | **Software testing** |
| **S182** | 2009 | Conference | 46 | N/A | 2000-2008 | 1.5 | Requirement engineering |
| **S183** | 2008 | Journal | 20 | PFMM | 1999-2007 | 2.5 | Software Product line engineering |
| **S184** | 2008 | Conference | 33 | N/A | 1999-2007 | 0.5 | Model driven engineering |
| **S185** | 2007 | Conference | 138 | BKBT, Ki | 2000-2006 | 2 | Software Design |
| **S186** | 2013 | Conference | 53 | PFMM | 1997-2012 | 2.5 | Requirement Engineering |
| **S187** | 2013 | Conference | 27 | PW, KC | 1996-2013 | 3 | Software Process Model |
| **S188** | **2014** | **Conference** | **34** | **PFMM** | **2004-2014** | **3.5** | **Software ecosystems** |
| **S189** | **2014** | **Journal** | **83** | **PFMM, KC** | **1985-2012** | **4** | **Method Engineering** |
| **S190** | 2015 | Conference | 46 | PFMM | 2002-2015 | 2 | Agile software development |
| **S191** | **2013** | **Journal** | **100** | **PFMM** | **1999-2010** | **4** | **Global software development** |
| **S192** | **2012** | **Conference** | **38** | **KC, PFMM** | **2002-2011** | **3.5** | **Integration of Aspect Orientation, Model-Driven Engineering** |
| **S193** | 2012 | Journal | 115 | PFMM | 2003-2011 | 2.5 | Software Process Model |
| **S194** | 2008 | Conference | 35 | Ki | 1996-2007 | 1.5 | Software Testing |
| **S195** | 2014 | Conference | 173 | Ki | 2013-2014 | 2.5 | Software Engineering |
| **S196** | 2013 | Journal | 36 | Ki, KC | 1994-2011 | 5 | component-based software system metrics |
| **S197** | **2013** | **Journal** | **146** | **KC, PFMM** | **1992-2011** | **3.5** | **Software visualization** |
| **S198** | **2011** | **Journal** | **32** | **KC, BKBTK, BTBK** | **2001-2010** | **4** | **Agile software development, Software Product Line** |
| **S199** | **2012** | **Journal** | **70** | **Ki** | **1997-2009** | **4.5** | **Distributed software development** |
| **S200** | **2015** | **Journal** | **113** | **PFMM** | **2006-2014** | **3.5** | **Crowdsourcing** |
| **S201** | **2013** | **Journal** | **130** | **KC, PFMM** | **1996-2011** | **4** | **Software Development** |
| **S202** | **2014** | **Conference** | **19** | **Ki** | **2004-2013** | **4.5** | **Human Computer Interation** |
| **S203** | 2014 | Conference | 51 | BKBTK | 1998-2013 | 2 | Software Requirements |
| **S204** | 2015 | Conference | 28 | KC | 2010-2014 | 2 | Software Ecosystem |
| **S205** | **2010** | **Conference** | **24** | **N/A** | **1999-2010** | **3.5** | **Global software development** |
| **S206** | **2013** | **Conference** | **53** | **KC, BTBK** | **1994-2014** | **4** | **Software Engineering Education** |
| **S207** | 2016 | Journal | 215 | KC | 2012-2015 | 2.5 | Software Development |
| **S208** | 2011 | Workshop | 44 | Ki | 2002-2011 | 2.5 | Software Ecosystem |
| **S209** | **2014** | **Workshop** | **57** | **BKBTK, KBB** | **1990-2013** | **3.5** | **Software Quality** |

| S210 | 2013 | Journal | 120 | BKBTK, PR | 1995-2010 | 4 | Object oriented software |
|---|---|---|---|---|---|---|---|

***PFMM=Peterson *et al.* [1]; KC=Kitchenham & Charters [2]; KBB=Kitchenham *et al.* [3];Ki=Kitchenham *et al.* [4]; BTBK=Budgen *et al.* [7]; KBBTBL=Kitchenham *et al.* [6]; PVK=Peterson *et al.* [10]; KPBBTNL=Kitchenham *et al.* [8]; KDJ=Kitchenham *et al.* [5]; BCR=Basili *et al.* [11]; Bio=Biolchini *et al.* [12]; DDH=;Dyba *et al.* [13]; Petticrew *et al.* [14]**

Table A2. Inclusion statistics for identified databases

| S.no | Online Database | Number | Percentage (%) |
|---|---|---|---|
| 1 | IEEEXplore | 183 | 87.14 |
| 2 | ACM | 177 | 84.28 |
| 3 | Science Direct | 117 | 55.71 |
| 4 | SpringerLink | 98 | 46.66 |
| 5 | Scopus | 95 | 45.23 |
| 6 | Web of Science | 56 | 26.66 |
| 7 | Google Scholar | 51 | 24.28 |
| 8 | Compendex | 44 | 20.95 |
| 9 | Inspec | 36 | 17.14 |
| 10 | Wiley | 29 | 13.80 |
| 11 | CiteSeerX | 10 | 10.0 |
| 12 | Elsevier | 10 | 10.0 |
| 13 | DBLP | 12 | 5.71 |
| 14 | Engineering Village | 6 | 1.85 |
| 15 | JSTOR | 5 | 1.42 |
| 16 | ProQuest | 3 | 1.42 |
| 17 | AIS | 3 | 1.42 |
| 18 | MAS | 3 | 1.42 |
| 19 | Scirus | 3 | 1.42 |
| 20 | ERIC | 2 | 0.95 |
| 21 | EBSCOhost | 2 | 0.95 |
| 22 | SAE | 1 | 0.47 |
| 23 | IET Digital Library | 1 | 0.47 |
| 24 | ArXiv | 1 | 0.47 |
| 25 | Inderscience | 1 | 0.47 |
| 26 | Kluwer Online | 1 | 0.47 |
| 27 | EMBASE | 1 | 0.47 |
| 28 | SciVerse | 1 | 0.47 |

Table A 3. Parameter-wise breakdown of Quality scores

| S.no | Inclusion and exclusion criteria | Search adequacy | Synthesis Method | Quality Criteria | Basic information about primary studies | Quality Sum |
|---|---|---|---|---|---|---|
| **S1** | 1 | 1 | 1 | 0 | 0 | 3 |
| **S2** | 1 | 1 | 0.5 | 0 | 0 | 2.5 |
| **S3** | **1** | **1** | **1** | **0** | **0.5** | **3.5** |
| **S4** | **1** | **1** | **1** | **0** | **0** | **3** |
| **S5** | 0.5 | 0.5 | 0.5 | 0 | 1 | 2.5 |
| **S6** | 1 | 0.5 | 0.5 | 0 | 1 | 3 |
| **S7** | 1 | 1 | 0.5 | 0 | 0.5 | 3 |
| **S8** | 1 | 0.5 | 0.5 | 0 | 0.5 | 2.5 |
| **S9** | **1** | **1** | **1** | **1** | **1** | **5** |
| **S10** | 1 | 0 | 0.5 | 1 | 0.5 | 3 |
| **S11** | 1 | 1 | 1 | 0 | 0 | 3 |
| **S12** | 1 | 1 | 0.5 | 0 | 0.5 | 3 |
| **S13** | 1 | 0 | 1 | 0.5 | 0.5 | 3 |
| **S14** | **1** | **1** | **1** | **1** | **1** | **5** |
| **S15** | **0.5** | **0.5** | **1** | **0.5** | **1** | **3.5** |
| **S16** | 0.5 | 1 | 0.5 | 0 | 0.5 | 2.5 |
| **S17** | **1** | **1** | **1** | **0** | **1** | **4** |
| **S18** | **1** | **1** | **1** | **0** | **0.5** | **3.5** |
| **S19** | 1 | 1 | 0.5 | 0 | 0.5 | 3 |
| **S20** | **1** | **1** | **1** | **0** | **1** | **4** |
| **S21** | 1 | 1 | 0.5 | 0 | 0.5 | 3 |
| **S22** | **1** | **1** | **1** | **0** | **1** | **4** |
| **S23** | **0.5** | **1** | **1** | **1** | **1** | **4.5** |
| **S24** | 0.5 | 0 | 0.5 | 0 | 0 | 1 |
| **S25** | **1** | **1** | **1** | **1** | **1** | **5** |
| **S26** | **1** | **0** | **1** | **1** | **1** | **4** |
| **S27** | 1 | 0 | 0.5 | 0 | 1 | 2.5 |
| **S28** | 0 | 0.5 | 0.5 | 0 | 1 | 2 |
| **S29** | 1 | 0 | 0.5 | 0 | 0.5 | 2 |
| **S30** | 0 | 0 | 0.5 | 0 | 1 | 1.5 |
| **S31** | 0.5 | 1 | 0.5 | 0 | 1 | 3 |
| **S32** | 0 | 1 | 0.5 | 0 | 1 | 2.5 |
| **S33** | 1 | 0.5 | 0.5 | 0 | 0.5 | 2.5 |
| **S34** | **1** | **1** | **0.5** | **0** | **1** | **3.5** |
| **S35** | 1 | 0.5 | 1 | 0 | 0.5 | 3 |
| **S36** | **1** | **1** | **1** | **0.5** | **1** | **4.5** |
| **S37** | 1 | 0.5 | 0.5 | 0.5 | 0.5 | 3 |
| **S38** | 1 | 1 | 0.5 | 0.5 | 0 | 3 |
| **S39** | 0.5 | 1 | 0.5 | 0 | 0 | 2 |
| **S40** | 1 | 0.5 | 1 | 0 | 0 | 2.5 |
| **S41** | **1** | **1** | **0.5** | **1** | **1** | **4.5** |
| **S42** | 0.5 | 1 | 0 | 0 | 1 | 2.5 |
| **S43** | **1** | **1** | **1** | **0.5** | **1** | **4.5** |
| **S44** | 1 | 0.5 | 0.5 | 0 | 0 | 2 |
| **S45** | 1 | 0 | 1 | 0 | 0.5 | 2.5 |
| **S46** | 1 | 0.5 | 0.5 | 0 | 1 | 3 |
| **S47** | **1** | **1** | **1** | **0.5** | **1** | **4.5** |
| **S48** | **1** | **1** | **0.5** | **1** | **1** | **4.5** |
| **S49** | 0.5 | 1 | 0.5 | 0 | 0 | 2 |
| **S50** | 1 | 1 | 1 | 0 | 0 | 3 |
| **S51** | **1** | **1** | **1** | **0.5** | **1** | **4.5** |
| **S52** | **1** | **1** | **1** | **0** | **1** | **4** |
| **S53** | **1** | **1** | **1** | **0.5** | **1** | **4.5** |
| **S54** | **0.5** | **1** | **0.5** | **0** | **1** | **3** |
| **S55** | **1** | **1** | **1** | **0.5** | **1** | **4.5** |
| **S56** | **1** | **1** | **0.5** | **0** | **1** | **3.5** |
| **S57** | 1 | 0 | 0.5 | 0 | 1 | 2.5 |
| **S58** | 0.5 | 1 | 0.5 | 0 | 1 | 3 |
| **S59** | **1** | **1** | **1** | **1** | **1** | **5** |
| **S60** | 0.5 | 1 | 0.5 | 0 | 1 | 3 |
| **S61** | 0.5 | 0 | 0.5 | 0 | 0.5 | 1.5 |
| **S62** | **1** | **1** | **1** | **1** | **1** | **5** |
| **S63** | 1 | 0 | 0 | 0 | 1 | 2 |
| **S64** | 0.5 | 1 | 0.5 | 0 | 1 | 3 |
| **S65** | **1** | **1** | **1** | **1** | **1** | **5** |
| **S66** | **1** | **1** | **1** | **0.5** | **0.5** | **4** |
| **S67** | **1** | **1** | **0.5** | **0** | **1** | **3.5** |
| **S68** | **1** | **1** | **1** | **1** | **1** | **5** |
| **S69** | **1** | **1** | **1** | **1** | **1** | **5** |
| **S70** | 0.5 | 0 | 1 | 0 | 0.5 | 2 |
| **S71** | **1** | **1** | **1** | **1** | **1** | **5** |
| **S72** | 1 | 0 | 0 | 0 | 0.5 | 1.5 |
| **S73** | 0.5 | 1 | 0 | 0 | 0 | 1.5 |
| **S74** | **1** | **1** | **1** | **0.5** | **1** | **4.5** |
| **S75** | 0.5 | 1 | 0.5 | 0 | 0.5 | 2.5 |
| **S76** | **1** | **1** | **0.5** | **1** | **1** | **4.5** |
| **S77** | 1 | 0.5 | 0.5 | 0 | 0.5 | 2.5 |
| **S78** | **1** | **1** | **0.5** | **0** | **1** | **3.5** |
| **S79** | **1** | **1** | **0.5** | **1** | **1** | **4.5** |
| **S80** | 0 | 1 | 0 | 0 | 0 | 1 |
| **S81** | **1** | **1** | **1** | **1** | **1** | **5** |
| **S82** | **1** | **1** | **0.5** | **1** | **1** | **4.5** |
| **S83** | **1** | **1** | **0.5** | **0** | **1** | **3.5** |
| **S84** | **1** | **1** | **0.5** | **0** | **1** | **3.5** |
| **S85** | **1** | **1** | **1** | **1** | **1** | **5** |
| **S86** | 1 | 1 | 0.5 | 0 | 0.5 | 3 |
| **S87** | **1** | **1** | **0.5** | **0** | **1** | **3.5** |
| **S88** | **1** | **1** | **1** | **1** | **1** | **5** |
| **S89** | 1 | 1 | 0.5 | 0 | 0 | 2.5 |
| **S90** | 1 | 1 | 1 | 0 | 0 | 3 |
| **S91** | 0 | 1 | 1 | 0 | 0 | 2 |
| **S92** | 1 | 1 | 1 | 0 | 0 | 3 |
| **S93** | **1** | **1** | **1** | **1** | **1** | **5** |
| **S94** | **1** | **1** | **1** | **1** | **1** | **5** |
| **S95** | **1** | **1** | **1** | **1** | **0.5** | **4.5** |
| **S96** | 0 | 1 | 0.5 | 0 | 0 | 1.5 |
| **S97** | 0 | 1 | 0.5 | 0 | 1 | 2.5 |
| **S98** | **1** | **1** | **1** | **1** | **1** | **5** |
| **S99** | 0 | 0.5 | 0.5 | 0.5 | 0.5 | 2 |
| **S100** | 1 | 1 | 0.5 | 0 | 0 | 2.5 |
| **S101** | 1 | 0.5 | 1 | 0 | 0 | 2.5 |
| **S102** | **1** | **1** | **1** | **0** | **1** | **4** |
| **S103** | **1** | **1** | **0.5** | **0** | **1** | **3.5** |
| **S104** | **1** | **1** | **1** | **0** | **0.5** | **3.5** |
| **S105** | 0 | 0.5 | 0 | 0.5 | 0 | 1 |
| **S106** | 1 | 1 | 0.5 | 0 | 0 | 2.5 |
| **S107** | **1** | **1** | **1** | **0.5** | **0.5** | **4** |
| **S108** | **1** | **1** | **0.5** | **0** | **1** | **3.5** |
| **S109** | 1 | 1 | 0.5 | 0 | 0 | 2.5 |
| **S110** | **1** | **1** | **0.5** | **0** | **1** | **3.5** |
| **S111** | **1** | **1** | **1** | **0** | **1** | **4** |
| **S112** | 1 | 1 | 0.5 | 0 | 0.5 | 3 |
| S113 | 1 | 1 | 1 | 0 | 1 | 4 |
| S114 | 1 | 1 | 1 | 1 | 1 | 5 |
| S115 | 1 | 1 | 1 | 0 | 0.5 | 3.5 |
| S116 | 1 | 1 | 0.5 | 0 | 0.5 | 3 |
| S117 | 1 | 1 | 1 | 0 | 1 | 4 |
| S118 | 1 | 1 | 0.5 | 0.5 | 1 | 4 |
| S119 | 1 | 1 | 1 | 1 | 0.5 | 4.5 |
| S120 | 1 | 1 | 0 | 0.5 | 1 | 3.5 |
| S121 | 1 | 0.5 | 0 | 0 | 1 | 2.5 |
| S122 | 1 | 1 | 1 | 1 | 1 | 5 |
| S123 | 1 | 0.5 | 0 | 0 | 0.5 | 2 |
| S124 | 1 | 1 | 1 | 0 | 1 | 4 |
| S125 | 1 | 1 | 1 | 1 | 1 | 5 |
| S126 | 1 | 1 | 0.5 | 0 | 0.5 | 3 |
| S127 | 1 | 1 | 0.5 | 0 | 1 | 3.5 |
| S128 | 1 | 1 | 0.5 | 0 | 0 | 2.5 |
| S129 | 1 | 1 | 1 | 0 | 1 | 4 |
| S130 | 1 | 1 | 0 | 0 | 0 | 2 |
| S131 | 0.5 | 1 | 1 | 1 | 0 | 3.5 |
| S132 | 1 | 0 | 0.5 | 0 | 1 | 2.5 |
| S133 | 1 | 1 | 0.5 | 0 | 1 | 3.5 |
| S134 | 1 | 1 | 0.5 | 0 | 1 | 3.5 |
| S135 | 1 | 1 | 1 | 1 | 1 | 5 |
| S136 | 1 | 1 | 1 | 0 | 1 | 4 |
| S137 | 1 | 1 | 0.5 | 0 | 1 | 3.5 |
| S138 | 1 | 1 | 1 | 0 | 1 | 4 |
| S139 | 1 | 1 | 0.5 | 0 | 1 | 3.5 |
| S140 | 1 | 1 | 0.5 | 0 | 1 | 3.5 |
| S141 | 1 | 1 | 1 | 0 | 1 | 4 |
| S142 | 1 | 1 | 0 | 1 | 1 | 4 |
| S143 | 1 | 1 | 1 | 1 | 1 | 5 |
| S144 | 1 | 1 | 0 | 1 | 1 | 4 |
| S145 | 0 | 1 | 0 | 0 | 0 | 1 |
| S146 | 1 | 0.5 | 1 | 0 | 1 | 3.5 |
| S147 | 1 | 1 | 0.5 | 0 | 0.5 | 3 |
| S148 | 1 | 1 | 0 | 0 | 0 | 2 |
| S149 | 1 | 0 | 0 | 0 | 1 | 2 |
| S150 | 1 | 1 | 0.5 | 0 | 1 | 3.5 |
| S151 | 1 | 0.5 | 0 | 0 | 0 | 1.5 |
| **S152** | 1 | 0.5 | 0 | 0 | 0 | 1.5 |
| **S153** | **1** | **1** | **1** | **0** | **1** | **4** |
| **S154** | 1 | 1 | 0 | 0 | 0 | 2 |
| **S155** | **1** | **1** | **1** | **0** | **1** | **4** |
| **S156** | **1** | **1** | **1** | **0** | **1** | **4** |
| **S157** | 1 | 1 | 0.5 | 0 | 0.5 | 3 |
| **S158** | **1** | **1** | **1** | **0** | **1** | **4** |
| **S159** | **1** | **1** | **1** | **0** | **1** | **4** |
| **S160** | 1 | 0 | 0.5 | 0 | 0.5 | 2 |
| **S161** | 1 | 0 | 0.5 | 0 | 0 | 1.5 |
| **S162** | **1** | **1** | **0.5** | **0** | **1** | **3.5** |
| **S163** | 1 | 0 | 0.5 | 0 | 0 | 1.5 |
| **S164** | **1** | **1** | **1** | **0** | **0.5** | **3.5** |
| **S165** | **1** | **1** | **1** | **1** | **1** | **5** |
| **S166** | 1 | 1 | 0 | 0 | 0.5 | 2.5 |
| **S167** | **1** | **1** | **1** | **1** | **0.5** | **4.5** |
| **S168** | **1** | **0** | **1** | **0** | **1** | **3** |
| **S169** | **1** | **1** | **0.5** | **0** | **1** | **3.5** |
| **S170** | 1 | 1 | 0 | 0 | 0 | 2 |
| **S171** | **1** | **1** | **1** | **0** | **0.5** | **3.5** |
| **S172** | **1** | **1** | **0.5** | **0** | **0.5** | **3** |
| **S173** | 0.5 | 1 | 0.5 | 0 | 0 | 2 |
| **S174** | **1** | **1** | **1** | **1** | **1** | **5** |
| **S175** | 1 | 1 | 0.5 | 0 | 0.5 | 3 |
| **S176** | **1** | **1** | **1** | **0** | **1** | **4** |
| **S177** | **1** | **0** | **1** | **1** | **1** | **4** |
| **S178** | 0.5 | 1 | 0.5 | 0 | 0.5 | 2.5 |
| **S179** | **1** | **1** | **1** | **0** | **1** | **4** |
| **S180** | 1 | 1 | 0.5 | 0 | 0.5 | 3 |
| **S181** | **1** | **1** | **1** | **1** | **1** | **5** |
| **S182** | 1 | 0.5 | 0 | 0 | 0 | 1.5 |
| **S183** | 0.5 | 1 | 0.5 | 0 | 0.5 | 2.5 |
| **S184** | 0 | 0.5 | 0 | 0 | 0 | 0.5 |
| **S185** | 0 | 1 | 1 | 0 | 0 | 2 |
| **S186** | 0 | 1 | 1 | 0 | 0.5 | 2.5 |
| **S187** | 1 | 1 | 0.5 | 0 | 0.5 | 3 |
| **S188** | **1** | **1** | **0.5** | **0** | **1** | **3.5** |
| **S189** | **1** | **1** | **1** | **0** | **1** | **4** |
| **S190** | 0.5 | 0 | 1 | 0 | 0.5 | 2 |
| **S191** | **1** | **1** | **1** | **0** | **1** | **4** |
| **S192** | **1** | **1** | **0.5** | **0** | **1** | **3.5** |
| **S193** | 0.5 | 0 | 1 | 0 | 1 | 2.5 |
| **S194** | 0.5 | 1 | 0 | 0 | 0 | 1.5 |
| **S195** | 1 | 1 | 0.5 | 0 | 0 | 2.5 |
| **S196** | **1** | **1** | **1** | **1** | **1** | **5** |
| **S197** | **1** | **1** | **1** | **0** | **0.5** | **3.5** |
| **S198** | **1** | **1** | **1** | **0** | **1** | **4** |
| **S199** | **1** | **1** | **1** | **0.5** | **1** | **4.5** |
| **S200** | **1** | **1** | **0.5** | **1** | **0** | **3.5** |
| **S201** | **1** | **1** | **1** | **0** | **1** | **4** |
| **S202** | **1** | **1** | **1** | **0.5** | **1** | **4.5** |
| **S203** | 0.5 | 0.5 | 0.5 | 0 | 0.5 | 2 |
| **S204** | 0.5 | 0.5 | 0.5 | 0 | 0.5 | 2 |
| **S205** | **1** | **1** | **0.5** | **0** | **1** | **3.5** |
| **S206** | **1** | **1** | **1** | **0** | **1** | **4** |
| **S207** | 0.5 | 1 | 0.5 | 0 | 0.5 | 2.5 |
| **S208** | 0.5 | 0.5 | 0.5 | 0 | 1 | 2.5 |
| **S209** | **1** | **1** | **0.5** | **0** | **1** | **3.5** |
| **S210** | **1** | **0** | **1** | **1** | **1** | **4** |